\documentclass[12pt,axodraw]{article}
\usepackage{axodraw}

\newcommand{\ds}{\displaystyle}

\newcommand{\cA}{{\cal A}}
\newcommand{\cB}{{\cal B}}

\newcommand{\cJ}{{\cal J}}

\newcommand{\cP}{{\cal P}}

\def\bc{\begin{center}}
\def\ec{\end{center}}
\def\bi{\begin{itemize}}
\def\ei{\end{itemize}}
\def\be{\begin{equation}}
\def\ee{\end{equation}}
\def\bea{\begin{eqnarray}}
\def\eea{\end{eqnarray}}

\def\bdm{\begin{displaymath}}
\def\edm{\end{displaymath}}
\def\ds{\displaystyle}

\begin{document}

\author{ Rogalyov R.N.\footnote{e-mail: rogalyov@th1.ihep.su }\\
        {\it Institute for High Energy Physics}\\
         {\it Protvino, Moscow region, 142281, Russia}}
\title{ One-Loop Diagrams in Lattice QCD with Wilson Fermions \\
         }

\maketitle
\begin{abstract}
A comprehensive number of integrals emerging in
one-loop computations in a gauge perturbation theory on a lattice
with Wilson fermions at $r=1$ is computed using the 
Burgio--Caracciolo--Pelissetto algorithm and the FORM package.
An explicit analytical form of the recursion relations needed for such computations
is presented.
\end{abstract}

\thispagestyle{empty}

\newpage

\section{Introduction}

Perturbation theory provides a useful tool in the studies 
of the weak coupling limit of field theories on a lattice \cite{Capitani}.
To simplify perturbative computations it is helpful to derive analytical 
expressions for the Feynman integrals.
Typical integrals emerging in the one-loop approximation have the form
\be\label{GenOneLoopInt}
\int_{BZ} {dk \over (2\pi)^4} {\cP(\hat k_{\mu_0},\widehat {(k-p_1)}_{\mu_1},...,\widehat {(k-p_1-...-p_n)}_{\mu_1} ) \over D_{latt}(k) D_{latt}(p_1-k) ... D_{latt}(p_1+...+p_n-k) }
\ee
where $\ds \hat p_\mu = {2\over a} \sin \left( {p_\mu a \over 2}\right)$,
$\ D_{latt}(k)= \sum_{\mu=1}^4 \hat k_\mu^2 +m^2 $ (in the boson case), symbol $\cP $ in the numerator
means ``some polynomial of'' and symbol $BZ$ indicates that integration is performed 
over the Brillouin zone, that is the domain $\ds -\;{\pi\over a}< k_\mu < {\pi\over a}$.

Such integrals  cannot be calculated analytically at finite values of $a$.
In the limit $a\to 0$ they can be evaluated by the Kawai--Nakayama--Seo method \cite{Kawai}.

The above integrand can be represented in the form 
\be\label{Kawai001}
I(k,\tilde p,m^2;a)=I(k,0,0;a)\ +\ (I(k,\tilde p,m^2;a)-I(k,0,0;a)),
\ee
where $\tilde p$ is the set of all relevant external momenta, $m$ is the mass.
$I(k,\tilde p,m^2;a)-I(k,0,0;a)$ has a smooth continuum limit ($pa\to 0$ and $ma\to 0$)
and involves no ultraviolet (UV) divergencies.
It can be computed with some continuum regularization
such as dimensional (DR) or with fictitious mass\footnote{The regularization by fictitious mass
is obtained by adding the term $\mu_R^2 = 2\mu_B^2$ to the denominator of each boson or fermion propagator.} (FMP).

Both $I(k,0,0;a)$ and $(I(k,\tilde p,m^2;a)-I(k,0,0;a))$
involve infrared (IR) divergencies.

Though these IR divergencies cancel each other, IR regularization is needed:
$$
I(k,\tilde p,m^2;a) = \lim_{\mu_R^2 \to 0} I(k,\tilde p,m^2;a,\mu_R^2).
$$
In this work, we use the infrared regularization introduced below in the formulas
(\ref{ScalarBosonPropLatt}) for bosons and (\ref {DenomFermPropLatt}) for fermions.
Thus the sought-for integral 
is represented as the sum of the integral over the Euclidean momentum space
(which is readily calculated by well-known method)
and the "zero-momentum" integral over the Brillouin zone.
Calculation of the latter integrals forms the subject of the present study.

In recent years, considerable study was given to computations with 
rather complicated actions (see, for example \cite{Holger1} and \cite{Holger2}). 
In so doing, one is confronted with an integrand involving products of 
$(\hat k_\mu)^n$ at large values of $n$.
Algorithms for computation of such integrals with both
bosonic and fermionic denominators were proposed in \cite{BCP} and \cite{Melnikov}.

An outline of this paper is as follows.
In this work, the algorithm proposed in \cite{BCP} is employed to obtain 
a comprehensive set of the integrals needed in computations
of various matrix elements. For the reader's convenience, a detailed exposition of the Burgio--Caracciolo--Pelissetto
(BCP) algorithm is given. In Section~1, we deal with the bosonic case.
Making use of the BCP algorithm, we derive an explicit form of the recursion relations:
formulas (\ref{FMR_BasBosInt_FinPart_min}), (\ref{FMR_BasBosInt_FinPart_min_vspom}),
and (\ref{RRforJfuncBosNeg}). 
In Section~2 we describe computations in the fermionic case.
In this Section, we also begin with the exposition of the BCP algorithm 
and use it to find an explicit form of the recursion relations for the 
functions $B(p,q)$ and $J(p,q)$ related to the functions ${\cal F}_\delta(p,q)$ used in \cite{BCP}.
These relations are presented in the Appendices. As a matter of fact,
they provide a computer program for a calculation of the general fermionic integrals (\ref{eq:GenFermInt}).
The results obtained in this way are discussed in the Conclusions.

\section{Boson Integrals}\label{BosonIntegrals}

We compute the bosonic `zero-momentum' integrals of the type (we set $a=1$)
\be\label{BosIntInitDef}
F(q,n_1,n_2,n_3,n_4) = \lim_{\delta\to 0} \int dk\;{(\cos k_1 )^{n_1} (\cos k_2 )^{n_2} (\cos k_3 )^{n_3} (\cos k_4 )^{n_4} \over \Delta_B^{(q+\delta)}}
\ee
where
\be\label{ScalarBosonPropLatt}
\Delta_B = 4+\mu_B^2-\cos(k_1 )-\cos(k_2 )-\cos(k_3 )-\cos(k_4 )
\ee
is the scalar boson propagator,
$\mu_B$ is the infrared regulator mass, and $\delta$ is an infinitesimal
parameter needed for an additional intermediate regularization.
All integrals in one-loop calculations with the boson propagators can be reduced to the integrals of this type.

Since $F(q;n_1,n_2,n_3,n_4) $ is symmetric in the arguments 
$ n_1, n_2, n_3,$ and $ n_4$, we consider only the case
\be 
n_1 \geq n_2 \geq n_3 \geq n_4.
\ee

\subsection{Formulas of Reduction}
A computation of the massless\footnote{Here $\mu_B$ is the regulator mass, 
thus we consider the massless limit at the end of computations.} boson integrals 
over the Brillouin zone is based on the following algorithm:
\bea\label{ReducFormulasGT2}
&\mbox{ if} &\qquad n_4\geq 2\qquad \mbox{ then}  \qquad   F(q,n_1,n_2,n_3,n_4) = F(q,n_1,n_2,n_3,n_4-2)\\ \nonumber
&& -\; {1\over q-1+\delta}  \left((n_4-1) F(q-1,n_1,n_2,n_3,n_4-1) -
(n_4-2) F(q-1,n_1,n_2,n_3,n_4-3)  \right),\\ \nonumber
 &  \mbox{else if}& \quad n_3\geq 2 \qquad \mbox{ then} \qquad  F(q,n_1,n_2,n_3,0) = F(q,n_1,n_2,n_3-2,0)\\ \nonumber
 && -\; {1\over q-1+\delta}  \left((n_3-1) F(q-1,n_1,n_2,n_3-1,0) -
(n_3-2) F(q-1,n_1,n_2,n_3-3,0)  \right),\\ \nonumber
& \mbox{ else if} &\quad n_2\geq 2 \qquad \mbox{ then} \qquad F(q,n_1,n_2,0,0) = F(q,n_1,n_2-2,0,0)\\ \nonumber
&& -\; {1\over q-1+\delta}  \left((n_2-1) F(q-1,n_1,n_2-1,0,0) -
(n_2-2) F(q-1,n_1,n_2-3,0,0)  \right),\\ \nonumber
& \mbox{ else if} &\quad n_1\geq 2 \qquad  \mbox{ then} \qquad F(q,n_1,0,0,0) = F(q,n_1-2,0,0,0)\\ \nonumber
&& -\; {1\over q-1+\delta}  \left((n_1-1) F(q-1,n_1-1,0,0,0) -
(n_1-2) F(q-1,n_1-3,0,0,0)  \right). \nonumber
\eea
\bea\label{ReducFormulasOne}
F(q,n_1,n_2,n_3,1) &=& (4+\mu_B^2) F(q,n_1,n_2,n_3,0) - F(q-1,n_1,n_2,n_3,0) \\ \nonumber
 &-& F(q,n_1+1,n_2,n_3,0) - F(q,n_1,n_2+1,n_3,0) - F(q,n_1,n_2,n_3+1,0), \\ \nonumber
F(q,n_1,n_2,1,0) &=& {1\over 2} \left( (4+\mu_B^2) F(q,n_1,n_2,0,0) - F(q-1,n_1,n_2,0,0)\right. \\ \nonumber
 && \left. - F(q,n_1+1,n_2,0,0) - F(q,n_1,n_2+1,0,0)\right), \\ \nonumber
F(q,n_1,1,0,0) &=& {1\over 3} \left( (4+\mu_B^2) F(q,n_1,0,0,0) - F(q-1,n_1,0,0,0)\right. \\ \nonumber
 && \left. - F(q,n_1+1,0,0,0) \right), \\ \nonumber
F(q,1,0,0,0) &=& {1\over 4} \left( (4+\mu_B^2) F(q,0,0,0,0) - F(q-1,0,0,0,0) \right).  \nonumber
\eea
The above identities can be obtained using integration by parts \cite{BCP};
order $O(\delta^2)$ terms should be omitted. 

Thus we obtain an expression for each integral $F(q,n_1,n_2,n_3,n_4)$
in terms of the functions\footnote{In what follows, $\ds G(q, \mu_B^2) = lim_{\delta\to 0} G_\delta (q,\mu_B^2)$.}
\be
G_\delta (q,\mu_B^2)=\int {dk \over (2\pi)^4 } {1\over (\Delta_B)^{q+\delta}},
\ee
it has the form
\bea\label{BosFtoGgen}
F(q,n_1,n_2,n_3,n_4) &=& \sum_{r=q-n_1-n_2-n_3-n_4}^{q} a_{qr}(\delta,\mu_B^2,\tilde n) 
G_\delta (r,\mu_B^2) \\ \nonumber
&=&
\sum_{r=q-n_1-n_2-n_3-n_4}^{0} a_{qr}(\delta,0,\tilde n) G_\delta (r,0) \;+\;
\sum_{r=1}^{q} a_{qr}(0,\mu_B^2,\tilde n) G(r,\mu_B^2)\;+\; {\cal O}(\mu_B^2), \nonumber
\eea
where $\tilde n$ is short-hand notation for $n_1,n_2,n_3,n_4$.
In this sum, the terms with $r\leq 0$ and those with $r>0$ should be considered separately:
\begin{itemize}
\item  The coefficients $a_{qr}(\delta,\mu_B^2,\tilde n)$ at $r\leq 0$ involve the pole $\ds {1\over \delta}$:
\[
a_{qr}(\delta,\mu_B^2,\tilde n) = a_{qr}^{(sing)}(\mu_B^2,\tilde n)\;{1\over \delta} + a_{qr}^{(reg)}(\mu_B^2,\tilde n) + O(\delta),
\]
so that $ G_\delta (r,0)$ must be expanded to the order ${\cal O}(\delta)$.
Since $ G_\delta (r,\mu_B^2)$ has no infrared divergencies at $r \leq 0$
and the coefficients $a_{qr}(\delta,\mu_B^2,\tilde n)$ are polynomials in $\mu_B^2$, 
the values of $a_{qr}^{(sing)}$ and $a_{qr}^{(reg)}$ should be evaluated at $\mu_B=0$.

\item At $r > 0$, the coefficients $a_{qr}(\delta,\mu_B^2,\tilde n)$ involve no 
poles in $\delta$ and, therefore, $\delta$ can be safely set to zero. However, at $r>0$,
$G_\delta (r,\mu_B^2)$ involves infrared divergencies, so that
the $\mu_B$ dependence of the coefficients $a_{qr}$ should be kept.
\end{itemize}

From these properties it follows that we should compute the quantities 
that appear in the right-hand sides of the formulas
\bea\label{DefJ}
\mbox{for} \quad r\leq 0, \qquad && G_\delta (r,\mu_B^2) = {\cal B}_{-r} + J(r)\delta +O(\delta^2)\;+\; {\cal O}(\mu_B^2), \\[1mm] \nonumber
r=1 \qquad && G (1,\mu_B^2) = J(1)\;+\; {\cal O}(\mu_B^2), \\[1mm] \nonumber
r=2 \qquad && G (2,\mu_B^2) = J(2) + (\ln \mu_B^2 + C)\;+\; {\cal O}(\mu_B^2), \\[1mm] \nonumber
\mbox{for} \quad r > 2, \qquad && G (r,\mu_B^2) = J(r) + D_{r0}(\ln \mu_B^2 + C) + \sum_{k=1}^{r-2} D_{rk}/(\mu_B)^{2k}\;+\; {\cal O}(\mu_B^2),
\eea
where $C=0.577...$ is the Euler-Mascheroni constant, $D_{rn}$ and $J(r)$ are some constants to be determined and 
${\cal B}_r$ are given by 
\be\label{Bdirect}
{\cal B}_r=\lim_{\mu_B\to 0}\ \int_{BZ} {dk\over (2\pi)^4}\; \Delta_B^r
\ee
(note that $r>0$), some of them can be found in Appendix~1.

\subsection{Computation of the Divergent Part \\ (Fictitious Mass Regularization)}

We consider the representation of $G_\delta(q,\mu_B^2)$
in terms of the modified Bessel function:
\be \label{GdeltaReprBessel}
G_\delta (q,\mu_B^2)=\int {dk \over (2\pi)^4 } {1\over (\Delta_B)^{q+\delta}}
={1\over \Gamma(q+\delta)} \int_0^\infty t^{q-1+\delta}\; dt\ \left[ e^{-4t-\mu_B^2 t} I_0^4(t)\right]
\ee
and divide the domain of integration into two parts:
$\int_0^\infty = \int_0^1 + \int_1^\infty $. The integral over
the segment $[0,1]$ converges. The divergent part arises from the
latter integral and can be isolated by subtracting $q-1$ terms of the
asymptotic expansion at $z\to \infty$ of the function 
\be\label{InfeldAsExp0}
\exp(-4z) I_0^4(z) \simeq {1\over (2\pi z)^2}
\left(1+ {b_1\over z} + {b_2\over z^2} + ... \right);
\ee
$b_i$ at $i\leq 20$ are given in Appendix~1.
We isolate the divergent part $\bar G_{div}^{M} (q,\delta,\mu_B)$ as follows:
\bea \label{G_FMRdef1}
G_\delta (q,\mu_B^2)= \bar G_{div}^{M} (q,\delta,\mu_B) + \bar J_\delta(q) &=& {1\over \Gamma(q+\delta)} \left\{ 
\int_0^1 t^{q-1+\delta}\; dt\ \left[ e^{-4t-\mu_B^2 t} I_0^4(t)\right]\right. + \\ \nonumber
&&  + \int_1^\infty t^{q-1+\delta}\; dt\ e^{-\mu_B^2 t} \left[ e^{-4t} I_0^4(t)- {1\over (2\pi t)^2} 
\sum_{n=0}^{q-2} {b_n \over t^n} \right]  \\ \nonumber
&& + \int_1^\infty t^{q-1+\delta}\; dt\ \left. {1\over (2\pi t)^2}\; e^{-\mu_B^2 t} \
\sum_{n=0}^{q-2} {b_n \over t^n}  \right\}, \nonumber
\eea
where the first and second lines are designated by $\bar J_\delta(q)$ and
the third---by $\bar G_{div}^{M}(q,\delta,\mu_B)$:
\bea\label{FMR_DivPart_nonmin}
&& \bar G_{div}^{M}(q,\delta,\mu_B) =
{1\over (2\pi )^2\; \Gamma(q+\delta)} \; \sum_{n=0}^{q-2}\left( \int_0^\infty - \int_0^1 \right)\;
dt \ b_n  t^{q-3-n+\delta}\; e^{-\mu_B^2 t} \\ \nonumber
&& = {1\over (2\pi)^2 \Gamma(q)} 
\left[ \; - \; b_{q-2} l_C
+ \sum_{k=1}^{q-2}  b_{q-k-2} \left({\Gamma (k) \over (\mu_B^2)^k} 
- {1\over k}  \right)\right] + \\ \nonumber
&+& {\delta \over (2\pi)^2 \Gamma(q)} \ \left[ b_{q-2} \left( {1\over 2} l_C^2 + {\pi^2\over 12}
+\psi(q) l_C \right)\right. 
\left. +\sum_{n=1}^{q-2} b_{q-2-n} \left( {\Gamma(n)\over (\mu_B^2)^n }
(\psi(n) - \psi(q) - l_C + C) + {1\over n^2} \right) \right], \nonumber
\eea
where $\ds \psi(n)=\sum_{k=1}^{n-1} {1\over k} \ - C$ and $l_C=(\ln \mu_B^2 +C)$.

In the case of purely boson integrals, $O(\delta)$ terms can be omitted;
however, they are needed  for a computation of the divergent part of the integrals
(\ref{eq:GenFermInt}) involving fermion denominators.

We can also isolate the divergent part in the so called ``minimal way''
\be\label{FMR_DivPart_min}
G_{div}^{M}(q,\delta,\mu_B) = {1\over (2\pi)^2 \Gamma(q)} \left\{
\left[-\; b_{q-2} l_C + \sum_{k=1}^{q-2}  b_{q-k-2} {\Gamma (k) \over (\mu_B^2)^k} 
\right] \ + \right. 
\ee
\[
+ \delta \left. \left[ b_{q-2} \left( {1\over 2} l_C^2 
+ \psi(q) l_C \right) +\sum_{n=1}^{q-2} b_{q-2-n} {\Gamma(n)\over (\mu_B^2)^n }
\left(\psi(n) - \psi(q) - l_C + C \right) \right] \right\}; \nonumber
\]

\be\label{DeltaJ}
\!\! \bar G_{div}^{M}(q,\delta,\mu_B) = G_{div}^{M}(q,\delta,\mu_B) - {1\over (2\pi)^2 \Gamma(q)} 
 \sum_{k=1}^{q-2}  {b_{q-k-2} \over k}
 + {\delta \over (2\pi)^2 \Gamma(q)} \ \left[ b_{q-2} {\pi^2\over 12}
+\sum_{n=1}^{q-2} {b_{q-2-n} \over n^2}\right].
\ee

Note that $J(q)$ that appears in (\ref{DefJ}) is connected with the quantity 
$\bar J(q)=\lim_{\delta\to 0} \bar J_\delta(q)$ (see (\ref{G_FMRdef1})) by the relations
\be
G (q,\mu_B^2) = \lim_{\delta\to 0} G_\delta (q, \mu_B^2) = 
\bar G_{div}^{M}(q,0,\mu_B^2) + \bar J(q) =  G_{div}^{M} (q,0,\mu_B^2) + J(q),
\ee
so that
\be
\bar J(q) - J(q) = {1\over (2\pi)^2 \Gamma(q)} \sum_{k=1}^{q-2}  {b_{q-k-2} \over k}
\ee
and the coefficients $D_{rk}$ from the formula (\ref{DefJ}) are determined from the
the equation (\ref{FMR_DivPart_min}):
\be
D_{rk} = {b_{r-k-2} \Gamma(k)\over (2\pi)^2 \Gamma(r)} \ \ \mbox{at} \ \ 1\leq k\leq r-2;
\qquad D_{r0} = -\;{ b_{r-2} \over (2\pi)^2 \Gamma(r)}.
\ee

\subsection{Computation of the Finite Parts \\ (Fictitious Mass Regularization)}

Using the reduction formulas (\ref{ReducFormulasGT2}) and (\ref{ReducFormulasOne}), 
we obtain expressions for the integrals
$F(q;n_1,n_2,n_3,n_4) \equiv F(q,\tilde n)$ 
in terms of the quantities $J(r), {\cal B}_r$, and $D_{rn}$ determined above (see formula (\ref{BosFtoGgen})).

The next step is to use recursion relations 
making it possible to express $J(r)$ at $r\geq 4$
in terms of the basic boson constants $J(1)$, $J(2)$, and $J(3)$
and at $r\leq 0$ in terms of $J(1)$, $J(2)$, $J(3)$, and $J(0)$.
The recursion relations are obtained by making use of the trivial identity
\be
\Delta_B-4 - \mu_B^2 +\sum_{\mu=1}^{4} \cos(k_\mu)=0.
\ee
Inserting this identity in the integrals
\bea
F(q,1,1,1,1) &=& \int dk\;{\cos(k_1) \cos(k_2) \cos(k_3) \cos(k_4) \over \Delta_B^{(q+\delta)}}  \nonumber
\eea
we arrive at
\be
(4+\mu_B^2) F(q,1,1,1,1)= 4 F(q,2,1,1,1)+4 F(q-1,1,1,1,1),
\ee
Now we express $F(q,1,1,1,1)$ etc. in terms of 
the values $J(r)$ and thus obtain the sought for relations between them.
With these relations, $J(r)$ at $r\leq 0$ and $r\geq 4$ is readily 
expressed in terms of $J(1)$, $J(2)$, and $J(3)$.
We consider FMR with the finite part (\ref{FMR_DivPart_min}) defined in the ``minimal''  way.
Given
\be\label{FMR_DivPart_min0}
G_{div}^{M}(q,\mu_B^2) = {1\over (2\pi)^2 \Gamma(q)} 
\left[b_{q-2} l_C + \sum_{k=1}^{q-2}  b_{q-k-2} {\Gamma (k) \over (\mu_B^2)^k} \right],
\ee
we derive the recursion relations for $J(q)\quad (q>0)$ as follows:

\bea\label{FMR_BasBosInt_FinPart_min}
J(q) &=& {1 \over 384 (q-1) (q-2)^2 (q-3)} \\ \nonumber
&& \left\{ 16 (q-2) (q-3) \left[12 + 25 (q-2) (q-3) \right] J(q-1) \right. \\  \nonumber
&& +\; 4(q-3)^2 \left[ -17 - 35(q-3)^2 \right]\; J(q-2) \\ \nonumber
&& +\; 4\;\left[1+ 5(q-3)^3(q-4)-5(q-3)(q-4)^2 \right] \;J(q-3)\\ \nonumber
&& \left. -(q-4)^4\;J(q-4) \right\}\\ \nonumber
&& +\; {1\over (q-2)}\; D(q)\\ \nonumber
&& -\; {25\over 24 (q-1) (q-2)}\; (2q-5)\; D(q-1)\\ \nonumber
&& + \; {1\over 96 (q-1)(q-2)^2}\; \left[17+105(q-3)^2\right] \; D(q-2)\\ \nonumber
&& +\; {5\over 96 (q-1)(q-2)^2(q-3)}\; \left[-1-4(q-3)^2(q-4)+2(q-4)^2\right]\; D(q-3)\\ \nonumber
&& + \; {5\over 384 (q-1) (q-2)^2 (q-3)}\; (q-4)^3 \; D(q-4);\nonumber
\eea
where $\ds D(q)={b_{q-2}\over (q-1)\!}$; this being so, $D(q)$
satisfies the recurrent relation
\bea\label{FMR_BasBosInt_FinPart_min_vspom}
 D(q) &=& {1\over 384 (q-1) (q-2)^2 (q-3)} \\ \nonumber
&& \left\{  16 (q-2)(q-3) \left[12+25(q-2)(q-3) \right]\; D(q-1) \right. \\ \nonumber
&&  +\; 4(q-3)^2 \left[ -17 - 35(q-3)^2\right]\; D(q-2)\\ \nonumber
&&  +\; 4 \left[1+ 5(q-3)^3(q-4)-5(q-3)(q-4)^2 \right]\; D(q-3)\\ \nonumber
&& \left. -\;(q-4)^4\;D(q-4)\right\};
\eea
with the initial conditions

\bea\label{J0123boson}
J(0)&=& J_0; \\ \nonumber
J(1)&=& 2 Z_0; \\ \nonumber
J(2)&=& {F_0\over (2\pi)^2}; \\ \nonumber
J(3)&=& {Z_1\over 32}\;+\; {1\over (2\pi)^2} {F_0\over 4}
\;-\; {1\over (2\pi)^2}\,{13 \over 48} \;-\; {1\over 128}; \nonumber
\eea
and

\be\label{IniCondforDIVboson}
D(1)=0; \qquad \qquad
D(2)= {1\over (2\pi)^{2}}; \qquad \qquad
D(3)= {1\over (2\pi)^{2}}\ {1\over 4}. 
\ee

Recurrent relations for $J(q)$ at $q<0$ can be
derived by the same token, they have the form
%
%
\bea\label{RRforJfuncBosNeg}
J(q)&=& -\; {1\over q^4}\ \left[-4 \left( 1+5(q+1)q+5(q+1)^2 q^2\right) J(q+1) \right. \\ \nonumber
	&+& 4 (q+1)^2 (17+35(q+1)^2) J(q+2) \\ \nonumber
	&-& 16 (q+2) (q+1) (25 (q+2) (q+1)+12) J(q+3) \\ \nonumber
	&+& 384 (q+3) (q+2)^2 (q+1) J(q+4) \\[1mm] \nonumber
	&+&{2 q^3 (3+5 (q+3)^2 (q+1)) \over (q+4)(q+3)(q+2)(q+1)}\; {\cal B}_{-q} \\[1mm] \nonumber
	&+& 4 \;{  (-40 q^6-330 q^5-985 q^4-1376 q^3 -1015 q^2-410 q-70)\over (q+4)(q+3)(q+2)(q+1)}{\cal B}_{-q-1}\\[1mm] \nonumber
	&+& 8 \;{ (q+1) (105 q^4+788 q^3+1998 q^2+2052 q +788)\over (q+4)(q+3)(q+2)}{\cal B}_{-q-2}\\[1mm] \nonumber
	&+& 32 \;{ (q+2) (q+1) (31 (q+4)^2 -81 (q+3)^2)\over (q+4)(q+3)} {\cal B}_{-q-3} \\[1mm] \nonumber
	&+&\left. 768\; {(q+3) (q+2) (q+1)\over(q+4)} {\cal B}_{-q-4} \right] \qquad \mbox{for}\quad  q\leq -3. \nonumber
\eea
The values of ${\cal B}_q$ can be computed either directly by the formula 
(\ref{Bdirect}) or with the use of the recurrent relations

\bea\label{RecRelForBbosonic}
{\cal B}_{q}&=& {1\over q^4}\left[ -\; 384 (q-1) (q-2)^2 (q-3) \;{\cal B}_{q-4} \right. \\ \nonumber
          &&\qquad +\; 16 (q-1) (q-2) \big(25 (q-1) (q-2) + 12\big) \;{\cal B}_{q-3} \\ \nonumber
          &&\qquad -\; 4 (q-1)^2 \big( 35 (q-1)^2 + 17\big) \;{\cal B}_{q-2} \\ \nonumber
          &&\qquad \left. +\; 4 (q^5 - (q-1)^5) \;  {\cal B}_{q-1} \right]  \nonumber
\eea


with the initial conditions

\be\label{IniCondForBbosonic}
 {\cal B}_{0} = 1; \qquad
 {\cal B}_{1} = 4; \qquad
 {\cal B}_{2} = 18;\qquad
 {\cal B}_{3} = 88.
\ee

The values of $J(-1), J(-2)$, and $J(-3)$ can be determined 
in the same way, the respective identities have the form
\bea 
 J(-4) &=& -9/16 (13/9-781/36 J(-3)+83 J(-2)-108 J(-1)+ 32 J(0) ); \\ \nonumber
 J(-3) &=& 16/27 (-11/2+211/12 J(-2)-157/3 J(-1)+124/3 J(0)+16 J(1));\\ \nonumber
 J(-2) &=& -3 \left(-3/2+{8\over (2\pi )^2} -31/12 J(-1)+13/3 J(0)+ 4 J(1)\right);\\ \nonumber
 J(-1) &=& 144 \left(-1/36-{ 13\over 9 (2\pi)^{2}}+1/36 J(0)+4/3 J(2)-16/3 J(3)\right); \nonumber
\eea

The integrals (\ref{BosIntInitDef}) can also be expressed 
in therms of the quantities $l_C$, and
\bea\label{BasBosConstNum}
Z_0 &\approx& 0.154933390231060214084837208 \\ \nonumber
Z_1 &\approx& 0.107781313539874001343391550 \\ \nonumber
F_0 &\approx& F_0^C -\ln 2 = 4.369225233874758 -\ln 2 \nonumber 
\eea
determined from the relations\footnote{In the review \cite{Capitani}, the constant $F_0^C \approx 4.369225233874758$ 
is designated by $F_0$.}
\bea
F(1,0,0,0,0) &=& 2Z_0+O(\mu_B^2) \\ \nonumber
F(2,0,0,0,0) &=& - \; {l_C\over (2\pi)^2}\;+\;{F_0\over (2\pi)^2} \; +O(\mu_B^2)\\ \nonumber
F(3,0,0,0,0) &=& {1\over (2\pi)^2}\left({1\over 2\;\mu_B^2} - \; {l_C\over 4} - {13\over 48} + {F_0 \over 4} \right)
-{1\over 128}\;+\;{Z_1\over 32} +O(\mu_B^2). \nonumber
\eea
This being so, the initial conditions for the recurrent relations
are given by the formula (\ref{J0123boson}).

It should be noted that $J_0$ does not appear 
in the ultimate expressions for the integrals of the type (\ref{BosIntInitDef}),
therefore, its numerical value is not needed.

%
%
%
%

\subsection{Dimensional Regularization \label{sec:DRbos}}

First we introduce the quantity $\bar J(q;\tilde n)$
analogous to $\bar J_\delta(q)$ defined in (\ref{G_FMRdef1}):
\bea \label{defJbasic1}
 \bar J(q;\tilde n) &=& \lim_{\mu_B \to 0} \lim_{\delta \to 0}{1\over \Gamma(q+\delta)} \left\{ 
\int_0^1 t^{q-1+\delta}\; dt\ \left[ e^{-(4+\mu_B^2) t} {{\cal T}(\tilde n)} \right]\right. + \\ \nonumber
&+& \left. \int_1^\infty t^{q-1+\delta}\; dt\ e^{-\mu_B^2 t} \left[ e^{-4t} {{\cal T}(\tilde n)} - {1\over (2\pi t)^2} 
\sum_{k=0}^{q-2} {b_k (\tilde n) \over t^k} \right]\right\},  \nonumber
\eea
where
\be 
{\cal T}(\tilde n)= \left[ \left( {\partial\over \partial t} \right)^{n_1} I_0(t)\right]\ 
\left[ \left( {\partial\over \partial t} \right)^{n_2} I_0(t)\right]\ 
\left[ \left( {\partial\over \partial t} \right)^{n_3} I_0(t)\right]\ 
\left[ \left( {\partial\over \partial t} \right)^{n_4} I_0(t)\right].
\ee
It represents the finite part of the general boson integral (\ref{BosIntInitDef})
provided that the divergent part is defined by the formula 
similar to (\ref{FMR_DivPart_nonmin}).
We omit here $O(\delta)$ terms because they are only needed for the calculation of fermion integrals
in the fictitious mass regularization. Thus we set $\delta=0$.
Then it should be noted that
\bea 
 \bar J(q;\tilde n) &=& \lim_{\epsilon \to 0} {1\over \Gamma(q)} \left\{ 
\int_0^1 t^{q-1}\; dt\ \left[ e^{-(4-2\epsilon)t} I_0^{-2\epsilon}(t) {{\cal T}(\tilde n)} \right]\right. + \\ \nonumber
&+& \int_1^\infty t^{q-1}\; dt\, \left[ e^{-(4-2\epsilon)t} I_0^{-2\epsilon}(t) {{\cal T}(\tilde n)} - {1\over (2\pi t)^{2-\epsilon}} 
\sum_{k=0}^{q-2} {\tilde b_k(\tilde n) \over t^k} \right],  \nonumber
\eea
where
\be
\tilde b_k (\tilde n) = b_k(\tilde n) -2\epsilon d_k (\tilde n),
\ee
where 
$b_k(\tilde n)$ are the coefficients of the asymptotic expansion at $t \to \infty$
\be
(2\pi t)^2\; e^{-4t} {{\cal T}(\tilde n)} \;\simeq \sum_{k=1}^\infty {b_k (\tilde n) \over t^k} 
\ee
and $d_k(\tilde n)$ are the coefficients of the asymptotic expansion
\be\label{d_coeff_def}
(2\pi t)^2 \; e^{-4t} {{\cal T}(\tilde n)}\;\ln \left[ e^{-t} I_0(t) \sqrt{2\pi t}\right] \simeq 
\sum_{n=1}^\infty {d_n (\tilde n) \over t^n} .
\ee
\vskip 1mm
Now we {\bf define} the general boson integral (\ref{BosIntInitDef}) in the 
dimensional regularization by the formula
\be
F(q;\tilde n) = \bar J(q;\tilde n) + F^{DR}_{div}(q;\tilde n),
\ee
where
\be 
F^{DR}_{div}(q;\tilde n) = \int_1^\infty t^{q-1}\; dt\  {1\over (2\pi t)^{2-\epsilon}} \
\sum_{k=0}^{q-2} {\tilde b_k(\tilde n) \over t^k}
\ee
and the dimensional regularization implies that
\be
\int_1^\infty dt\ t^{n+\epsilon} =0 \quad\mbox{at}\ \ n\neq -1, \qquad \qquad  
\int_1^\infty {dt \over t^{1-\epsilon}} = -\ {1\over \epsilon}.
\ee
This being so,
\be 
F^{DR}_{div}(q;\tilde n) = {1\over (2\pi )^{2}}  {1\over \Gamma(q)} 
\left\{ -\ {1\over \epsilon}\; b_{q-2}(\tilde n) - \ln(2\pi)\, b_{q-2}(\tilde n) +2d_{q-2}(\tilde n) \right\}.
\ee
Now we isolate the "canonical" divergent part in the dimensional regularization; 
that is, 
\be
\bar F^{DR}_{div}(q;\tilde n, \mu^2) =-\; {1\over (2\pi )^{2}} \, {1\over \Gamma(q)} 
\left[ {1\over \epsilon}\;-C+\ln\left({4\pi\over\mu^2}\right)\right] b_{q-2}(\tilde n).
\ee
where $\mu$ is the parameter of dimensional regularization\footnote{In this subsection it is considered that intergation in (\ref{BosIntInitDef}) is performed over the $4-2\epsilon$ dimensional space; the integral
under consideration should be multiplied by $\mu^{2\epsilon}$};
the "canonical" divergent part is needed to compensate for the 
infrared divergent part in the respective continuum integral
with nonvanishing external momenta. We see that
\bea\label{GDR-barGDR}
F^{DR}_{div}(q;\tilde n,) &=& \bar F^{DR}_{div}(q;\tilde n, \mu^2) \\ \nonumber
&+& {1\over (2\pi )^{2}} \, {1\over \Gamma(q)} 
\left[\left(-C+\ln {2\over\mu^2}\right) b_{q-2}(\tilde n)+2d_{q-2}(\tilde n) \right]. \nonumber
\eea
The respective finite parts can be determined by the formula
\be\label{2repr-of-FDR}
F^{DR}(q;\tilde n) = \bar J(q;\tilde n) + F^{DR}_{div}(q;\tilde n) = \bar J^{DR}(q;\tilde n) + \bar F^{DR}_{div}(q;\tilde n),
\ee
\vskip 1mm
Now we express $\bar J^{DR}(q;\tilde n)$ in terms of the quantity $J(q;\tilde n)$
which can be calculated in the fictitious mass regularization by the method described below.
First we note that $\bar J(q;\tilde n)$ is connected with $\bar J^{DR}(q;\tilde n)$ by
the relation
\be
\bar J^{DR}(q;\tilde n)= \bar J(q;\tilde n) + {1\over (2\pi )^{2}} \, {1\over \Gamma(q)} 
\left[\left(-C+\ln {2}\right) b_{q-2}(\tilde n)+2d_{q-2}(\tilde n) \right].
\ee
The relation between $\bar J(q;\tilde n)$ and $J(q;\tilde n)$ is
derived from the formula
\be
F^{FMR}(q;\tilde n) = \bar J(q;\tilde n) + \bar F^M_{div}(q;\tilde n) = J(q;\tilde n) + F^{M}_{div}(q;\tilde n),
\ee
(it is the definition of $J(q;\tilde n)$). From the formula analogous to (\ref{DeltaJ}) it follows that
\bea\label{FFMR-barFFMR}
\bar J(q;\tilde n) &=& J(q;\tilde n) + (F^M_{div}(q;\tilde n) - \bar F^{M}_{div}(q;\tilde n))\\ \nonumber
&=& J(q;\tilde n)\; + \;{1\over (2\pi)^2 \Gamma(q)}  \sum_{k=1}^{q-2}  {b_{q-k-2} (\tilde n)\over k},
\eea
where $\bar F^{M}_{div}(q;\tilde n)$ and $ F^{M}_{div}(q;\tilde n)$ are natural
analogs of the quantities $\bar G^{M}_{div}(q)$ and $ G^{M}_{div}(q)$ introduced 
above. Combining formula (\ref{FFMR-barFFMR}) with (\ref{GDR-barGDR}) and (\ref{2repr-of-FDR}), 
we arrive at
\bea
F^{DR}(q;\tilde n) &=& \bar F^{DR}_{div}(q;\tilde n;\mu)\;+\; \bar J^{DR}(q;\tilde n;\mu),\ \ \ \mbox{where} \\ \nonumber
\bar F^{DR}_{div}(q;\tilde n, \mu) &=& -\ {1\over (2\pi )^{2}} \, {1\over \Gamma(q)} 
\left[ {1\over \epsilon}\;-C+\ln\left({4\pi\over\mu^2}\right)\right] b_{q-2}(\tilde n), \\ \nonumber
\bar J^{DR}(q;\tilde n;\mu) &=&  J(q;\tilde n)\; + \;{1\over (2\pi)^2 \Gamma(q)}  \sum_{k=1}^{q-2}  {b_{q-k-2}(\tilde n) \over k}  \\ \nonumber
&& + {1\over (2\pi )^{2}} \, {1\over \Gamma(q)} 
\left[(-C+\ln 2) b_{q-2}(\tilde n)+2d_{q-2}(\tilde n)\right], \nonumber
\eea
 where $J(q;\tilde n)$ can be calculated as follows:
using the relations (\ref{ReducFormulasGT2}) and 
(\ref{ReducFormulasOne}), $F(q,\tilde n)$ is transformed to 
a linear combination of the quantities $G_\delta(r,\mu_B^2)$ (\ref{BosFtoGgen})
and the substitutions (\ref{DefJ}) are employed. 
In the resulting expression, $\mu_B^{-1}$ and $l_C$
are formally set equal to zero, all that remains represents
the sought-for $J(q;\tilde n)$.

It should be noticed that  $\ds {1\over \epsilon}$ appears in $F^{DR}(q;\tilde n)$
only in the combination $\ds {1\over \epsilon} -\ln 2 - F_0 + \ln \left({4\pi\over \mu^2}\right)$
(the Euler-Mascheroni constant $C$ cancels in the total expression).


\section{Fermion Integrals}

Here we consider the integrals (remember that $a=1$)
\be\label{eq:GenFermInt}
F(p,q;\tilde n)=\lim_{\delta \to 0}\int {d^4k\over (2\pi)^4} 
{\cos^{n_1}(k_1) \cos^{n_2}(k_2) \cos^{n_3}(k_3) \cos^{n_4}(k_4) \over \Delta_B^q
\Delta_F^{p+\delta}}
\ee
where $p>0$, $\delta$ is a regularization parameter, and 
\be\label{DenomFermPropLatt}
\Delta_F = 10-4\;\sum_{\mu=1}^4 \cos(k_\mu) + \sum_{1\leq \mu < \nu \leq 4} \cos(k_\mu) \cos(k_\nu) + \mu_B^2
\ee
is the denominator of the fermionic propagator.
Making use of the recursion relations 
\bea\label{RecFermnnnn}
\hspace*{-5mm} F(p,q,...,l,...) &=&  F(p,q,...,l-2,...) \\ \nonumber
&&\hspace*{-22mm} + \ \mu_B^2 \left(F(p,q,...,l-1,...)-F(p,q,...,l-3,...)\right) \\   \nonumber
&&\hspace*{-22mm} -\ \left(F(p,q-1,...,l-1,...)-F(p,q-1,...,l-3,...)\right) \\ \nonumber
&&\hspace*{-22mm} -\ {q\over p-1+\delta} \; \left(F(p-1,q+1,...,l-1,...) - F(p-1,q+1,...,l-3,...)\right) \\ \nonumber
&&\hspace*{-22mm} -\ {1\over p-1+\delta}\; \left( (l-2) F(p-1,q,...,l-2,...) - (l-3)F(p-1,q,...,l-4,...)\right);
\eea 
\bea\label{RecFerm21} 
F(p,q;n1,n2,n3,2)&=&F(p,q-2,n1,n2,n3,0)-2\,\mu_B^2\,F(p,q-1,n1,n2,n3,0)\\  \nonumber
&&	-2\,F(p-1,q,n1,n2,n3,0)+(4+2\,\mu_B^2+\mu_B^4)\,F(p,q,n1,n2,n3,0)\\  \nonumber
&&	-F(p,q,n1+2,n2,n3,0)-F(p,q,n1,n2+2,n3,0) \\ \nonumber
&& -F(p,q,n1,n2,n3+2,0), \\  \nonumber
F(p,q,n1,n2,n3,1)&=&(\mu_B^2+4)\,F(p,q,n1,n2,n3,0)\\  \nonumber
&&	-F(p,q-1,n1,n2,n3,0)-F(p,q,n1+1,n2,n3,0)\\ \nonumber
&&	-F(p,q,n1,n2+1,n3,0)-F(p,q,n1,n2,n3+1,0)\\ \nonumber
F(p,q;n1,n2,2,0) &=& {1\over 2} \left( F(p,q-2,n1,n2,0,0) - 2 \mu_B^2 F(p,q-1;n1,n2,0,0)\right. \\ \nonumber
&& - 2 F(p-1,q;n1,n2,0,0) + (4+2 \mu_B^2+\mu_B^4) F(p,q,n1,n2,0,0) \\ \nonumber
&& \left. - F(p,q;n1+2,n2,0,0)-F(p,q;n1,n2+2,0,0)\right); \\ \nonumber
F(p,q;n1,n2,1,0) &=& {1\over 2} \left( (\mu_B^2+4) F(p,q;n1,n2,0,0) - F(p,q-1;n1,n2,0,0)\right. \\ \nonumber
&& \left. - F(p,q;n1+1,n2,0,0) - F(p,q;n1,n2+1,0,0) \right); \\ \nonumber
F(p,q;n1,2,0,0) &=& {1\over 3} \Big( F(p,q-2;n1,0,0,0) - 2 \mu_B^2 F(p,q-1;n1,0,0,0) \\ \nonumber
&& - 2 F(p-1,q;n1,0,0,0) + (4 + 2 \mu_B^2 + \mu_B^4) F(p,q;n1,0,0,0)\\ \nonumber
&& - F(p,q;n1+2,0,0,0) \Big); \\ \nonumber
F(p,q;n1,1,0,0) &=& {1\over 3} \Big( (\mu_B^2+4) F(p,q;n1,0,0,0) - F(p,q-1;n1,0,0,0)\\ \nonumber
&&  - F(p,q;n1+1,0,0,0)\Big); \\ \nonumber
F(p,q;2,0,0,0) &=& {1\over 4} \Big( F(p,q-2;0,0,0,0) - 2 \mu_B^2 F(p,q-1;0,0,0,0)\\ \nonumber
&& - 2 F(p-1,q,0,0,0,0) + ( 4 + 2 \mu_B^2 + \mu_B^4) F(p,q;0,0,0,0)); \\  \nonumber
F(p,q;1,0,0,0) &=& {1\over 4} \Big( (\mu_B^2 + 4) F(p,q;0,0,0,0) -F (p,q-1;0,0,0,0)\Big);\nonumber
\eea 

we can express the quantities (\ref{eq:GenFermInt}) in terms of the integrals
\be\label{eq:BasFermInt}
G_\delta(p,q;\mu_B^2)=\int {d^4k\over (2\pi)^4} {1 \over \Delta_B^q \Delta_F^{p+\delta}}
\ee
as follows:
\be\label{FtoGfirst}
F(p,q;n_1, n_2, n_3, n_4)=\sum_{r=p-n}^{p} \ \sum_{s=q+2p-n-2r}^{q+p-r} C_{pq;\tilde n}^{rs}(\mu_B^2,\delta) G_\delta(r,s;\mu_B^2),
\ee
where $n=n_1+n_2+n_3+n_4$. The coefficients $C_{pq;n_1 n_2 n_3 n_4}^{rs}$ 
are polynomials in $\mu_B^2$ (however, at $r+s\leq 2$ $C_{pq;\tilde n}^{rs}(\mu_B^2,\delta)$
can be replaced by $C_{pq;\tilde n}^{rs}(0,\delta)$); at $r\leq 0$ they involve
the singularity $\ds {1\over \delta}$; that is, they can be represented in the form
\be 
C_{pq;n_1 n_2 n_3 n_4}^{rs} = {1\over \delta} S_{pq;n_1 n_2 n_3 n_4}^{rs}(\mu_B^2) + R_{pq;n_1 n_2 n_3 n_4}^{rs}(\mu_B^2)
+ O(\delta),
\ee
where $S_{pq;n_1 n_2 n_3 n_4}^{rs}=0$ at $r\leq 0$ or $p\leq 0$.
A straightforward calculation of both $S_{pq;n_1 n_2 n_3 n_4}^{rs}(\mu_B^2)$ 
and $R_{pq;n_1 n_2 n_3 n_4}^{rs}(\mu_B^2)$ by employing the above relations is rather simple.

To compute the basic integrals $G_\delta(p,q;\mu_B^2)$, we consider the cases
$p>0$ and $p\leq 0$ separately. At $p>0$, only zeroth order of the expansion 
of $G_\delta(p,q)$ in a power series in $\delta$ gives a nonvanishing contribution, 
whereas at $p\leq 0$ one should also keep the term linear in $\delta$.

It is convenient\footnote{It should be noted that the functions $G_\delta(p,q)$ are related to the functions ${\cal F}_\delta(p,q)$ used in \cite{BCP} by the formulas ${\cal F}_\delta(p,q)= 2^{-p-q} G_\delta(p,q)$.} to represent $G_\delta(p,q)$ in the form 
\bea\label{GdeltaExpansion}
G_\delta(p,q)&=& D(p,q;\mu_B^2) + B(p,q) + \delta \;(L(p,q;\mu_B^2)+J(p,q)) + O(\delta^2) , \qquad p\leq 0; \\ \nonumber
G_\delta(p,q)&=& D(p,q;\mu_B^2) + J(p,q) + O(\delta), \qquad p > 0.  \nonumber
\eea
where the quantities $B(p,q)$, $D(p,q;\mu_B^2)$, $L(p,q;\mu_B^2)$, and $J(p,q)$ 
are defined as follows:\\
$D(p,q;\mu_B^2)+\delta L(p,q;\mu_B^2)$ is 
the divergent part of $G_\delta(p,q;\mu_B^2)$ at $p\leq 0$ (up to terms $O(\delta^2)$),\\
$D(p,q;\mu_B^2)$ is 
the divergent part of $G_\delta(p,q;\mu_B^2)$ at $p > 0$ (up to terms $O(\delta)$),\\
$B(p,q)+ \delta \;J(p,q)$ and $J(p,q)$ are the respective finite parts\footnote{Note
that $J(p,q)$ designates the finite part in the order $O(1)$ at $p>0$ and in the order
$O(\delta$ at $p\leq 0$}.

The finite and divergent parts are unambiguously fixed by the requirement that 
$D(p,q;\mu_B^2)$ and $L(p,q;\mu_B^2)$ can be represented in the form
\bea\label{DPcoeffOne}
D(p,q;\mu_B^2) &=& D_{0}(p,q) (\ln \mu_B^2 +C) + \sum_{r=1}^{p+q-2} {D_{r}(p,q) \over (\mu_B^2)^r} \\ \nonumber
L(p,q;\mu_B^2) &=& {1\over 2} L_{0}^{(2)}(p,q) (\ln \mu_B^2 +C)^2 + L_{0}^{(1)}(p,q) (\ln \mu_B^2 +C) \\ \nonumber
&& + \sum_{r=1}^{p+q-2} {L_{r}^{(2)}(p,q) (\ln \mu_B^2 + C) \over (\mu_B^2)^r} 
+\sum_{r=1}^{p+q-2} {L_{r}^{(1)}(p,q) \over (\mu_B^2)^r}. \nonumber
\eea

In the domain $p\leq 0$ we also use the quantities
\be\label{BBandJJat_pleq0}
\cB(p,q;\mu_B^2)= B(p,q)+ D(p,q;\mu_B^2) \qquad \mbox{and} \qquad
\cJ(p,q;\mu_B^2) = L(p,q;\mu_B^2)+J(p,q).  \nonumber
\ee
At $p> 0$, 
\be\label{BBandJJat_pgeq0}
\cB(p,q;\mu_B^2)= B(p,q) = 0 \qquad \mbox{and} \qquad
\cJ(p,q;\mu_B^2) = D(p,q;\mu_B^2)+J(p,q).  \nonumber
\ee
Note that, at $q < 2-p,\ \ $ $D(p,q,\mu_B^2)=L(p,q,\mu_B^2)=0$, thus
$\cB(p,q;\mu_B^2)=B(p,q)$ and $\cJ(p,q;\mu_B^2)=J(p,q)$ and one can use both
designations.


\subsection{Divergent Part in the Fictitious Mass Regularization \label{DPFIFMR}}


First we note that (symbol ${\cal D\!P}$ means `divergent part of')
\bea\label{CalcDPcoeff}
p\leq 0, q\geq2-p \qquad  &&  D (p,q;\mu_B^2) = {\cal D\!P} \int {dk\over (2\pi)^4} \; {\Delta_F^{-p} \over \Delta_B^{q}}, \\ \nonumber
p = 0, q\geq2 \qquad  &&  L (0,q;\mu_B^2) = {d\over d\delta}\left|_{\delta=0} {\cal D\!P}
\int {dk\over (2\pi)^4}\; { 1 \over \Delta_B^{q+\delta}} \right. \\ \nonumber
&& +\sum_{l=1}^{q-2} {(-1)^l\over l} \; 
{\cal D\!P} \int {dk\over (2\pi)^4}\; {\Delta^l \over \Delta_B^{l+q}} , \\ \nonumber
p>0, q\geq2-p \qquad  &&  D (p,q;\mu_B^2) = \sum_{l=0}^{p+q-2} {(-1)^l\; (p+l-1)!\over l! (p-1)!} 
{\cal D\!P} \int {dk\over (2\pi)^4} \; {\Delta^l \over \Delta_B^{p+q+l}},  \nonumber
\eea

The divergent parts of $\delta$-independent integrals 
in formulas (\ref{CalcDPcoeff}) can be 
calculated as follows. First one employs the recursion relations 
(\ref{ReducFormulasGT2}) and (\ref{ReducFormulasOne})
for boson integrals to transform the integrand to a linear combination 
of the basic boson integrals (\ref{GdeltaReprBessel}) and then 
evaluates the divergent part of each integral by the formula 
\be	
{\cal D\!P} \int {dk\over (2\pi)^4} \; {1 \over \Delta_B^{q}}
= {1\over (2\pi)^2 \Gamma(q)} 
\left[-\; b_{q-2} l_C + \sum_{k=1}^{q-2}  b_{q-k-2} {\Gamma (k) \over (\mu_B^2)^k} 
\right] 
\ee
(see derivation of the formula (\ref{FMR_DivPart_min})), according to the MS prescription
in the FMR.
The $\delta$-dependent divergent parts that appears in formula (\ref{CalcDPcoeff})
can determined by the same token, however, with the use of the formula
\bea 
&& {d\over d\delta}\left|_{\delta=0} {\cal D\!P} 
\int {dk\over (2\pi)^4}\; { 1 \over \Delta_B^{q+\delta}} \right. = \\ \nonumber
&& ={1\over (2\pi)^2 \Gamma(q)}
  \left[ b_{q-2} \left( {1\over 2} l_C^2 
+ \psi(q) l_C \right) +\sum_{n=1}^{q-2} b_{q-2-n} {\Gamma(n)\over (\mu_B^2)^n }
\left(\psi(n) - \psi(q) - l_C + C \right) \right];  \nonumber
\eea
(see  (\ref{FMR_DivPart_min})). 
The divergent parts at $p\leq 0$ can be obtained by the recursion relations, see below.

The divergent parts $D(p,q;\mu_B^2)$ and $L(p,q;\mu_B^2)$ introduced in the
formula (\ref{GdeltaExpansion}) are presented in Appendix~2 at $p+q\leq 8$;
at other values it can be readily calculated by the above formulas.

\subsection{Finite Parts}

Given the divergent part, we use the recurrent relations (\ref{RecFermnnnn})
and (\ref{RecFerm21}) to express any integral of the type (\ref{eq:GenFermInt})
in terms of the functions $B(p,q)$ and $J(p,q)$, which, in their turn, can be found
by making use of the two types of the recursion relations \cite{BCP}.

The relations of the first type (the so called $T$-identities) can be obtained 
by inserting the expression $ \Delta_B-4 - \mu_B^2 +\sum_{\mu=1}^{4} \cos(k_\mu)$
which is identically equals zero, in the integrand
\bea
F(q,1,1,1,1) &=& \int dk\;{\cos(k_1) \cos(k_2) \cos(k_3) \cos(k_4) \over\Delta_F^{(p+\delta)}\; \Delta_B^{q}}  \nonumber
\eea
The relations of the second type ($S$-identities) can be obtained by inserting 
$$
10-4\;\sum_{\mu=1}^4 \cos(k_\mu) + \sum_{1\leq \mu < \nu \leq 4} \cos(k_\mu) \cos(k_\nu) + \mu_B^2 - \Delta_F,
$$
which is also identically equals zero, in the same integrand.
In so doing, we arrive at
\bea \label {TandSidentities}
(4+\mu_B^2) F(p,q,1,1,1,1) - F(p,q-1,1,1,1,1) - 4 F(p,q,2,1,1,1)) &=& 0, \\ \nonumber
(4+2 \mu_B^2+\mu_B^4) F(p+1,q,1,1,1,1) - 2 \mu_B^2 F(p+1,q-1,1,1,1,1)&& \\ \nonumber
- 4 F(p+1,q,3,1,1,1) + F(p+1,q-2,1,1,1,1)-2 F(p,q,1,1,1,1) &=& 0.
\eea
Then we express the functions $F(p,q;\tilde n)$ emerging here in terms of $B(p,q)$
and $J(p,q)$ and obtain the sought for identities. Due to the terms singular in $\delta$,
the cases $p=1,2,3$ should be considered separately and, 
in the case $p\leq 0$, one should remember that only the order $O(\delta)$ part is nontrivial.

In what follows, we indicate how to use the explicit form of the derived recurrent relations
for the $\cB$ and $\cJ$ functions presented in Appendices 3-11 in order to compute\footnote{For more detail, 
see http://www.lattice.itep.ru/$\sim$pbaivid/lattpt/ or contact me via e-mail
}
$B(p,q)$ and $J(p,q)$ at $p<9$ and any values of $q$.
Note that some of the relations deal with $B(p,q)$ and $J(p,q)$,
whereas the other---with $\cB(p,q)$ and $\cJ(p,q)$.
In the appendices, $B(p,q)$ and $J(p,q)$ are designated by
{\tt B(p,q)} and {\tt J(p,q)}, respectively, whereas $\cB(p,q)$ and $\cJ(p,q)$
are designated by {\tt BB(p,q)} and {\tt JJ(p,q)}.

Both the divergent and the finite parts as well as the recurrent relations were 
obtained using the FORM \cite{Vermaseren} and (partially) REDUCE \cite{HEARN} packages.

\subsubsection{$\cB$ functions}

First we compute the $\cB$ functions defined by the formula
(\ref{BBandJJat_pleq0}) (see also (\ref {GdeltaExpansion})). 

\bi 
\item{\bf $\cB$ functions at $p=0$, $q>0$}
\ei
\be
\cB(0,q,\mu_B^2) = G_{div}^{M}(q,\mu_B^2)+J(q), 
\ee
where $G_{div}^{M}(q)$ is defined by the equation (\ref{FMR_DivPart_min0})
($G_{div}^{M}(q)\neq 0$ only at $q\geq 2$)\\
and $J(q)$ at $q\geq 4$---by the recurrent relations\footnote{Note that $J(q)$
has nothing to do with $J(p,q)$ or $J(0,q)$.} (\ref{FMR_BasBosInt_FinPart_min}).
The initial conditions are provided by $J(q)$ (and, therefore, $\cB(0,q,\mu_B^2)$) 
at $1 \leq q \leq 3$ (see the formulas (\ref{J0123boson}), (\ref{IniCondforDIVboson}), 
and (\ref{BasBosConstNum})).

\bi 
\item{\bf $\cB$ functions at $p=0$, $q\leq0$}
\ei
\be
\cB(0,q,\mu_B^2) = \cB_{-q}, 
\ee
where $\cB_{-q}$ are defined in the formula (\ref{Bdirect})
and can also be determined from the recurrent relations (\ref{RecRelForBbosonic})
with the initial conditions (\ref{IniCondForBbosonic}) 
(some values are given in Appendix~1).

\bi 
\item{\bf $\cB$ functions at $p\leq -1$, $q\leq 0$}
\ei
In this domain, $\cB(p,q;\mu_B^2)$ functions
involve no divergencies: $\cB(p,q;\mu_B^2) = B(p,q) = \cB(p,q;0)$. Provided that
$B(0,q) = \cB_{-q}$ at $q\leq 0$ are known, $B(p,q)$ at $p<0, q\leq 0$
can be determined by the recurrent relations 
for the domain $p<0, q \geq 4$ (see below), however, with $\mu_B=0$,
and, therefore, with $B(p,q)$ instead of $\cB(p,q)$.

\bi 
\item{\bf $\cB$ functions over the strip $p\leq -1$, $1\leq q \leq 3$}
\ei

In this domain, $\cB(p,q) = B(p,q)$ unless $(p,q)=(-1,3)$.

\noindent $\cB$ functions at $-2\leq p \leq -1$, $1\leq q \leq 3$ can be 
calculated by the formula 
\be
\cB(p,q)=\int dk\ { \Delta_F^{-p}(k,\mu_B^2) \over \Delta^q_B(k,\mu_B^2)},
\ee
the result is as follows:
\bea\label{Binput}
B(-1,1) &=& 1 + 12 \;Z_1;\\ \nonumber
B(-1,2) &=& 4 \;Z_0;\\ \nonumber
B(-1,3) &=& {1 \over (2\pi)^2}\ \left({1\over 2} + F_0  \right) \; + \; {1\over 2} \;Z_0,
\qquad \qquad \left[\cB (-1,3) = B(-1,3) - {l_C \over (2\pi)^2}\right]; \\[1mm] \nonumber
B(-2,1) &=&   188/3 - \;{1 \over (2\pi)^2}\; {736\over 9}  - {632\over 3} \;Z_1 - {224\over 3} \;Z_0;\\ \nonumber
B(-2,2) &=&  - 11 + \;{24 \over (2\pi)^2} + 114 \;Z_1 + 24 \;Z_0;\\ \nonumber
B(-2,3) &=&   {3\over 4} - 3 \;Z_1 + 6 \;Z_0. \nonumber
\eea

\begin{figure}
\bc
\begin{picture}(400,400)(0,0)
\LongArrow(0,200)(400,200)
\LongArrow(200,0)(200,400)
%
\SetColor{Orange}
\Line(40,400)(400,40)
\SetColor{Blue}
%
%
\Vertex(260,280){2}
\Vertex(240,300){2}
\Vertex(260,300){2}
\Vertex(220,320){2}
\Vertex(240,320){2}
\Vertex(260,320){2}
\Vertex(200,340){2}
\Vertex(220,340){2}
\Vertex(240,340){2}
\Vertex(260,340){2}
\Vertex(200,360){2}
\Vertex(220,360){2}
\Vertex(240,360){2}
\Vertex(260,360){2}
\Vertex(200,380){2}
\Vertex(220,380){2}
\Vertex(240,380){2}
\Vertex(260,380){2}
\Vertex(20,220){2}
\Vertex(40,220){2}
\Vertex(60,220){2}
\Vertex(80,220){2}
\Vertex(100,220){2}
\Vertex(20,240){2}
\Vertex(40,240){2}
\Vertex(60,240){2}
\Vertex(80,240){2}
\Vertex(100,240){2}
\Vertex(20,260){2}
\Vertex(40,260){2}
\Vertex(60,260){2}
\Vertex(80,260){2}
\Vertex(100,260){2}
\Vertex(200,20){2}
\Vertex(220,20){2}
\Vertex(240,20){2}
\Vertex(260,20){2}
\Vertex(200,40){2}
\Vertex(220,40){2}
\Vertex(240,40){2}
\Vertex(260,40){2}
\Vertex(200,60){2}
\Vertex(220,60){2}
\Vertex(240,60){2}
\Vertex(260,60){2}
%
%
\SetColor{Green}
\Vertex(20,20){2}
\Vertex(40,20){2}
\Vertex(60,20){2}
\Vertex(80,20){2}
\Vertex(100,20){2}
\Vertex(120,20){2}
\Vertex(140,20){2}
\Vertex(160,20){2}
\Vertex(180,20){2}
\Vertex(20,40){2}
\Vertex(40,40){2}
\Vertex(60,40){2}
\Vertex(80,40){2}
\Vertex(100,40){2}
\Vertex(120,40){2}
\Vertex(140,40){2}
\Vertex(160,40){2}
\Vertex(180,40){2}
\Vertex(20,60){2}
\Vertex(40,60){2}
\Vertex(60,60){2}
\Vertex(80,60){2}
\Vertex(100,60){2}
\Vertex(120,60){2}
\Vertex(140,60){2}
\Vertex(160,60){2}
\Vertex(180,60){2}
\Vertex(20,80){2}
\Vertex(40,80){2}
\Vertex(60,80){2}
\Vertex(80,80){2}
\Vertex(100,80){2}
\Vertex(120,80){2}
\Vertex(140,80){2}
\Vertex(160,80){2}
\Vertex(180,80){2}
\Vertex(20,100){2}
\Vertex(40,100){2}
\Vertex(60,100){2}
\Vertex(80,100){2}
\Vertex(100,100){2}
\Vertex(120,100){2}
\Vertex(140,100){2}
\Vertex(160,100){2}
\Vertex(180,100){2}
\Vertex(20,120){2}
\Vertex(40,120){2}
\Vertex(60,120){2}
\Vertex(80,120){2}
\Vertex(100,120){2}
\Vertex(120,120){2}
\Vertex(140,120){2}
\Vertex(160,120){2}
\Vertex(180,120){2}
\Vertex(20,140){2}
\Vertex(40,140){2}
\Vertex(60,140){2}
\Vertex(80,140){2}
\Vertex(100,140){2}
\Vertex(120,140){2}
\Vertex(140,140){2}
\Vertex(160,140){2}
\Vertex(180,140){2}
\Vertex(20,160){2}
\Vertex(40,160){2}
\Vertex(60,160){2}
\Vertex(80,160){2}
\Vertex(100,160){2}
\Vertex(120,160){2}
\Vertex(140,160){2}
\Vertex(160,160){2}
\Vertex(180,160){2}
\Vertex(20,180){2}
\Vertex(40,180){2}
\Vertex(60,180){2}
\Vertex(80,180){2}
\Vertex(100,180){2}
\Vertex(120,180){2}
\Vertex(140,180){2}
\Vertex(160,180){2}
\Vertex(180,180){2}
\Vertex(20,200){2}
\Vertex(40,200){2}
\Vertex(60,200){2}
\Vertex(80,200){2}
\Vertex(100,200){2}
\Vertex(120,200){2}
\Vertex(140,200){2}
\Vertex(160,200){2}
\Vertex(180,200){2}
\Vertex(20,280){2}
\Vertex(40,280){2}
\Vertex(60,280){2}
\Vertex(80,280){2}
\Vertex(100,280){2}
\Vertex(120,280){2}
\Vertex(140,280){2}
\Vertex(160,280){2}
\Vertex(180,280){2}
\Vertex(20,300){2}
\Vertex(40,300){2}
\Vertex(60,300){2}
\Vertex(80,300){2}
\Vertex(100,300){2}
\Vertex(120,300){2}
\Vertex(140,300){2}
\Vertex(160,300){2}
\Vertex(180,300){2}
\Vertex(20,320){2}
\Vertex(40,320){2}
\Vertex(60,320){2}
\Vertex(80,320){2}
\Vertex(100,320){2}
\Vertex(120,320){2}
\Vertex(140,320){2}
\Vertex(160,320){2}
\Vertex(180,320){2}
\Vertex(20,340){2}
\Vertex(40,340){2}
\Vertex(60,340){2}
\Vertex(80,340){2}
\Vertex(100,340){2}
\Vertex(120,340){2}
\Vertex(140,340){2}
\Vertex(160,340){2}
\Vertex(180,340){2}
\Vertex(20,360){2}
\Vertex(40,360){2}
\Vertex(60,360){2}
\Vertex(80,360){2}
\Vertex(100,360){2}
\Vertex(120,360){2}
\Vertex(140,360){2}
\Vertex(160,360){2}
\Vertex(180,360){2}
\Vertex(20,380){2}
\Vertex(40,380){2}
\Vertex(60,380){2}
\Vertex(80,380){2}
\Vertex(100,380){2}
\Vertex(120,380){2}
\Vertex(140,380){2}
\Vertex(160,380){2}
\Vertex(180,380){2}
\Vertex(280,20){2}
\Vertex(300,20){2}
\Vertex(320,20){2}
\Vertex(340,20){2}
\Vertex(360,20){2}
\Vertex(380,20){2}
\Vertex(280,40){2}
\Vertex(300,40){2}
\Vertex(320,40){2}
\Vertex(340,40){2}
\Vertex(360,40){2}
\Vertex(380,40){2}
\Vertex(280,60){2}
\Vertex(300,60){2}
\Vertex(320,60){2}
\Vertex(340,60){2}
\Vertex(360,60){2}
\Vertex(380,60){2}
\Vertex(280,80){2}
\Vertex(300,80){2}
\Vertex(320,80){2}
\Vertex(340,80){2}
\Vertex(360,80){2}
\Vertex(380,80){2}
\Vertex(280,100){2}
\Vertex(300,100){2}
\Vertex(320,100){2}
\Vertex(340,100){2}
\Vertex(360,100){2}
\Vertex(380,100){2}
\Vertex(280,120){2}
\Vertex(300,120){2}
\Vertex(320,120){2}
\Vertex(340,120){2}
\Vertex(360,120){2}
\Vertex(380,120){2}
\Vertex(280,140){2}
\Vertex(300,140){2}
\Vertex(320,140){2}
\Vertex(340,140){2}
\Vertex(360,140){2}
\Vertex(380,140){2}
%
\Vertex(280,200){2}
\Vertex(300,200){2}
\Vertex(320,200){2}
\Vertex(340,200){2}
\Vertex(360,200){2}
\Vertex(380,200){2}
\Vertex(280,220){2}
\Vertex(300,220){2}
\Vertex(320,220){2}
\Vertex(340,220){2}
\Vertex(360,220){2}
\Vertex(380,220){2}
\Vertex(280,240){2}
\Vertex(300,240){2}
\Vertex(320,240){2}
\Vertex(340,240){2}
\Vertex(360,240){2}
\Vertex(380,240){2}
\Vertex(280,260){2}
\Vertex(300,260){2}
\Vertex(320,260){2}
\Vertex(340,260){2}
\Vertex(360,260){2}
\Vertex(380,260){2}
\Vertex(280,280){2}
\Vertex(300,280){2}
\Vertex(320,280){2}
\Vertex(340,280){2}
\Vertex(360,280){2}
\Vertex(380,280){2}
\Vertex(280,300){2}
\Vertex(300,300){2}
\Vertex(320,300){2}
\Vertex(340,300){2}
\Vertex(360,300){2}
\Vertex(380,300){2}
\Vertex(280,320){2}
\Vertex(300,320){2}
\Vertex(320,320){2}
\Vertex(340,320){2}
\Vertex(360,320){2}
\Vertex(380,320){2}
\Vertex(280,340){2}
\Vertex(300,340){2}
\Vertex(320,340){2}
\Vertex(340,340){2}
\Vertex(360,340){2}
\Vertex(380,340){2}
\Vertex(280,360){2}
\Vertex(300,360){2}
\Vertex(320,360){2}
\Vertex(340,360){2}
\Vertex(360,360){2}
\Vertex(380,360){2}
\Vertex(280,380){2}
\Vertex(300,380){2}
\Vertex(320,380){2}
\Vertex(340,380){2}
\Vertex(360,380){2}
\Vertex(380,380){2}
\SetColor{Gray}
\Vertex(200,80){2}
\Vertex(220,80){2}
\Vertex(240,80){2}
\Vertex(260,80){2}
\Vertex(200,100){2}
\Vertex(220,100){2}
\Vertex(240,100){2}
\Vertex(260,100){2}
\Vertex(200,120){2}
\Vertex(220,120){2}
\Vertex(240,120){2}
\Vertex(200,140){2}
\Vertex(220,140){2}
\Vertex(240,140){2}
\Vertex(200,160){2}
\Vertex(200,180){2}
\Vertex(260,180){2}
%
%
\Vertex(280,160){2}
\Vertex(300,160){2}
\Vertex(320,160){2}
\Vertex(340,160){2}
\Vertex(360,160){2}
\Vertex(380,160){2}
\Vertex(280,180){2}
\Vertex(300,180){2}
\Vertex(320,180){2}
\Vertex(340,180){2}
\Vertex(360,180){2}
\Vertex(380,180){2}
%
%
\Vertex(160,220){2}
\Vertex(160,240){2}
\Vertex(160,260){2}
\Vertex(120,220){2}
\Vertex(120,240){2}
\Vertex(120,260){2}
\Vertex(140,220){2}
\Vertex(140,240){2}
\Vertex(140,260){2}
\Vertex(260,200){2}
\Vertex(240,220){2}
\Vertex(260,220){2}
\Vertex(220,240){2}
\Vertex(240,240){2}
\Vertex(260,240){2}
\Vertex(200,260){2}
\Vertex(220,260){2}
\Vertex(240,260){2}
\Vertex(260,260){2}
\Vertex(200,280){2}
\Vertex(220,280){2}
\Vertex(240,280){2}
\Vertex(200,300){2}
\Vertex(220,300){2}
\Vertex(200,320){2}
%
\SetColor{Red}
\Vertex(180,240){2}
\Vertex(200,220){2}
\Vertex(220,220){2}
\Vertex(220,200){2}
\Vertex(240,200){2}
\Vertex(220,180){2}
\Vertex(240,180){2}
\Vertex(220,160){2}
\Vertex(240,160){2}
\Vertex(260,160){2}
\Vertex(260,140){2}
\Vertex(260,120){2}
\SetColor{Yellow}
\Vertex(200,200){2}
\Vertex(200,240){2}
\Vertex(180,220){2}
\Vertex(180,260){2}
\Text(220,197)[tr]{1} 
\Text(240,197)[tr]{2} 
\Text(260,197)[tr]{3} 
\Text(280,197)[tr]{4} 
\Text(300,197)[tr]{5} 
\Text(320,197)[tr]{6} 
\Text(340,197)[tr]{7} 
\Text(360,197)[tr]{8}
\Text(380,197)[tr]{9} 
\Text(198,224)[tr]{1} 
\Text(198,264)[tr]{3} 
\Text(198,284)[tr]{4} 
\Text(198,304)[tr]{5} 
\Text(198,324)[tr]{6} 
\Text(198,344)[tr]{7} 
\Text(198,364)[tr]{8} 
\Text(198,184)[tr]{-1} 
\Text(198,164)[tr]{-2} 
\Text(198,144)[tr]{-3} 
\Text(198,124)[tr]{-4} 
\Text(198,104)[tr]{-5} 
\Text(198,84)[tr]{-6} 
\Text(196,399)[tr]{\Large $q$} 
\Text(400,197)[tr]{\Large $p$} 
\end{picture}
\ec
\caption{Values of $(p,q)$ at which the 
functions $G(p,q;\mu_B^2)$ are calculated are shown by dots.
The 12 fermionic basic constants are determined from
the functions associated with red dots; yellow dots 
show the constants $X_0 \div X_3$ that appear in the 
expressions for $G(p,q;\mu_B^2)$ but cancel in the
expressions for Feynman integrals. Functions shown by gray
are calculated explicitly and given in the formulas (\ref{Binput}),
(\ref{Ydef1}), (\ref{Ydef2}), (\ref{eq:JcrossDOWN}), (\ref{eq:JcrossLEFT}), 
and (\ref{eq:JcrossUP}), which provide the initial conditions for the recurrent 
relations. The order of implementation
of the recurrent relations is as follows:
1. $\cB$ functions are calculated at $p=0$, then at $p<0,q\leq 0$, then 
over the left strip, then at $p<0, q\geq 4$.
2. $\cJ$ functions are calculated in the ``down'' strip, then
lower-left domain (shown by green), then ``left'' strip, then ``up'' strip
then up-left domain, then up-right domain, then down-right domain. }
\end{figure}
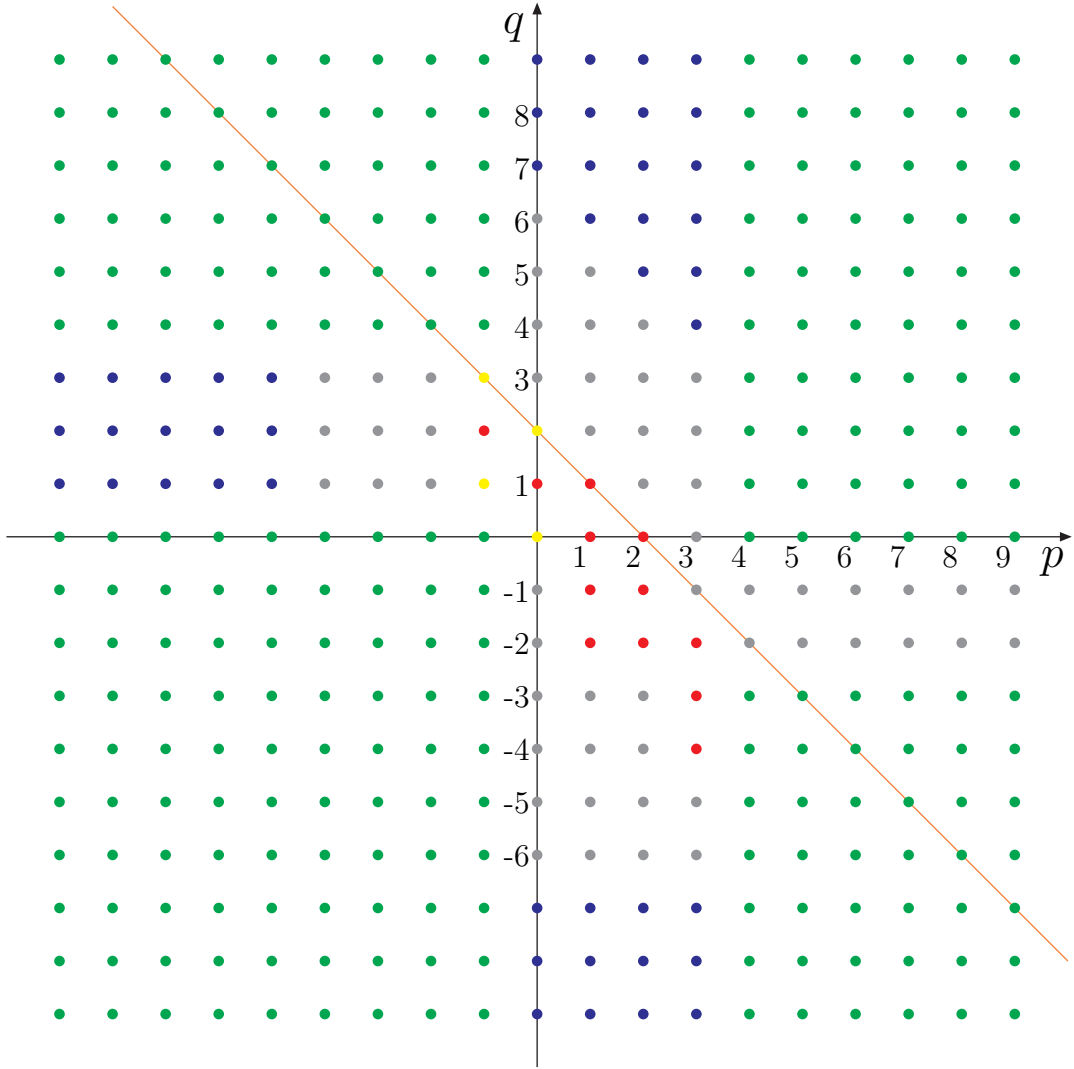

At $p\leq 3$, $\cB(p,q;\mu_B^2)$
functions involve no divergencies: $\cB(p,q;\mu_B^2) = B(p,q) = \cB(p,q;0)$. 
They can be determined by the recurrent relations presented in Appendix~3.
The initial conditions are provided by the formulas (\ref{Binput})
and $B$ functions calculated previously.

To express $B(r,1)$ or $B(r,2)$ or $B(r,3)$ at $r\leq 3$ in terms of $F_0, Z_0$, and $Z_1$, 
one should apply these relations beginning with $p=r$ and ending with $p=-3$.
In so doing, one must know $B(p,q)$ at $q=0,-1,-2$ and $r\leq p \leq 0$

\bi 
\item{\bf $\cB$ functions at $p\leq -1$, $q\geq 4$}
\ei
$\cB$ functions in this domain 
are computed by the recurrent relations presented in the Appendix~4.
The initial conditions for these relations  are provided by $\cB(0,q;\mu_B^2)$ 
as well as $\cB(p,q;\mu_B^2)$ at $p\leq -1, q\leq 3$ presented above.


\subsubsection{$\cJ$ functions.}

Here, the quantities $J(p,q)$ introduced in (\ref{GdeltaExpansion})
are expressed in terms of the boson constants $Z_1$, $Z_0$, $F_0$  
introduced in the previous Section, and the quantities $Y_0 \div Y_{11}$ defined by the relations
\bea\label{Ydef1}
Y_4={J(1,0)\over 2}, & \qquad Y_5 = J(1,-1), & \qquad Y_6 = 2 J(1,-2), \\ \nonumber
Y_7={J(2,-1)\over 2}, & \qquad Y_8 = J(2,-2), & \qquad Y_9 = {J(3,-2)\over 2}, \\ \nonumber
Y_{10} = J(3,-3), & \qquad Y_{11}= 2 J(3,-4), & \qquad Y_0= {J(2,0)\over 4}\; - \; {F_0 \over 16\pi^2},  \nonumber
\eea
and
\bea\label{Ydef2}
Y_1 &=& {1 \over 48} - {1 \over 4}\; Z_0 - {1 \over 24}\; J(-1,2) + 
{1 \over 12}\; J(0,1) + {1 \over 12}\; J(1,0); \\ \nonumber
Y_2 &=&  { 1 \over 6 } - {1\over \pi^2} - Z_0 - { 1 \over 6 }\; J(-1,2) + 
{1 \over 3}\; J(0,1) -{ 1 \over 24} \; J(1,-2) - {1 \over 12} \; J(1,-1) - \\ \nonumber
&& - 
{17 \over 8}\; J(1,0) + 4\; J(1,1) - {1 \over 48}\; J(2,-2) + {25 \over 6}\; J(2,-1)
 - 4\; J(2,0); \\ \nonumber
Y_3 &=&  - {1\over 384\pi^2} - F_0\; {1\over 128\pi^2} + {1 \over 96}\; Z_0 - 
{1 \over 48}\; J(-1,3) + {1 \over 192}  \; J(0,1) + {1 \over 48}\; J(0,2) + {1 \over 48}\; J(1,1);
\eea
and the quantities 
\bea
X_0= J(-1,1), &&  X_1= J(-1,3), \\ \nonumber
X_2= J(0,0), &&  X_3=J(0,2).  \nonumber
\eea
as well. The numerical values of the quantities $X_0, X_1, X_2$, and $X_3$
are not needed because expressions for the integrals (\ref{eq:GenFermInt}) 
in terms of $G(p,q)$ etc. do not involve them. As for the other constants,
they are well known (the table below is taken from the review \cite{Capitani},
notation ${\cal F}(p,q)$ is borrowed from there):
\begin{table}[h]
\begin{center}
\begin{tabular}{|c|c|}
\hline
$Y_0$            & $-$             0.01849765846791657356  \\
$Y_1$            & \hphantom{$-$}  0.00376636333661866811  \\
$Y_2$            & \hphantom{$-$}  0.00265395729487879354  \\
$Y_3$            & \hphantom{$-$}  0.00022751540615147107  \\
$Y_4= {\cal F}(1,0)$  & \hphantom{$-$}  0.08539036359532067914  \\
$Y_5= {\cal F}(1,-1)$ & \hphantom{$-$}  0.46936331002699614475  \\
$Y_6= {\cal F}(1,-2)$ & \hphantom{$-$}  3.39456907367713000586  \\
$Y_7= {\cal F}(2,-1)$ & \hphantom{$-$}  0.05188019503901136636  \\
$Y_8= {\cal F}(2,-2)$ & \hphantom{$-$}  0.23874773756341478520  \\
$Y_9= {\cal F}(3,-2)$ & \hphantom{$-$}  0.03447644143803223145  \\
$Y_{10}= {\cal F}(3,-3)$ & \hphantom{$-$}  0.13202727122781293085  \\
$Y_{11}= {\cal F}(3,-4)$ & \hphantom{$-$}  0.75167199030295682254  \\
\hline
\end{tabular}
\caption{New constants appearing in the general fermionic case.}
\label{tab:fermionicconstants}
\end{center}
\end{table}

\bi 
{\item Thus $\cJ$ functions over the domain
$
\cA = \left\{ (p,q): 0 \leq p \leq 3,\  -6 \leq p \leq 6-p \right\} \cup $\\ 
$\cup \left\{ (p,q): - 4\leq p \leq -1,\  1\leq q \leq 3 \right\}
$
calculated by the procedure indicated in \cite{BCP} can 
be represented in terms of the above constants as follows:}
\ei
\bea\label{eq:JcrossDOWN}
&&  J(0,0) = \; X_2; \\ \nonumber && 
  J(0,-1) = 4 \; X_2 + 315 \; Y_{10} - 1218 \; Y_9 - 134 \; Y_8 + 804 \; Y_7 - 2 \; Y_6 
+ {25 \over  2} \; Y_5  - 114 \; Y_4 + {15 \over (2\pi)^2 }; \\ \nonumber && 
  J(0,-2) = - 443/12 + 18 \; X_2 + 525/16 \; Y_{11} + 5265/4 \; Y_{10} - 7661/2 \; Y_9- 8173/12 \; Y_8 \\ \nonumber && 
\quad + 2085 \; Y_7 - 7/48 \; Y_6 + 1657/24 \; Y_5 - 339/2 \; Y_4 + 35/4 \;  (2\pi)^{-2}; \\ \nonumber && 
 J(0,-3) = - 1549/27 + 88\; X_2 + 595/4\; Y_{11} + 92395/6 \; Y_{10} - 487883/9\; Y_9 - 197384/27 \; Y_8 \\ \nonumber && 
\quad + 103330/3 \; Y_7 - 6569/108 \; Y_6 + 57949/108 \; Y_5 - 12829/3 \; Y_4 + 9029/18 \;  (2\pi)^{-2}; \\ \nonumber && 
  J(0,-4) = - 1024087/576 + 917/2 \; X_2 + 1919221/768 \; Y_{11} + 7343317/192 \; Y_{10} \\ \nonumber && 
\quad -  20889553/288 \; Y_9 - 14398667/576 \; Y_8 + 3075865/144 \; Y_7 + 325123/2304 \; Y_6 \\ \nonumber && 
\quad +  2538989/1152 \; Y_5 + 501607/96 \; Y_4 - 910433/576 \;  (2\pi)^{-2}; \\ \nonumber && 
  J(0,-5) =   22816157/32400 + 2514 \; X_2 + 14234765/4032 \; Y_{11} + 6270202009/8400 \; Y_{10} \\ \nonumber && 
\quad - 171277520509/63000 \; Y_9 - 11273402747/32400 \; Y_8 + 33415313987/18900 \; Y_7 \\ \nonumber && 
\quad - 402215297/129600 \; Y_6 + 52200958189/2268000 \; Y_5 - 4825909001/21000 \; Y_4 \\ \nonumber && 
\quad + 3567631667/126000 \;  (2\pi)^{-2}; \\ \nonumber && 
   J(0,-6) =  - 59060175671/583200 + 14376 X_2 - 456897556151/2268000\; (2pi)^{-2}   \\ \nonumber && 
\quad + 290748296317/1814400 Y_{11} - 46846789343/252000 Y_{10} + 1099242273317/226800\; Y_9  \\ \nonumber && 
\quad  - 1709413667089/4082400\; Y_8 - 8051167629571/1701000\; Y_7 + 41141268191/2332800\; Y_6 \\ \nonumber && 
\quad  + 1776831395699/40824000\; Y_5 + 61514640887/54000\ Y_4; \\  \nonumber && 
  J(1,0) = 2 \; Y_4; \\ \nonumber && 
  J(1,-1) = \; Y_5; \\ \nonumber && 
  J(1,-2) =  1/2 \; Y_6; \\ \nonumber && 
  J(1,-3) =   52/3 + 683/2 \; Y_{10} - 1342 \; Y_9 - 439/3 \; Y_8 + 924 \; Y_7\\ \nonumber && 
\quad - 13/3 \; Y_6 + 89/12 \; Y_5  - 132 \; Y_4 + 17 \;  (2\pi)^{-2}; \\ \nonumber && 
  J(1,-4) = - 761/9 + 683/12 \; Y_{11} - 10807/3 \; Y_{10} + 48538/3 \; Y_9 + 12539/9 \; Y_8 \\ \nonumber && 
\quad - 35284/3 \; Y_7 + 1481/36 \; Y_6 - 887/18 \; Y_5 + 1874 \; Y_4 - 803/3 \;  (2\pi)^{-2}; \\ \nonumber && 
  J(1,-5) =    34745/27 - 52147/63 \; Y_{11} + 6069519/140 \; Y_{10} - 27087257/140 \; Y_9 \\ \nonumber && 
\quad - 450260/27 \; Y_8 + 88135577/630 \; Y_7 - 42845/108 \; Y_6 + 2730857/3780 \; Y_5 \\ \nonumber &&
\quad - 3125867/140 \; Y_4  + 541505/168 \;  (2\pi)^{-2}; \\ \nonumber && 
  J(1,-6) = - 3470071/324 + 123774917/10800 \; Y_{11} - 15164760473/31500 \; Y_{10} \\ \nonumber && 
\quad + 34075577647/15750 \; Y_9 + 10210300231/56700 \; Y_8 - 36997574549/23625 \; Y_7 \\ \nonumber && 
\quad + 5205091/1296 \; Y_6 - 4977314411/567000 \; Y_5 + 1327628299/5250 \; Y_4 \\ \nonumber && 
\quad - 232435501/6300 \;  (2\pi)^{-2}; \\ \nonumber && 
  J(2,0) =  4 \; Y_0 + F_0 \;  (2\pi)^{-2}; \\ \nonumber && 
  J(2,-1) = 2 \; Y_7; \\ \nonumber && 
  J(2,-2) = \; Y_8; \\ \nonumber && 
  J(2,-3) = - \; Y_{10} + 90 \; Y_9 - 2 \; Y_8 - 84 \; Y_7 + 5/2 \; Y_5 + 18 \; Y_4 - 3 \;  (2\pi)^{-2}; \\ \nonumber && 
  J(2,-4) =  - 13/3 - 1/4 \; Y_{11} + \; Y_{10} - 326 \; Y_9 + 13/3 \; Y_8 + 300 \; Y_7 + 31/12 \; Y_6 \\ \nonumber && 
\quad - 19/6 \; Y_5 - 66 \; Y_4 + 13 \;  (2\pi)^{-2}; \\ \nonumber && 
  J(2,-5) = 1012/9 + 53/63 \; Y_{11} + 519787/210 \; Y_{10} - 316251/35 \; Y_9 - 3217/3 \; Y_8 \\ \nonumber && 
\quad + 214918/35 \; Y_7  - 343/9 \; Y_6 + 17621/420 \; Y_5 - 29051/35 \; Y_4 + 3859/42 \;  (2\pi)^{-2}; \\ \nonumber && 
  J(2,-6) =  - 48007/36 + 8466287/15120 \; Y_{11} - 272218199/6300 \; Y_{10} + 593087111/3150 \; Y_9 \\ \nonumber && 
\quad + 64920421/3780 \; Y_8 - 30568297/225 \; Y_7 + 70867/144 \; Y_6 - 22014731/37800 \; Y_5 \\ \nonumber && 
\quad + 7424029/350 \; Y_4 - 3722933/1260 \;  (2\pi)^{-2}; \\ \nonumber && 
  J(3,0) =  - 35/256 \; Y_{11} - 57/64 \; Y_{10} - 301/2304 \; Y_9 + 85/96 \; Y_8 + 1405/1152 \; Y_7\\ \nonumber && 
\quad - 1/384 \; Y_5 - 461/768 \; Y_4 - 2 \; Y_0 + (1433/1536 - 1/2 F_0) \; (2\pi)^{-2}; \\ \nonumber && 
  J(3,-1) = 155/96 \; Y_9 - 83/48 \; Y_7 + 11/32 \; Y_4 + 4 \; Y_0  + (F_0 - 31/64 )\;  (2\pi)^{-2}; \\ \nonumber && 
  J(3,-2) =  2 \; Y_9; \\ \nonumber && 
  J(3,-3) = \; Y_{10}; \\ \nonumber && 
  J(3,-4) = 1/2 \; Y_{11}; \\ \nonumber && 
  J(3,-5) = - 85/42 \; Y_{11} - 1649/140 \; Y_{10} + 15548/35 \; Y_9 + 7/6 \; Y_8 - 14264/35 \; Y_7 \\ \nonumber && 
\quad + 5609/840 \; Y_5 + 2943/35 \; Y_4 - 106/7 \;  (2\pi)^{-2}; \\ \nonumber && 
  J(3,-6) = - 224/9 + 2812/945 \; Y_{11} + 47669/1575 \; Y_{10} - 595726/225 \; Y_9 + 6854/945 \; Y_8 \\ \nonumber && 
\quad + 3804196/1575 \; Y_7 + 101/9 \; Y_6 - 173539/9450 \; Y_5 - 90298/175 \; Y_4 + 33203/315 \; (2\pi)^{-2}; \nonumber  
\eea 
\bea\label{eq:JcrossLEFT}
&&  J(-1,1) = \; X_0; \\ \nonumber && 
  J(-2,1) = - 1135/9 - 158/9 \; X_0 + 448/3 \; X_3 + 722/9 \; X_2 - 448/3 \; X_1 + 1295/12 \; Y_{11} \\ \nonumber && 
\quad + 56237/9 \; Y_{10} - 61066/3 \; Y_9 - 83036/27 \; Y_8 + 108068/9 \; Y_7 - 283/18 \; Y_6 \\ \nonumber && 
\quad + 15217/54 \; Y_5 - 1274 \; Y_4 - 19456/3 \; Y_3 - 32/3 \; Y_2 + 1472/3 \; Y_1 + 896/3 \; Y_0 \\ \nonumber && 
\quad + 865/9 \;  (2\pi)^{-2} + 64/9 F_0 \;  (2\pi)^{-2} + 536/3 \; Z_1 + 5488/27 \; Z_0; \\ \nonumber && 
 J(-3,1) = 11845003/15000 + 105068/375 \; X_0 - 371584/125 \; X_3 - 200897/375 \; X_2 \\ \nonumber && 
\quad + 371584/125 \; X_1 + 204127/160 \; Y_{11} + 19090453/600 \; Y_{10} - 133042333/1500 \; Y_9  \\ \nonumber && 
\quad - 175667117/9000 \; Y_8+ 39223253/750 \; Y_7 - 214589/12000 \; Y_6 + 8996881/18000 \; Y_5 \\ \nonumber && 
\quad - 207851/60 \; Y_4  + 15388672/125 \; Y_3 + 27568/125 \; Y_2 - 1398848/125 \; Y_1  \\ \nonumber && 
\quad - 640512/125 \; Y_0 + 6221759/15000 \;  (2\pi)^{-2} - 153472/375 F_0 \;  (2\pi)^{-2}  \\ \nonumber && 
\quad - 2696248/625 \; Z_1 - 24764192/5625 \; Z_0; \\ \nonumber && 
  J(-4,1) = - 3329832752387/64827000 - 117549296/25725 \; X_0 + 446453248/8575 \; X_3 \\ \nonumber && 
\quad + 566923184/25725 \; X_2 - 446453248/8575 \; X_1 + 737380289/14112 \; Y_{11} \\ \nonumber && 
\quad + 242892361651/123480 \; Y_{10} - 5552574235627/926100 \; Y_9 \\ \nonumber && 
\quad - 640505149891/617400 \; Y_8 + 1563368068291/463050 \; Y_7 - 23776321823/7408800 \; Y_6 \\ \nonumber && 
\quad + 321309701383/3704400 \; Y_5 - 89421147379/308700 \; Y_4 - 17930338304/8575 \; Y_3 \\ \nonumber && 
\quad - 92940608/25725 \; Y_2 + 1782287616/8575 \; Y_1 + 2138417152/25725 \; Y_0 \\ \nonumber && 
\quad + 991072905679/64827000 \;  (2\pi)^{-2} + 4988416/525 F_0 \;  (2\pi)^{-2} \\ \nonumber && 
\quad + 25248697568/300125 \; Z_1 + 73082913664/900375 \; Z_0; \\ \nonumber && 
 J(-1,2) = 2/3 - 8 \; X_3 + 8\; X_1 - 1/24\; Y_8 + 50/3 \; Y_7 - 1/24\; Y_6 - 1/6\; Y_5 - 35/6 \; Y_4 \\ \nonumber && 
\quad + 384 \; Y_3 - 2 \; Y_2 - 16 \; Y_1 - 32 \; Y_0  + 4 (F_0 - 1) \;  (2\pi)^{-2} - 10 \; Z_0; \\ \nonumber && 
 J(-2,2) = 135/4 + 19/2 \; X_0 - 48 \; X_3 - 41/2 \; X_2 + 48 \; X_1 - 105/16\; Y_{11} - 1851/4 \; Y_{10} \\ \nonumber && 
\quad + 3075/2 \; Y_9 + 827/4 \; Y_8 - 841 \; Y_7 + 11/8 \; Y_6 - 269/8 \; Y_5 + 175/2 \; Y_4 + 2304 \; Y_3 \\ \nonumber && 
\quad + 12 \; Y_2 - 96 \; Y_1 - 96 \; Y_0 - 45/4 \;  (2\pi)^{-2} - 114 \; Z_1 - 76 \; Z_0; \\ \nonumber && 
 J(-3,2) = - 29723/16 - 2365/9 \; X_0 + 7712/3 \; X_3 + 32881/36 \; X_2 - 7712/3 \; X_1 \\ \nonumber && 
\quad + 56805/64 \; Y_{11} + 6355039/144 \; Y_{10} - 5035139/36 \; Y_9 - 4682939/216 \; Y_8  \\ \nonumber && 
\quad + 1411847/18 \; Y_7 - 4543/48 \; Y_6 + 2051831/864 \; Y_5 - 68044/9 \; Y_4 - 330752/3 \; Y_3\\ \nonumber && 
\quad  - 252 \; Y_2 + 26992/3 \; Y_1 + 13952/3 \; Y_0 + (54773/72 + 2336/9 F_0) \;  (2\pi)^{-2} \\ \nonumber && 
\quad + 13706/3 \; Z_1 + 111704/27 \; Z_0; \\ \nonumber && 
 J(-4,2) = 492009959/12000 + 87748/15 \; X_0 - 324224/5 \; X_3 - 275752/15 \; X_2 + 324224/5 \; X_1 \\ \nonumber && 
\quad - 3344999/384 \; Y_{11} - 240354739/480 \; Y_{10} + 5720706571/3600 \; Y_9 \\ \nonumber && 
\quad + 329440181/1440 \; Y_8 - 302115271/360 \; Y_7 + 2092337/1920 \; Y_6  - 518442119/14400 \; Y_5 \\ \nonumber && 
\quad + 285938617/3600 \; Y_4 + 2650112 \; Y_3 + 75952/15 \; Y_2 - 1259712/5 \; Y_1 - 318976/3 \; Y_0 \\ \nonumber && 
\quad - (423392089/36000 + 159872/15 F_0) \; (2\pi)^{-2} - 44526248/375 \; Z_1 - 40322528/375 \; Z_0; \\ \nonumber && 
  J(-1,3) = \; X_1; \\ \nonumber && 
  J(-2,3) = 1/12 - 1/4 \; X_0 - 12 \; X_3 + \; X_2 + 12 \; X_1 + 105/2 \; Y_{10} - 203 \; Y_9 - 1075/48 \; Y_8 \\ \nonumber && 
 + 159 \; Y_7 - 19/48 \; Y_6 + 23/12 \; Y_5 - 103/4 \; Y_4 + 576 \; Y_3 - 7 \; Y_2 - 60 \; Y_1 \\ \nonumber && 
 - 48 \; Y_0 - 7/2 \;  (2\pi)^{-2} + 6 F_0 \;  (2\pi)^{-2} + 12 \; Z_1 - 19 \; Z_0; \\ \nonumber && 
  J(-3,3) =  871/2 + 629/9 \; X_0 - 1696/3 \; X_3 - 1778/9 \; X_2 + 1696/3 \; X_1 - 175/2 \; Y_{11} \\ \nonumber && 
 - 50840/9 \; Y_{10} + 167119/9 \; Y_9 + 279089/108 \; Y_8 - 91946/9 \; Y_7 + 187/12 \; Y_6 \\ \nonumber && 
 - 10154/27 \; Y_5 + 9446/9 \; Y_4 + 77824/3 \; Y_3 + 116 \; Y_2 - 4784/3 \; Y_1 - 3136/3 \; Y_0 \\ \nonumber && 
 - 1739/6 \;  (2\pi)^{-2} - 304/9 F_0 \;  (2\pi)^{-2} - 1502 \; Z_1 - 28456/27 \; Z_0; \\ \nonumber && 
  J(-4,3) =  - 40263778/1875 - 2997542/1125 \; X_0 + 10588096/375 \; X_3 + 20165461/2250 \; X_2 \\ \nonumber && 
 - 10588096/375 \; X_1 + 602637/80 \; Y_{11} + 161298737/450 \; Y_{10} \\ \nonumber && 
 - 5056485013/4500 \; Y_9 - 4717685677/27000 \; Y_8 + 1381579093/2250 \; Y_7 \\ \nonumber && 
 - 8668453/12000 \; Y_6 + 284825729/13500 \; Y_5 - 50984761/900 \; Y_4 \\ \nonumber && 
 - 444755968/375 \; Y_3 - 347064/125 \; Y_2 + 39247712/375 \; Y_1 + 17916928/375 \; Y_0 \\ \nonumber && 
 + 182925853/15000 \;  (2\pi)^{-2} + 4427968/1125 F_0 \;  (2\pi)^{-2} + 38504584/625 \; Z_1 \\ \nonumber && 
 + 860640608/16875 \; Z_0;  \nonumber 
\eea
\bea\label{eq:JcrossUP}
&&  J(0,1) =  1/12 - 4 \; X_3 + 4 \; X_1 - 1/48 \; Y_8 + 25/3 \; Y_7 - 1/48 \; Y_6 - 1/12 \; Y_5 - 59/12 \; Y_4 \\ \nonumber && 
\quad + 192 \; Y_3 - \; Y_2 + 4 \; Y_1 - 16 \; Y_0  + (2 F_0 - 2) \;  (2\pi)^{-2} - 2 \; Z_0; \\ \nonumber && 
 J(0,2) =  \; X_3; \\ \nonumber && 
 J(0,3) = 91/1024 + 1/384 \; X_0 + 1/4 \; X_3 - 1/96 \; X_2 - 35/768 \; Y_{11} - 27/32 \; Y_{10} + 109/72 \; Y_9 \\ \nonumber && 
\quad + 19229/36864 \; Y_8 + 383/2304 \; Y_7 - 11/4096 \; Y_6 - 443/9216 \; Y_5 - 1243/3072 \; Y_4 \\ \nonumber && 
\quad - 3/2 \; Y_3 - 33/256 \; Y_2 + 17/64 \; Y_1 - 33/16 \; Y_0 - 13/72 \;  (2\pi)^{-2} \\ \nonumber && 
\quad - 5/48 F_0 \;  (2\pi)^{-2} - 1/4 \; Z_1 - 49/192 \; Z_0; \\ \nonumber && 
 J(0,4) = 80441/2211840 + 31/27648 \; X_0 + 17/288 \; X_3 - 31/6912 \; X_2 + 1/288 \; X_1 \\ \nonumber && 
\quad + 161/24576 \; Y_{11} - 17767/92160 \; Y_{10} + 23/32 \; Y_9 + 161077/2949120 \; Y_8 \\ \nonumber && 
\quad - 10667/552960 \; Y_7 - 14369/8847360 \; Y_6 - 16271/737280 \; Y_5 - 343291/2211840 \; Y_4 \\ \nonumber && 
\quad - 1379/1440 \; Y_3 - 14369/184320 \; Y_2 + 9409/46080 \; Y_1 - 8609/11520 \; Y_0 \\ \nonumber && 
\quad +( 539/17280 F_0 - 18031/86400 ) \;  (2\pi)^{-2} - 237/2560 \; Z_1 - 965/9216 \; Z_0; \\ \nonumber && 
 J(0,5) = 7632781/594542592 + 523/1327104 \; X_0 + 209/13824 \; X_3 - 523/331776 \; X_2   \\ \nonumber && 
\quad + 25/13824 \; X_1 + 77651/42467328 \; Y_{11} - 1253633/17694720 \; Y_{10} + 56974703/ \\ \nonumber && 
\quad      159252480 \; Y_9 + 782803057/35672555520 \; Y_8 + 34653559/2229534720 \; Y_7 -  \\ \nonumber && 
\quad      10633541/11890851840 \; Y_6 - 80793859/8918138880 \; Y_5 - 291120383/2972712960 \; Y_4  \\ \nonumber && 
\quad - 262597/645120 \; Y_3 - 10633541/247726080 \; Y_2 + 8290501/61931520 \; Y_1 - 5150021/15482880 \; Y_0\\ \nonumber && \quad + (43529/5806080 F_0 - 8543128133/78033715200 )\;  (2\pi)^{-2} \\ \nonumber &&
\quad  - 569479/20643840 \; Z_1 - 2292133/61931520 \; Z_0; \\ \nonumber && 
  J(0,6) =  1133309347/237817036800 + 1429/10616832 \; X_0 + 2371/552960 \; X_3 - 1429/2654208 \; X_2 \\ \nonumber && 
\quad + 401/552960 \; X_1 + 10571051/10192158720 \; Y_{11} - 639402031/29727129600 \; Y_{10} \\ \nonumber && 
\quad + 48006407249/267544166400 \; Y_9 + 5666277343/1223059046400 \; Y_8 + 2239163267/107017666560 \; Y_7 \\ \nonumber && \quad - 470661551/951268147200 \; Y_6 - 1645375871/428070666240 \; Y_5 - 40565004503/713451110400 \; Y_4 \\ \nonumber && \quad  - 23258381/154828800 \; Y_3 - 470661551/19818086400 \; Y_2 + 406642351/4954521600 \; Y_1 \\ \nonumber && 
 \quad - 153477551/1238630400 \; Y_0 
+( 1169977/232243200 F_0 - 1234692078509/18728091648000)\; (2\pi)^{-2} \\ \nonumber && 
\quad - 117647549/14863564800 \; Z_1 - 202376477/14863564800 \; Z_0; \\ \nonumber &&
  J(1,1) = - 1/48 + 1/192 \; Y_8 - 25/12 \; Y_7 + 1/192 \; Y_6 + 1/48 \; Y_5 + 59/48 \; Y_4\\ \nonumber && 
\quad + 1/4 \; Y_2 - \; Y_1 + 4 \; Y_0 + (1 + F_0) \;  (2\pi)^{-2}; \\ \nonumber && 
 J(1,2) = - 19/1536 - 35/768 \; Y_{11} - 19/64 \; Y_{10} - 173/288 \; Y_9 + 5497/18432 \; Y_8 \\ \nonumber && 
\quad - 293/1152 \; Y_7 + 19/6144 \; Y_6 + 53/4608 \; Y_5 + 187/512 \; Y_4 + \; Y_3 + 19/128 \; Y_2  \\ \nonumber && 
\quad - 19/32 \; Y_1 + 3/8 \; Y_0 + 307/576 \;  (2\pi)^{-2} + 1/96 \; Z_0; \\ \nonumber && 
  J(1,3) = - 5491/737280 + 721/73728 \; Y_{11} + 1957/30720 \; Y_{10} - 21361/55296 \; Y_9 \\ \nonumber && 
\quad - 542443/8847360 \; Y_8 - 266629/552960 \; Y_7 + 5959/2949120 \; Y_6 + 18289/2211840 \; Y_5 \\ \nonumber && 
\quad + 274901/737280 \; Y_4 + 149/480 \; Y_3 + 5959/61440 \; Y_2 - 5959/15360 \; Y_1 \\ \nonumber && 
\quad + 1799/3840 \; Y_0 + (438551/2764800 + 11/160 F_0) \;  (2\pi)^{-2} - 13/5120 \; Z_1 + 35/3072 \; Z_0; \\ \nonumber && J(1,4) = - 135181/27525120 - 1519/3538944 \; Y_{11} - 4123/1474560 \; Y_{10}- 2532143/13271040 \; Y_9 \\ \nonumber && 
\quad + 12187051/2972712960 \; Y_8 - 12605351/37158912 \; Y_7 + 1307897/990904320 \; Y_6 \\ \nonumber && 
\quad + 783523/148635648 \; Y_5 + 18292649/82575360 \; Y_4 - 533/161280 \; Y_3 \\ \nonumber && 
\quad + 1307897/20643840 \; Y_2 - 1307897/5160960 \; Y_1 + 93817/1290240 \; Y_0 \\ \nonumber && 
\quad + (346641047/2167603200 - 1/1680 F_0) \;  (2\pi)^{-2} - 22817/15482880 \; Z_1 + 146059/15482880 \; Z_0; \\ \nonumber && 
  J(1,5) = - 32271257/9512681472 + 9401131/2038431744 \; Y_{11} + 178621489/5945425920 \; Y_{10} \\ \nonumber && 
\quad - 1051530899/53508833280 \; Y_9 - 49631503063/1712282664960 \; Y_8 \\ \nonumber && 
\quad - 5652301063/15288238080 \; Y_7 + 167849953/190253629440 \; Y_6 \\ \nonumber && 
\quad + 1548254101/428070666240 \; Y_5 + 26366380697/142690222080 \; Y_4 \\ \nonumber && 
\quad - 4481117/30965760 \; Y_3 + 167849953/3963617280 \; Y_2 - 167849953/990904320 \; Y_1 \\ \nonumber && 
\quad + 9544673/247726080 \; Y_0 + (260638894247/3745618329600 + 1339/215040 F_0) \; (2\pi)^{-2} \\ \nonumber && 
\quad - 541139/990904320 \; Z_1 + 806521/110100480 \; Z_0; \\ \nonumber && 
  J(2,1) = 23/4608 - 35/384 \; Y_{11} - 19/32 \; Y_{10} - 173/144 \; Y_9 + 3619/6144 \; Y_8 \\ \nonumber && 
\quad + 2827/1152 \; Y_7 - 23/18432 \; Y_6 - 31/4608 \; Y_5 - 3949/4608 \; Y_4 + \; Y_3 - 23/384 \; Y_2 \\ \nonumber && 
\quad + 23/96 \; Y_1 - 19/24 \; Y_0 + ( 127/288 - 1/4 F_0) \;  (2\pi)^{-2} - 1/96 \; Z_0; \\ \nonumber && 
 J(2,2) = 7043/1105920 + 343/12288 \; Y_{11} + 931/5120 \; Y_{10} - 3731/4608 \; Y_9 \\ \nonumber && 
\quad - 806147/4423680 \; Y_8 + 339959/276480 \; Y_7 - 6467/4423680 \; Y_6 - 5879/1105920 \; Y_5 \\ \nonumber && 
\quad - 365953/1105920 \; Y_4 + 221/240 \; Y_3 - 6467/92160 \; Y_2 + 6467/23040 \; Y_1 \\ \nonumber && 
\quad + 4373/5760 \; Y_0  + (23/160 F_0 - 25657/76800) \;  (2\pi)^{-2} - 1/480 \; Z_1 - 7/512 \; Z_0; \\ \nonumber && 
  J(2,3) = 4611371/743178240 - 73045/3538944 \; Y_{11} - 39653/294912 \; Y_{10} \\ \nonumber && 
\quad - 12722017/13271040 \; Y_9 + 393115913/2972712960 \; Y_8 + 103603669/61931520 \; Y_7 \\ \nonumber && 
\quad - 4248887/2972712960 \; Y_6 - 504563/82575360 \; Y_5 - 79536305/148635648 \; Y_4 \\ \nonumber && 
\quad + 155081/161280 \; Y_3 - 4248887/61931520 \; Y_2 + 4248887/15482880 \; Y_1 \\ \nonumber && 
\quad + 66473/3870720 \; Y_0 + 493568683/6502809600 \;  (2\pi)^{-2} \\ \nonumber && 
\quad - 1093/26880 F_0 \;  (2\pi)^{-2} - 10069/5160960 \; Z_1 - 74813/5160960 \; Z_0; \\ \nonumber && 
 J(2,4) =  196602193/35672555520 + 3693635/509607936 \; Y_{11} + 14035813/297271296 \; Y_{10} \\ \nonumber && 
\quad - 12442566511/13377208320 \; Y_9 - 20639077343/428070666240 \; Y_8 \\ \nonumber && 
\quad + 36967799143/26754416640 \; Y_7 - 181900981/142690222080 \; Y_6 \\ \nonumber && 
\quad - 530928403/107017666560 \; Y_5 - 2755058051/7134511104 \; Y_4 + 2424601/2580480 \; Y_3 \\ \nonumber && 
\quad - 181900981/2972712960 \; Y_2 + 181900981/743178240 \; Y_1 + 46110859/185794560 \; Y_0 \\ \nonumber && 
\quad - 79216495853/936404582400 \;  (2\pi)^{-2} + 4141/215040 F_0 \;  (2\pi)^{-2} \\ \nonumber && 
\quad - 408367/247726080 \; Z_1 - 125077/9175040 \; Z_0; \\ \nonumber && 
 J(3,1) = - 2473/2211840 + 1687/24576 \; Y_{11} + 4579/10240 \; Y_{10} + 28153/27648 \; Y_9 \\ \nonumber && 
\quad - 3930323/8847360 \; Y_8 - 866969/552960 \; Y_7 + 2797/8847360 \; Y_6 + 5689/2211840 \; Y_5 \\ \nonumber && 
\quad + 1137263/2211840 \; Y_4 - 211/480 \; Y_3 + 2797/184320 \; Y_2 - 2797/46080 \; Y_1 \\ \nonumber && 
\quad + 12317/11520 \; Y_0  + ( 23/80 F_0 - 323129/460800) \; (2\pi)^{-2} - 3/5120 \; Z_1 + 11/3072 \; Z_0; \\ \nonumber && 
  J(3,2) =  - 326465/148635648 - 206605/3538944 \; Y_{11} - 112157/294912 \; Y_{10} \\ \nonumber && 
\quad + 2493365/2654208 \; Y_9 + 225122965/594542592 \; Y_8 - 27027877/37158912 \; Y_7 \\ \nonumber && 
\quad + 336725/594542592 \; Y_6 + 19049/16515072 \; Y_5 + 6739903/148635648 \; Y_4 \\ \nonumber && 
\quad - 22955/32256 \; Y_3 + 336725/12386304 \; Y_2 - 336725/3096576 \; Y_1 - 483563/774144 \; Y_0 \\ \nonumber && 
\quad + (429664553/1300561920 - 331/2688 F_0) \;  (2\pi)^{-2} - 95/344064 \; Z_1 + 6823/1032192 \; Z_0; \\ \nonumber && 
  J(3,3) = - 69355327/23781703680 + 4168243/339738624 \; Y_{11} + 79196617/990904320 \; Y_{10} \\ \nonumber && 
\quad + 3687114839/2972712960 \; Y_9 - 22463493247/285380444160 \; Y_8 \\ \nonumber && 
\quad - 1869533095/1189085184 \; Y_7 + 70582891/95126814720 \; Y_6 \\ \nonumber && 
\quad + 45684329/14269022208 \; Y_5 + 362000483/880803840 \; Y_4 - 1586311/1720320 \; Y_3 \\ \nonumber && 
\quad + 70582891/1981808640 \; Y_2 - 70582891/495452160 \; Y_1 + 7231211/123863040 \; Y_0 \\ \nonumber && 
\quad - 131572132177/624269721600 \;  (2\pi)^{-2} + 12241/215040 F_0 \;  (2\pi)^{-2} \\ \nonumber && 
\quad - 34099/165150720 \; Z_1 + 1460153/165150720 \; Z_0; \nonumber 
\eea

\bi 
\item{\bf $\cJ(p,q;\mu_B^2)$ functions  at $0\leq p\leq 3$, $q\leq -7$}
\ei

Provided that $\cJ(p,q;\mu_B^2)$ over the domain $\cA$ (at $q\leq 0$)
and $B(0,q)$ at $q\leq 0$ are known, $\cJ$ functions over this strip 
can be found by the recurrent relations in Appendix~5.
Note that there are no divergent terms here.

\bi 
\item{\bf $\cJ(p,q;\mu_B^2)$ functions at $ p\leq -1$, $q\leq 0$}
\ei

Given $\cJ$ functions over the strip $0\leq p \leq 3,\ q\leq 0$
and $\cB$ functions at $p\leq 0, q \leq 0$,
$\cJ$ functions at $ p\leq -1$, $q\leq 0$ can be
determined using the recurrent relations
in Appendix~8, however, with $\mu_B=0$ (in this domain, $\cJ(p,q;\mu_B^2)=J(p,q)$).

\bi 
\item{\bf $\cJ(p,q;\mu_B^2)$ functions at $ p\leq -5$, $1 \leq q\leq 3$}.
\ei

They can be determined using the recurrent relations presented in Appendix~6.
The initial conditions are provided by the formulas (\ref{eq:JcrossLEFT})
and $B$ functions determined previously.
In calculation of $J(p,q)$ one needs $B(r,s)$ at $r\geq p$ and $-3\leq q \leq 3$.

\bi
\item{\bf $\cJ(p,q;\mu_B^2)$ functions at $ 0\leq p\leq 3$ and $q\geq 6-p$.}
\ei

In this domain, the divergent part $D(p,q)$ is calculated separately,
the results are partially presented in (\ref{DivPart_PP_min_domain});
thus we calculate the functions $J(p,q)$ by the formulas

\be
J(n,q+1-n) = { 1 \over (2q-1)(2q-3)} \sum_{k=0}^3 M(n,k,q) Z_k(q),
\ee

where $M(n,k,q)$ are given by 
\bea
M(0,0,q) &=&(q + 1)/32/(q - 1); \\ \nonumber
M(0,1,q) &=& -(8q-3)(q-2)/32/q/(q - 1); \\ \nonumber
M(0,2,q) &=&-(q-2)/4/q/(q - 1); \\ \nonumber
M(0,3,q) &=&-(q-2)/2/q/(q-1)^2; \\ \nonumber
M(1,0,q) &=&(q + 1)/32; \\ \nonumber
M(1,1,q) &=&3/32; \\ \nonumber
M(1,2,q) &=& -(q-2)/4/q; \\ \nonumber
M(1,3,q) &=& -(q-2)/2/q/(q - 1); \\ \nonumber
M(2,0,q) &=& q(q + 1)/32; \\ \nonumber
M(2,1,q) &=& 3q/32; \\ \nonumber
M(2,2,q) &=&3/16/q; \\ \nonumber
M(2,3,q) &=& -(q-2)/2/(q - 1); \\ \nonumber
M(3,0,q) &=& q(q+1)^2/64; \\ \nonumber
M(3,1,q) &=&3q(q + 1)/64; \\ \nonumber
M(3,2,q) &=&3(q + 1)/32/q; \\ \nonumber
M(3,3,q) &=& - (8q^2-19q+3)/16/q/(q - 1); \\ \nonumber
\eea
and $Z_k(q)$ can be determined from the recurrent relations
presented in Appendix~7. The initial conditions are provided by the formulas 
(\ref{eq:JcrossDOWN}), (\ref{eq:JcrossLEFT}), and (\ref{eq:JcrossUP});
$\cB(0,q)$ at $q>0$ and $D(p,q,r)$ at $0\leq p \leq 3, q>1-p, 0\leq r\leq p+q-2$ are also needed.

\bi
\item{\bf $\cJ(p,q;\mu_B^2)$ functions at $ p\leq -1$, $q\geq 4$.}
\ei

The respective recurrent relations can be found in Appendix~8,
the initial conditions are provided by the $\cB(p,q)$ 
functions at $p\leq 0$ and $q\geq -2$ and $\cJ(p,q)$ functions
at $0\leq p\leq 3$ and $-2\leq q \leq 3$.

\bi
\item{\bf $\cJ(p,q)$ functions in the domain $p \geq 4, q\geq 0 $ }
\ei
The recurrent relations in Appendix~9 give expressions for the 
finite part $J(p,q)$, the divergent part should be found by the 
procedure described in the beginning of this Section,
see also (\ref{DivPart_PP_min_domain}).

It should be also noted that the formulas in Appendix~9 {\bf are valid 
only in the case $p>4$,} in order to use them at $p=4$ the quantities $J(0,q)$
that appear in the right-hand part should be replaced by $B(0,q)$.

\bi
\item{\bf $\cJ(p,q)$ functions in the domain $p \geq 4, q < 0 $ }
\ei
The explicit expressions for the finite parts $J(p,q)$ at $q=-1, -2$ and 
$p\leq 9$ are presented in the Appendix~10, the recurrent relations
valid at $p>4, q<-2$ are given in the Appendix~11.
To employ these relations at $p=4$, one should replace $J(0,q)$ that
appears in the right-hand side by the function $B(0,q)$.

The divergent parts are given by the formula (\ref{DivPart_PM_up_to_9}).

Therewith, it should be noted that the constants $X_0 \div X_3$ 
that appear in some expressions for $J(p,q)$ at $p\leq 0$ cancel 
in the expressions for the integrals (\ref{eq:GenFermInt}) at $p>0$
and thus their numerical values are not needed.

\vspace*{-1mm}
\subsection{Dimensional Regularization}

In the dimensional regularization, 
each integral $F(p,q;n1,n2,n3,n4)$ (see (\ref{eq:GenFermInt})) is associated with
the respective boson integral $B_F(p,q;n1,n2,n3,n4)$ that has the same divergent part. 
$B_F(p,q;n1,n2,n3,n4)$ is determined by the procedure 
similar to that indicated in subsection (\ref{DPFIFMR}).
For example, at $p>0, q\geq2-p$
\be\label{ABIexample1}
B_F (p,q;n1,n2,n3,n4) = \sum_{l=0}^{p+q-2} {(-1)^l\; (p+l-1)!\over l! (p-1)!} 
\int {dk\over (2\pi)^4} \; {\Delta^l  \cos^{n_1}(k_1)\ .... \cos^{n_4}(k_4)\ \over \Delta_B^{p+q+l}}. \\ \nonumber
\ee
Then we compute $B_F(p,q;n1,n2,n3,n4)$ in the dimensional regularization as 
it is described in subsection (\ref{sec:DRbos}) and 
$F(p,q;n1,n2,n3,n4)-B_F(p,q;n1,n2,n3,n4)$ (which is convergent) in the fictitious mass regularization.
The sum of these quantities provides the sought for result.\\[1mm]

\vspace*{-1mm}
\section{Conclusions}

The BCP algorithm described above and the explicit formulas obtained with it and presented
in the Appendices make it a straightforward matter to express an integral 
of the type (\ref{eq:GenFermInt}) at $p\leq 9$ and arbitrary values of 
$n_1, n_2, n_3, n_4$ and $q$ in terms of the constants $F_0, Z_0, Z_1$ and $Y_0 \div Y_{11}$.
In fact, these formulas provide a computer program,
which can easily be realized with various packages. 
Such program was written in FORM and performed, some of 
the results are presented on the web: \ \  
{\tt http://www.lattice.itep.ru/$\sim$pbaivid/lattpt/ }

These are
\bi 
\item the values of the functions $J(p,q)$ and $B(p,q)$
at $-26 \leq p \leq 0,\ \ -56 - 2p \leq q \leq 34 $ 
and the values of $J(p,q)$  at $1\leq p \leq 9, \ \ -28 \leq q \leq 33 - p$;
\item the expressions for the integrals of the type (\ref{eq:GenFermInt})
at some particular values of $p$ and $q$ and $n_1\leq 6$;
\item the program for the computation of the integrals 
(\ref{eq:GenFermInt}) at $0\leq p,q \leq 9$ and $n_\mu^{max}\leq 6$
that can be readily used by anyone.
\ei 

I hope that this work will facilitate using the BCP algorithm in practical computations.

{\large \bf Acknowledgments:} I am grateful to H.Perlt, A.Schiller, and V. Bornyakov for stimulating discussions,
to P.Buividovich for the help with the presentation on the web,
and to the Leipzig University, where this study was started, for hospitality.
This work was supported in part by the grant for scientific schools no. NSh-679.2008.2
and by the Russian Foundation for Basic Research (RFBR grant no. 07-02-0237).

\newpage

{\Large \bf Appendix 1.}\\[2mm]

Some values used in the text are listed below.

The integrals defined in (\ref{Bdirect}) at $4\leq q\leq 12$
(see also (\ref{IniCondForBbosonic}))
\bea\label{BBosonicTable1}
&& {\cal B}_{4} = 917/2; \\ \nonumber
&& {\cal B}_{5} = 2514; \\ \nonumber
&& {\cal B}_{6} = 14376; \\ \nonumber
&& {\cal B}_{7} = 85152; \\ \nonumber
&& {\cal B}_{8} = 16628949/32; \\ \nonumber
&& {\cal B}_{9} = 26026877/8; \\ \nonumber
&& {\cal B}_{10} = 333148183/16; \\ \nonumber
&& {\cal B}_{11} = 543325293/4; \\ \nonumber
&& {\cal B}_{12} = 14415564199/16;  \nonumber
\eea

The coefficients introduced in formula (\ref{InfeldAsExp0}) are
\bea
&&	b_{0}=1; \\ \nonumber
&&	b_{1}=1/2; \\ \nonumber
&&	b_{2}=3/8; \\ \nonumber
&&	b_{3}=13/32; \\ \nonumber
&&	b_{4}=77/128; \\ \nonumber
&&	b_{5}=297/256; \\ \nonumber
&&	b_{6}=5727/2048; \\ \nonumber
&&	b_{7}=66687/8192; \\ \nonumber
&&	b_{8}=912303/32768; \\ \nonumber
&&	b_{9}=3586545/32768; \\ \nonumber
&&	b_{10}=127448505/262144; \\ \nonumber
&&	b_{11}=2523924765/1048576; \\ \nonumber
&&	b_{12}=110207056005/8388608; \\ \nonumber
&&	b_{13}=657259273755/8388608; \\ \nonumber
&&	b_{14}=68022530602425/134217728; \\ \nonumber
&&	b_{15}=1897008475419225/536870912; \\ \nonumber
&&	b_{16}=56719614296927925/2147483648; \\ \nonumber
&&	b_{17}=226232753142332475/1073741824; \\ \nonumber
&&	b_{18}=15346146376168947675/8589934592; \\ \nonumber
&&	b_{19}=275641831899783381375/17179869184; \\ \nonumber
&&	b_{20}=41819089838429396989125/274877906944;  \nonumber
\eea
The coefficients $c_q(n_1,n_2,n_3,n_4)$ used 
in the dimensional regularization are introduced in (\ref{d_coeff_def}):
\bea
   c_0(n_1,n_2,n_3,n_4) &=& 0, \\ \nonumber
   c_1(n_1,n_2,n_3,n_4) &=& 1/8, \nonumber
\eea
these equations at $q=0,1$ are valid for all $n_\mu$.
The coefficients $c_q=c_q(0,0,0,0)$ at $q\leq 10$ are
\bea
   c_2(0,0,0,0) &=& 1/8; \\ \nonumber
   c_3(0,0,0,0) &=& 55/384; \\ \nonumber
   c_4(0,0,0,0) &=& 5/24; \\ \nonumber
   c_5(0,0,0,0) &=& 1973/5120; \\ \nonumber
   c_6(0,0,0,0) &=& 54583/61440; \\ \nonumber
   c_7(0,0,0,0) &=& 8558131/3440640; \\ \nonumber
   c_8(0,0,0,0) &=& 4727509/573440; \\ \nonumber
   c_9(0,0,0,0) &=& 652905649/20643840; \\ \nonumber
   c_{10}(0,0,0,0) &=& 2276619691/16515072;  \nonumber
\eea
And the values some other of the coefficients $c_q(n_1,n_2,n_3,n_4)$:
\bea
   c_2(1,0,0,0) &=& 1/16; \\ \nonumber
   c_2(1,1,0,0) &=& 0; \\ \nonumber
   c_2(1,1,1,0) &=& - 1/16; \\ \nonumber
   c_2(1,1,1,1) &=& - 1/8; \\ \nonumber
   c_2(2,0,0,0) &=& 0; \\ \nonumber
   c_2(2,1,0,0) &=& - 1/16; \\ \nonumber
   c_2(2,1,1,0) &=& - 1/8; \\ \nonumber
   c_2(2,1,1,1) &=& - 3/16; \\ \nonumber
   c_3(1,0,0,0) &=&  25/384; \\ \nonumber
   c_3(1,1,0,0) &=&  7/384; \\ \nonumber
   c_3(1,1,1,0) &=&  1/384; \\ \nonumber
   c_4(1,0,0,0) &=&  27/256; \\ \nonumber
   c_4(1,1,0,0) &=&  19/384; \\ \nonumber
   c_4(1,1,1,0) &=&  19/768; \\ \nonumber
   c_4(1,1,1,1) &=&  1/64; \\ \nonumber
   c_4(2,0,0,0) &=&  55/384; \\ \nonumber
   c_5(1,0,0,0) &=&  1143/5120; \\ \nonumber
   c_5(1,1,0,0) &=&  1999/15360; \\ \nonumber
   c_5(2,0,0,0) &=&  1433/5120; \\ \nonumber
   c_6(1,0,0,0) &=&  7085/12288; \\ \nonumber
   c_6(1,1,0,0) &=&  23387/61440; \\ \nonumber
   c_6(2,0,0,0) &=&  40867/61440; \nonumber
\eea
\pagebreak 

{\Large \bf Appendix 2.}

\bea\label{DPpartQ_eq_0_dlt}
 L(0,2) &=& ( \;l_C + 1/2 \;l_C^2  ) / (2\pi)^{2} ; \\ \nonumber 
 L(0,3) &=& ( - 3/4 \mu_B^{-2} - 1/2 \; \mu_B^{-2} \;l_C + 5/8 \;l_C + 1/8 \;l_C^2 ) / (2\pi)^{2} ; \\ \nonumber 
 L(0,4) &=& ( ( - 1/6 \; \mu_B^{-4} - 1/12 \; \mu_B^{-2} + 137/960 ) \;l_C \\ \nonumber 
  &&   - 5/36 \; \mu_B^{-4} - 31/144 \; \mu_B^{-2}   + 1/32  \;l_C^2  ) / (2\pi)^{2} ; \\ \nonumber 
 L(0,5) &=& (  ( - 1/12 \; \mu_B^{-6} - 1/48 \; \mu_B^{-4}  - 1/64 \; \mu_B^{-2} + 15527/322560 ) \;l_C \\ \nonumber 
  && - 7/144 \; \mu_B^{-6}  - 101/2880 \; \mu_B^{-4} - 151/3840 \; \mu_B^{-2}+ 13/1536 \;l_C^2  ) / (2\pi)^{2} ; \\ \nonumber  
 L(0,6) &=& ( ( - 1/20 \; \mu_B^{-8} - 1/120 \; \mu_B^{-6} - 1/320 \; \mu_B^{-4} - 13/3840 \; \mu_B^{-2} + 172241/12902400 ) \;l_C  \\ \nonumber 
  && - 9/400 \; \mu_B^{-8} - 77/7200 \; \mu_B^{-6} - 709/134400  \; \mu_B^{-4} - 3953/403200 \; \mu_B^{-2} \\ \nonumber 
  && + 77/30720 \;l_C^2 ) / (2\pi)^{2} ; \\ \nonumber 
 L(0,7) &=& ( (- 1/30 \; \mu_B^{-10} - 1/240 \; \mu_B^{-8} - 1/960 \; \mu_B^{-6} \\ \nonumber 
&& - 13/23040 \; \mu_B^{-4} - 77/92160 \; \mu_B^{-2}  + 457867/94617600 ) \;l_C  \\ \nonumber 
&&  - 11/900 \; \mu_B^{-10} - 439/100800 \; \mu_B^{-8} - 53/38400 \; \mu_B^{-6} \\ \nonumber 
&&  - 5371/4838400 \; \mu_B^{-4} - 183101/77414400 \; \mu_B^{-2} + 33/40960 \;l_C^2  ) / (2\pi)^{2} ; \\ \nonumber 
 L(0,8) &=& (  ( - 1/42 \; \mu_B^{-12} - 1/420 \; \mu_B^{-10} - 1/2240 \; \mu_B^{-8} - 13/80640 \;\mu_B^{-6}\\ \nonumber 
   && - 11/92160 \; \mu_B^{-4} - 33/143360 \; \mu_B^{-2} + 51135377/31791513600 )\;l_C \\ \nonumber 
  &&  - 13/1764 \; \mu_B^{-12} - 743/352800 \; \mu_B^{-10} - 2789/5644800 \; \mu_B^{-8} \\ \nonumber 
  &&  - 3371/13547520 \;\mu_B^{-6} - 28441/121651200 \; \mu_B^{-4} - 10159/14450688  \; \mu_B^{-2} \\ \nonumber 
  &&   + 1909/6881280 \;l_C^2  ) / (2\pi)^{2} ;  \nonumber 
\eea
\bea\label{DivPart_PP_min_domain}
 D(0,2) &=&  - \;l_C  / (2\pi)^{2} ; \\ \nonumber 
 D(0,3) &=& (   1/2 \; \mu_B^{-2} - 1/4 \;l_C ) / (2\pi)^{2} ; \\ \nonumber 
 D(0,4) &=& (   1/6 \; \mu_B^{-4} + 1/12 \; \mu_B^{-2} - 1/16 \;l_C ) / (2\pi)^{2} ; \\ \nonumber 
 D(0,5) &=& (   1/12 \; \mu_B^{-6} + 1/48 \; \mu_B^{-4} + 1/64 \; \mu_B^{-2}  - 13/768 \;l_C ) / (2\pi)^{2} ; \\ \nonumber 
 D(0,6) &=& (   1/20 \; \mu_B^{-8} + 1/120 \; \mu_B^{-6} + 1/320 \; \mu_B^{-4} + 13/3840 \; \mu_B^{-2} - 77/15360 \;l_C ) / (2\pi)^{2} ; \\ \nonumber 
 D(0,7) &=& ( 1/30 \; \mu_B^{-10} + 1/240 \; \mu_B^{-8} + 1/960 \; \mu_B^{-6} + 13/23040 \; \mu_B^{-4}\\ \nonumber 
&&    + 77/92160 \; \mu_B^{-2} - 33/20480 \;l_C ) / (2\pi)^{2} ; \\ \nonumber 
 D(0,8) &=& ( 1/42 \; \mu_B^{-12} + 1/420 \; \mu_B^{-10} + 1/2240 \; \mu_B^{-8} + 13/80640 \; \mu_B^{-6}\\ \nonumber 
&&   + 11/92160 \; \mu_B^{-4} + 33/143360 \; \mu_B^{-2} - 1909/3440640 \;l_C ) / (2\pi)^{2} ; \\ \nonumber 
   D(1,1) &=& (    - \;l_C ) / (2\pi)^{2} ; \\ \nonumber 
   D(1,2) &=& (   1/2 \; \mu_B^{-2} ) / (2\pi)^{2} ; \\ \nonumber 
   D(1,3) &=& ( 1/6 \; \mu_B^{-4} + 1/48 \; \mu_B^{-2} - 11/160 \;l_C ) / (2\pi)^{2} ; \\ \nonumber 
 D(1,4) &=& ( 1/12 \;\mu_B^{-6} + 1/120 \; \mu_B^{-4} + 7/480 \; \mu_B^{-2} + 1/1680 \;l_C ) / (2\pi)^{2}; \\ \nonumber 
   D(1,5) &=& (  1/20 \; \mu_B^{-8} + 1/240 \; \mu_B^{-6} + 3/1120 \; \mu_B^{-4} + 19/21504 \; \mu_B^{-2} - 1339/215040 \;l_C ) / (2\pi)^{2} ; \\ \nonumber 
   D(1,6) &=& (  1/30 \; \mu_B^{-10} + 1/420 \; \mu_B^{-8} + 23/26880 \; \mu_B^{-6} + 13/53760 \; \mu_B^{-4} \\ \nonumber 
 && + 181/215040 \; \mu_B^{-2} - 67/630784 \;l_C ) / (2\pi)^{2} ; \\ \nonumber 
  D(1,7) &=& ( 1/42 \;\mu_B^{-12} + 1/672 \;\mu_B^{-10} + 29/80640 \;\mu_B^{-8} + 277/3225600 \;\mu_B^{-6} \\ \nonumber 
&& + 15163/141926400 \;\mu_B^{-4} + 46523/567705600 \; \mu_B^{-2} - 52001/75694080 \;l_C ) / (2\pi)^{2} ; \\ \nonumber 
  D(2,1) &=& (  1/2 \; \mu_B^{-2} + 1/4 \;l_C ) / (2\pi)^{2} ; \\ \nonumber 
 D(2,2) &=& ( 1/6 \; \mu_B^{-4} - 1/24 \; \mu_B^{-2} - 23/160 \;l_C ) / (2\pi)^{2} ; \\ \nonumber 
 D(2,3) &=& ( 1/12 \;\mu_B^{-6} - 1/240 \; \mu_B^{-4} + 1/40 \;\mu_B^{-2} + 1093/26880 \;l_C ) / (2\pi)^{2}; \\ \nonumber 
 D(2,4) &=& ( 1/20 \;\mu_B^{-8} + 13/3360 \;\mu_B^{-4} - 37/8960 \;\mu_B^{-2} - 4141/215040 \;l_C )/(2\pi)^{2};\\ \nonumber 
 D(2,5) &=& ( 1/30 \; \mu_B^{-10} + 1/1680 \; \mu_B^{-8} + 29/26880 \; \mu_B^{-6} - 1/3072 \; \mu_B^{-4} \\ \nonumber 
&& + 31/15360 \; \mu_B^{-2} + 34689/6307840 \;l_C ) / (2\pi)^{2} ; \\ \nonumber 
   D(2,6) &=& ( 1/42 \;\mu_B^{-12} + 1/1680 \; \mu_B^{-10} + 11/26880 \; \mu_B^{-8} - 17/537600 \;\mu_B^{-6} \\ \nonumber
&&  + 4747/23654400 \; \mu_B^{-4} - 36473/94617600 \; \mu_B^{-2} - 128103/50462720 \;l_C ) / (2\pi)^{2} ; \\ \nonumber 
 D(3,1) &=& (  1/6 \; \mu_B^{-4} - 5/48 \; \mu_B^{-2} - 23/80 \;l_C ) / (2\pi)^{2} ; \\ \nonumber 
 D(3,2) &=& (  1/12 \; \mu_B^{-6} - 1/60 \; \mu_B^{-4} + 3/64 \; \mu_B^{-2}  + 331/2688 \;l_C ) / (2\pi)^{2} ; \\ \nonumber 
 D(3,3) &=& ( 1/20 \; \mu_B^{-8} - 1/240 \; \mu_B^{-6} + 3/448 \;\mu_B^{-4} - 19/1344 \; \mu_B^{-2} - 12241/215040 \;l_C ) / (2\pi)^{2} ; \\ \nonumber 
   D(3,4) &=& ( 1/30 \; \mu_B^{-10} - 1/840 \; \mu_B^{-8} + 23/13440 \;\mu_B^{-6} - 19/13440 \; \mu_B^{-4} \\ \nonumber 
&& + 481/86016 \; \mu_B^{-2} +  35879/1576960 \;l_C ) / (2\pi)^{2} ; \\ \nonumber 
  D(3,5) &=& ( 1/42 \; \mu_B^{-12} - 1/3360 \; \mu_B^{-10} + 1/1680 \;\mu_B^{-8} - 53/215040 \;\mu_B^{-6} \\ \nonumber 
&& + 4801/9461760 \; \mu_B^{-4} - 33797/18923520 \; \mu_B^{-2} - 492689/50462720 \;l_C ) / (2\pi)^{2} ;  \nonumber 
\eea
\bea
   D(4,1) &=& ( 1/12 \; \mu_B^{-6} - 7/240 \; \mu_B^{-4} + 77/960 \;\mu_B^{-2} + 7201/26880 \;l_C ) / (2\pi)^{2}; \\ \nonumber 
   D(4,2) &=& ( 1/20 \; \mu_B^{-8} - 1/120 \; \mu_B^{-6} + 5/448 \;\mu_B^{-4} - 85/2688 \; \mu_B^{-2} - 29671/215040 \;l_C ) / (2\pi)^{2} ; \\ \nonumber 
   D(4,3) &=& ( 1/30 \; \mu_B^{-10} - 1/336 \; \mu_B^{-8} + 37/13440 \;\mu_B^{-6} - 19/5760 \; \mu_B^{-4} \\ \nonumber 
&& + 17267/1290240 \; \mu_B^{-2} + 51223/788480 \;l_C ) / (2\pi)^{2} ; \\ \nonumber 
   D(4,4) &=& ( 1/42 \; \mu_B^{-12} - 1/840 \; \mu_B^{-10} + 37/40320 \; \mu_B^{-8} - 11/17920 \; \mu_B^{-6} \\ \nonumber 
&& + 16853/14192640 \; \mu_B^{-4}       - 923/177408 \; \mu_B^{-2} - 2284033/75694080 \;l_C ) / (2\pi)^{2} ; \\ \nonumber 
   D(5,1) &=& ( 1/20 \; \mu_B^{-8} - 1/80 \; \mu_B^{-6} + 29/1680 \; \mu_B^{-4}  \\ \nonumber 
&& - 2117/35840 \; \mu_B^{-2} - 30869/107520 \;l_C ) / (2\pi)^{2} ; \\ \nonumber 
   D(5,2) &=& ( 1/30 \;\mu_B^{-10} - 1/210 \;\mu_B^{-8} + 113/26880 \;\mu_B^{-6} - 1009/161280 \;\mu_B^{-4} \\ \nonumber 
&& + 35873/1290240  \; \mu_B^{-2} + 482633/3153920 \;l_C ) / (2\pi)^{2} ; \\ \nonumber 
  D(5,3) &=& ( 1/42 \;\mu_B^{-12} - 1/480 \; \mu_B^{-10} + 37/26880 \; \mu_B^{-8} - 767/645120 \; \mu_B^{-6} \\ \nonumber
&& + 69689/28385280 \;\mu_B^{-4} - 116749/9461760 \; \mu_B^{-2} - 5912603/75694080 \;l_C ) / (2\pi)^{2} ; \\ \nonumber 
   D(6,1) &=& ( 1/30 \; \mu_B^{-10} - 11/1680 \; \mu_B^{-8} + 163/26880 \; \mu_B^{-6} \\ \nonumber 
&& - 3407/322560 \; \mu_B^{-4} + 66911/1290240 \; \mu_B^{-2} + 400193/1261568 \;l_C ) / (2\pi)^{2} ; \\ \nonumber 
 D(6,2) &=& ( 1/42 \; \mu_B^{-12} - 1/336 \; \mu_B^{-10} + 53/26880 \; \mu_B^{-8} - 467/230400 \;\mu_B^{-6} \\ \nonumber 
 && + 162457/35481600 \; \mu_B^{-4} - 1214401/47308800 \; \mu_B^{-2} - 26852377/151388160 \;l_C ) / (2\pi)^{2} ; \\ \nonumber 
  D(7,1) &=& (  1/42 \; \mu_B^{-12} - 13/3360 \; \mu_B^{-10} + 109/40320 \; \mu_B^{-8} - 10267/3225600 \; \mu_B^{-6}\\ \nonumber 
&&  + 53257/6758400  \; \mu_B^{-4} - 27585673/567705600 \; \mu_B^{-2} - 18404583/50462720 \;l_C ) / (2\pi)^{2};\nonumber 
\eea
\bea\label{DivPart_PM_up_to_9}
   D(2,0) &=& ( - \;l_C ) / (2\pi)^{2} ; \\ \nonumber 
   D(3,0) &=& ( 1/2 \; \mu_B^{-2} + 1/2 \;l_C ) / (2\pi)^{2} ; \\ \nonumber 
   D(3,-1) &=& (  - \;l_C ) / (2\pi)^{2} ; \\ \nonumber 
  D(4,0) &=& ( 1/6 \; \mu_B^{-4} - 1/6 \; \mu_B^{-2} - 1/2 \;l_C ) / (2\pi)^{2} ; \\ \nonumber 
  D(4,-1) &=& ( 1/2 \; \mu_B^{-2} + 3/4 \;l_C ) / (2\pi)^{2} ; \\ \nonumber 
  D(4,-2) &=& ( - \;l_C ) / (2\pi)^{2} ; \\ \nonumber 
  D(5,0) &=& ( 1/12 \; \mu_B^{-6} - 1/24 \; \mu_B^{-4} + 1/8 \; \mu_B^{-2} + 95/192 \;l_C ) / (2\pi)^{2} ; \\ \nonumber 
  D(5,-1) &=& ( 1/6 \; \mu_B^{-4} - 11/48 \; \mu_B^{-2} - 25/32 \;l_C ) / (2\pi)^{2} ; \\ \nonumber 
  D(5,-2) &=& (  1/2 \; \mu_B^{-2} + \;l_C ) / (2\pi)^{2} ; \\ \nonumber 
  D(5,-3) &=& ( - \;l_C ) / (2\pi)^{2} ; \\ \nonumber 
  D(6,0) &=& ( 1/20 \; \mu_B^{-8} - 1/60 \; \mu_B^{-6} + 1/40 \; \mu_B^{-4} - 19/192 \; \mu_B^{-2} - 1027/1920 \;l_C ) / (2\pi)^{2} ; \\ \nonumber 
  D(6,-1) &=& ( 1/12 \;\mu_B^{-6} - 13/240 \;\mu_B^{-4} + 29/160 \;\mu_B^{-2} + 3163/3840 \;l_C )/(2\pi)^{2}; \\ \nonumber 
  D(6,-2) &=& ( 1/6 \; \mu_B^{-4} - 7/24 \; \mu_B^{-2} - 181/160 \;l_C ) / (2\pi)^{2} ; \\ \nonumber 
  D(6,-3) &=& ( 1/2 \; \mu_B^{-2} + 5/4 \;l_C ) / (2\pi)^{2} ; \\ \nonumber 
  D(6,-4) &=& (  - \;l_C ) / (2\pi)^{2} ; \\ \nonumber 
  D(7,0) &=& ( 1/30 \; \mu_B^{-10} - 1/120 \; \mu_B^{-8} + 1/120 \; \mu_B^{-6} - 19/1152 \; \mu_B^{-4} \\ \nonumber 
&& + 1027/11520 \; \mu_B^{-2} + 3067/5120 \;l_C ) / (2\pi)^{2} ; \\ \nonumber 
  D(7,-1) &=& ( 1/20 \; \mu_B^{-8} - 1/48 \; \mu_B^{-6} + 11/320 \; \mu_B^{-4} - 1181/7680 \; \mu_B^{-2} - 14099/15360 \;l_C ) / (2\pi)^{2} ; \\ \nonumber 
  D(7,-2) &=& ( 1/12 \; \mu_B^{-6} - 1/15 \; \mu_B^{-4} + 239/960 \;\mu_B^{-2} + 2447/1920 \;l_C ) / (2\pi)^{2}; \\ \nonumber 
  D(7,-3) &=& ( 1/6 \; \mu_B^{-4} - 17/48 \; \mu_B^{-2} - 31/20 \;l_C ) / (2\pi)^{2} ; \\ \nonumber 
  D(7,-4) &=& ( 1/2 \; \mu_B^{-2} + 3/2 \;l_C ) / (2\pi)^{2} ; \\ \nonumber 
  D(7,-5) &=& (  - \;l_C ) / (2\pi)^{2} ; \\ \nonumber 
  D(8,0) &=& ( 1/42 \; \mu_B^{-12} - 1/210 \; \mu_B^{-10} + 1/280 \; \mu_B^{-8} - 19/4032 \; \mu_B^{-6}  \\ \nonumber 
&& + 1027/80640 \; \mu_B^{-4} - 3067/35840 \; \mu_B^{-2} - 74609/107520 \;l_C ) / (2\pi)^{2} ; \\ \nonumber 
  D(8,-1) &=& ( 1/30 \; \mu_B^{-10} - 17/1680 \; \mu_B^{-8} + 37/3360  \; \mu_B^{-6} \\ \nonumber 
&&  - 3923/161280 \; \mu_B^{-4} + 9281/64512 \; \mu_B^{-2} + 150949/143360 \;l_C ) / (2\pi)^{2} ; \\ \nonumber 
  D(8,-2) &=& (1/20 \; \mu_B^{-8} - 1/40 \; \mu_B^{-6} + 61/1344 \; \mu_B^{-4} - 405/1792 \; \mu_B^{-2} - 79489/53760 \;l_C ) / (2\pi)^{2} ; \\ \nonumber 
  D(8,-3) &=& ( 1/12 \; \mu_B^{-6} - 19/240 \; \mu_B^{-4} + 21/64 \;\mu_B^{-2} + 10037/5376 \;l_C ) / (2\pi)^{2} ; \\ \nonumber 
  D(8,-4) &=& ( 1/6 \; \mu_B^{-4} - 5/12 \; \mu_B^{-2} - 163/80 \;l_C ) / (2\pi)^{2} ; \\ \nonumber 
   D(8,-5) &=& ( 1/2 \; \mu_B^{-2} + 7/4 \;l_C ) / (2\pi)^{2} ; \\ \nonumber 
   D(8,-6) &=& (    - \;l_C ) / (2\pi)^{2} ; \\ \nonumber 
   D(9,-1) &=& ( 1/42 \; \mu_B^{-12} - 19/3360 \; \mu_B^{-10} + 41/8960 \;\mu_B^{-8} - 4303/645120 \; \mu_B^{-6}  \\ \nonumber 
&& + 50513/2580480 \; \mu_B^{-4} - 163217/1146880 \; \mu_B^{-2} - 8535263/6881280 \;l_C ) / (2\pi)^{2} ; \\ \nonumber 
   D(9,-2) &=& ( 1/30 \; \mu_B^{-10} - 1/84 \; \mu_B^{-8} + 379/26880 \; \mu_B^{-6} \\ \nonumber 
&& - 791/23040 \; \mu_B^{-4} + 5087/23040 \; \mu_B^{-2} + 501267/286720 \;l_C ) / (2\pi)^{2} ; \\ \nonumber 
   D(9,-3) &=& (       1/20 \; \mu_B^{-8} - 7/240 \; \mu_B^{-6} + 13/224 \; \mu_B^{-4} - 6841/21504 \; \mu_B^{-2} - 487141/215040 \;l_C ) / (2\pi)^{2} ; \\ \nonumber 
   D(9,-4) &=& (   1/12 \; \mu_B^{-6} - 11/120 \; \mu_B^{-4} + 67/160 \;\mu_B^{-2} + 8807/3360 \;l_C ) / (2\pi)^{2} ; \\ \nonumber 
   D(9,-5) &=& (    1/6 \; \mu_B^{-4} - 23/48 \; \mu_B^{-2} - 83/32 \;l_C ) / (2\pi)^{2} ; \\ \nonumber 
   D(9,-6) &=& (   1/2 \; \mu_B^{-2} + 2 \;l_C ) / (2\pi)^{2} ; \\ \nonumber 
   D(9,-7) &=& (    - \;l_C ) / (2\pi)^{2} ;  \nonumber 
\eea

\pagebreak[1]
{\Large \bf Appendix 3.}
\begin{verbatim}
B(p,1) = ( 1536*B(p+2,2)*(80*p^4 + 596*p^3 + 1628*p^2 + 1935*p + 846)
 + 64*B(p+1,2)*(520*p^4 + 3404*p^3 + 8106*p^2+ 8413*p+3245)
 + 48*B( p,2)*(40*p^4 + 216*p^3 + 394*p^2 + 290*p + 75)
 + 73728*B(p+4,1)*(8*p^5 + 92*p^4 + 410*p^3 + 885*p^2 + 927*p + 378)
 + 1536*B(p+3,1)*(320*p^5 + 3220*p^4 + 12772*p^3 + 24943*p^2 + 23993*p + 9110)
 + 64*B(p+2,1)*(1080*p^5 + 8260*p^4 + 24470*p^3 + 35247*p^2 + 24823*p + 6920)
 + 48*B(p+1,1)*( - 40*p^5 - 480*p^4 - 1946*p^3 - 3508*p^2 - 2911*p - 915)
 + 1152*B(p+5,-3)*(4*p^5 + 48*p^4 + 221*p^3 + 489*p^2 + 522*p + 216)
 + 12*B(p+4,-3)*( - 724*p^5 - 6504*p^4 - 22567*p^3 - 37962*p^2 - 31033*p - 9858)
 + 9*B(p+3,-3)*( - 128*p^5 - 828*p^4 - 1924*p^3 - 1961*p^2 - 867*p - 130)
 + 6*B(p+2,-3)*(32*p^5 + 180*p^4 + 296*p^3 + 15*p^2 - 328*p - 195)
 + 7296*B(p+5,-2)*(4*p^5 + 48*p^4 + 221*p^3 + 489*p^2 + 522*p + 216)
 + 4*B(p+4,-2)*( - 13756*p^5 - 133560*p^4 - 508645*p^3 - 953406*p^2
    - 881659*p - 322086)
 + 2*B(p+3,-2)*( - 3588*p^5 - 25448*p^4 - 67995*p^3 - 85874*p^2
    - 52149*p - 12410)
 + 4*B(p+2,-2)*(244*p^5 + 1600*p^4 + 3635*p^3 + 3080*p^2 + 216*p - 585)
 + 3072*B(p+5,-1)*(4*p^5 + 48*p^4 + 221*p^3 + 489*p^2 + 522*p + 216)
 + 32*B(p+4,-1)*( - 5908*p^5 - 62136*p^4 - 256375*p^3 - 518814*p^2
    - 515197*p - 201018)
 + 8*B(p+3,-1)*( - 1508*p^5 - 9252*p^4 - 19927*p^3 - 16451*p^2 - 2040*p + 2084)
 + 2*B(p+2,-1)*(1712*p^5 + 13228*p^4 + 37256*p^3 + 46893*p^2 + 26223*p + 5070)
 + 3*B(p+1,-1)*( - 64*p^5 - 472*p^4 - 1072*p^3 - 422*p^2 + 1160*p + 975)
 + 73728*B(p+5,0)*( - 4*p^5 - 48*p^4 - 221*p^3 - 489*p^2 - 522*p - 216)
 + 768*B(p+4,0)*( - 400*p^5 - 4432*p^4 - 19118*p^3 - 40113*p^2 - 41003*p - 16386)
 + 8*B(p+3,0)*(10100*p^5 + 96976*p^4 + 370679*p^3 + 704416*p^2
    + 663875*p + 247102)
 + 4*B(p+2,0)*(3220*p^5 + 22680*p^4 + 62143*p^3 + 84494*p^2 + 57851*p + 16040)
 + 4*B(p+1,0)*( - 492*p^5 - 3268*p^4 - 7609*p^3 - 7163*p^2 - 2243*p - 120)
)/12/(32*p^5 + 176*p^4 + 336*p^3 + 280*p^2 + 106*p + 15);

B(p,2) = (  2304*B(p+2,2)*(2*p^2 + 7*p + 6)
 + 32*B(p+1,2)*(4*p^2 + 12*p + 11)
 + 768*B(p+3,1)*( - 10*p^3 - 59*p^2 - 110*p - 64)
 + 128*B(p+2,1)*( - 60*p^3 - 299*p^2 - 495*p - 271)
 + 24*B(p+1,1)*( - 40*p^3 - 144*p^2 - 172*p - 69)
 + 6*B(p+4,-3)*( - 10*p^3 - 61*p^2 - 115*p - 66)
 + 12*B(p+3,-3)*p*(10*p^2 + 31*p + 22)
 + 38*B(p+4,-2)*( - 10*p^3 - 61*p^2 - 115*p - 66)
 + B(p+3,-2)*(742*p^3 + 2795*p^2 + 3167*p + 1090)
 + 18*B(p+2,-2)*p*(2*p^2 + 5*p + 3)
 + 16*B(p+4,-1)*( - 10*p^3 - 61*p^2 - 115*p - 66)
 + 4*B(p+3,-1)*(590*p^3 + 2719*p^2 + 4057*p + 1958)
 + 6*B(p+2,-1)*(20*p^3 + 50*p^2 + 84*p + 59)
 + 384*B(p+4,0)*(10*p^3 + 61*p^2 + 115*p + 66)
 + 4*B(p+3,0)*(1178*p^3 + 6705*p^2 + 12285*p + 7174)
 + 2*B(p+2,0)*( - 302*p^3 - 591*p^2 + 69*p + 416)
 + 6*B(p+1,0)*( - 10*p^3 - 23*p^2 - 10*p + 3)
)/48/(2*p + 3)/(p+1);

B(p,3) = ( 4*B(p+1,2)*( - 20*p^2 - 46*p - 23) 
 + 192*B(p+3,1)*( - 2*p^3 - 11*p^2 - 19*p - 10) 
 + 2*B(p+2,1)*( - 138*p^3 - 589*p^2 - 827*p - 376)
 + 3*B(p+1,1)*( - 4*p^3 - 14*p^2 - 14*p - 3) 
 + 3*B(p+4,-3)*( - p^3 - 6*p^2 - 11*p - 6) 
 + 6*B(p+3,-3)*p*(p^2 + 3*p + 2) 
 + 19*B(p+4,-2)*( - p^3 - 6*p^2 - 11*p - 6) 
 + 2*B(p+3,-2)*(19*p^3 + 70*p^2 + 77*p + 26) 
 + 8*B(p+4,-1)*( - p^3 - 6*p^2 - 11*p - 6)
 + B(p+3,-1)*(124*p^3 + 559*p^2 + 809*p + 374)
 + 3*B(p+2,-1)*( - 2*p^3 - 8*p^2 - 5*p + 1)
 + 192*B(p+4,0)*(p^3 + 6*p^2 + 11*p + 6)
 + 2*B(p+3,0)*(89*p^3 + 472*p^2 + 793*p + 410)
 + 2*B(p+2,0)*( - 37*p^3 - 119*p^2 - 122*p - 40)
)/12/(2*p + 3);
\end{verbatim}

{\Large \bf Appendix 4.}

Here and below, the symbol {\tt BB(p,q)} in any Appendix
designates $\cB(p,q)$, whereas {\tt B(p,q)} designates $B(p,q)$.

\begin{verbatim}
BB(p,q)= ((  24*BB(p+1,q-1)*(5*p+3)*(q-2)*(q-3) 
 + 2*BB(p+1,q-2)*(q-3)*(3+p*(20*p +21*q-28)-3*(q-5)*(2*q-5))
 + 24*BB(p+2,q-2)*(p+1)*(q-3)*(11*p+14) 
 + 4*BB(p+2,q-3)*(p+1)*(10*p*(3*p+2*q+2) -15*q^2+121*q-175)
 + BB(p+2,q-4)*(p+1)*(2*p*(5*p-4*q+26) -3*(7*q-20)*(q-4))
 + 24*BB(p+3,q-3)*(p+2)*(p+1)*(15*p+31) 
 - 2*BB(p+3,q-4)*(p+2)*(p+1)*(73*p+ 66*q - 85) 
 - 4*BB(p+3,q-5)*(p+2)*(p+1)*(10*p + 10*q - 23)
 - 6*BB(p+3,q-6)*(p+2)*(p+1)*(p + q - 3) 
 + (-192*BB(p+4,q-4)+20*BB(p+4,q-5) +3*BB(p+4,q-6))*(p+3)*(p+2)*(p+1))/12 
+muB^2*( - 24*BB(p+1,q)*(q-1)*(q-2)*(q-3) 
 + 6*BB(p+1,q-1)*(-7*p+4*q-13)*(q-2)*(q-3)
 - 120*BB(p+2,q-1)*(p+1)*(q-2)*(q-3)
 + 2*BB(p+2,q-2)*(p+1)*(-80*p+39*q-204)*(q-3) 
 + BB(p+2,q-3)*(p+1)*(2*p*(-5*p+8*q-38)+3*(21*q^2 -131*q+203))
 + 264*BB(p+3,q-2)*(p+2)*(p+1)*(-q+3)
 + 2*BB(p+3,q-3)*(p+2)*(p+1)*( - 47*p + 92*q - 383)
 + 2*BB(p+3,q-4)*(p+2)*(p+1)*(30*p + 64*q - 183)
 + 6*BB(p+3,q-5)*(p+2)*(p+1)*(3*p + 4*(q-3))
 + (- 168*BB(p+4,q-3) + 106*BB(p+4,q-4) + 31*BB(p+4,q-5) 
     + 6*BB(p+4,q-6))*(p+3)*(p+2)*(p+1))/12 
+muB^4*( - 12*BB(p+1,q)*(q-1)*(q-2)*(q-3) 
 - 18*BB(p+2,q-1)*(p+1)*(q-2)*(q-3) 
 + BB(p+2,q-2)*(p+1)*(- 8*p - 63*q + 165)*(q-3) 
 - 12*BB(p+3,q-2)*(p+2)*(p+1)*(q-3) 
 + 2*BB(p+3,q-3)*(p+2)*(p+1)*( - 10*p - 68*q + 209) 
 - 18*BB(p+3,q-4)*(p+2)*(p+1)*(p + 2*(q-3)) 
 - (6*BB(p+4,q-3)+ 61*BB(p+4,q-4)
 + 18*BB(p+4,q-5))*(p+3)*(p+2)*(p+1))/12 
+muB^6*( 7*BB(p+2,q-1)*(p+1)*(q-2)*(q-3) 
 + 16*BB(p+3,q-2)*(p+2)*(p+1)*(q-3) 
 + 2*BB(p+3,q-3)*(p+2)*(p+1)*(p + 4*(q-3)) 
 + (9*BB(p+4,q-3)+6*BB(p+4,q-4))*(p+3)*(p+2)*(p+1))/4 
+ muB^8*( - BB(p+3,q-2)*(p+2)*(p+1)*(q-3) 
 - BB(p+4,q-3)*(p+3)*(p+2)*(p+1))/2)
/(-2)/(q-1)/(q-2)/(q-3);
\end{verbatim}

{\Large \bf Appendix 5.}

\begin{verbatim}
J(0,p) = ( 6144*J(3,3+p)*( - 4428972*p^9 - 96411618*p^8
 - 738214452*p^7 - 2743289766*p^6 - 5415151530*p^5
 - 5585181043*p^4 - 2539249652*p^3 - 67504115*p^2 + 233419148*p + 30299280)
 + 256*J(3,2+p)*(139751460*p^9 + 3070483974*p^8
 + 23818829328*p^7 + 89626132062*p^6 + 178818682872*p^5
 + 185987859019*p^4 + 85026655094*p^3 + 2165966651*p^2
 - 7934451500*p - 1033526160)
 + 32*J(3,1+p)*( - 53064072*p^9 - 1189331640*p^8
 - 9510540174*p^7 - 36782676126*p^6 - 74957974878*p^5
 - 78555090042*p^4 - 34288910369*p^3 + 1761870029*p^2
 + 4642726412*p + 613660240)
 + 72*J(3,p)*( - 682164*p^9 - 21229552*p^8 - 194740397*p^7
 - 812923675*p^6 - 1742215447*p^5 - 1924011565*p^4
 - 966215776*p^3 - 94303632*p^2 + 58837408*p + 8169600)
 + 18*J(3,-1+p)*( - 69084*p^9 + 722493*p^8 + 15902908*p^7
 + 89008228*p^6 + 226205991*p^5 + 276568093*p^4
 + 134439291*p^3 - 7926512*p^2 - 20882848*p - 2804160)
 + 3072*J(2,4+p)*(4428972*p^10 + 122985450*p^9
 + 1316684160*p^8 + 7172576478*p^7 + 21874890126*p^6
 + 38076090223*p^5 + 36050335910*p^4 + 15303002027*p^3
 + 171605542*p^2 - 1430814168*p - 181795680)
 + 128*J(2,3+p)*( - 113259276*p^10 - 3205411506*p^9
 - 35040435228*p^8 - 194508285894*p^7 - 602752373652*p^6
 - 1062870299215*p^5 - 1016468592718*p^4 - 434436290355*p^3
 - 4442421788*p^2 + 41237682352*p + 5258528640)
 + 32*J(2,2+p)*(130442076*p^10 + 3413592018*p^9
 + 34321240080*p^8 + 177812525358*p^7 + 522305372877*p^6
 + 884026310037*p^5 + 817641485963*p^4 + 335486857709*p^3
 - 4288900710*p^2 - 34908660968*p - 4428920480)
 + 8*J(2,1+p)*(93154860*p^10 + 2376526194*p^9
 + 22985820378*p^8 + 114170267016*p^7 + 322016440995*p^6
 + 526074477498*p^5 + 474729514373*p^4 + 196318386058*p^3
 + 4895262676*p^2 - 16149703888*p - 2017957440)
 + 36*J(2,p)*(2556162*p^10 + 61577829*p^9
 + 565688878*p^8 + 2669754107*p^7 + 7157734171*p^6
 + 11142549507*p^5 + 9630402307*p^4 + 3844859343*p^3
 + 101897232*p^2 - 294481376*p - 35482560)
 + 18*J(2,-1+p)*( - 115758*p^10 - 3068601*p^9
 - 28796323*p^8 - 133726220*p^7 - 342937513*p^6
 - 494864596*p^5 - 374444902*p^4 - 102361975*p^3
 + 32325936*p^2 + 22856992*p + 2804160)
 + 192*J(1,4+p)*(8878356*p^10 + 259596072*p^9
 + 2924912646*p^8 + 16568071920*p^7 + 51995667138*p^6
 + 92411299412*p^5 + 88851864277*p^4 + 38204834002*p^3
 + 531726629*p^2 - 3565245012*p - 454489200)
 + 16*J(1,3+p)*( - 9079560*p^11 - 430232670*p^10
 - 7372801566*p^9 - 61656609141*p^8 - 287230567893*p^7
 - 786492878195*p^6 - 1266898022654*p^5 - 1131366942039*p^4
 - 454714067077*p^3 + 2641691135*p^2 + 44246230980*p + 5521098480)
 + 4*J(1,2+p)*(59152572*p^11 + 1593555408*p^10
 + 16912911906*p^9 + 95861230578*p^8 + 324080159613*p^7
 + 679347181198*p^6 + 874596683962*p^5 + 642957167296*p^4
 + 207762312811*p^3 - 18032380112*p^2 - 24531833312*p - 2885392320)
 + 3*J(1,1+p)*( - 1971756*p^11 - 49219953*p^10
 - 469272444*p^9 - 2325010713*p^8 - 6687694463*p^7
 - 11583569599*p^6 - 11836532108*p^5 - 6263432419*p^4
 - 618308369*p^3 + 977759600*p^2 + 412312224*p + 42062400)
 + 36*J(1,p)*p*(1593*p^10 + 20446*p^9 + 98415*p^8
 + 200300*p^7 + 49659*p^6 - 470802*p^5 - 694935*p^4
 - 123400*p^3 + 456948*p^2 + 373456*p + 88320)
 + 54*J(1,-1+p)*p^2*(378*p^9 + 4783*p^8 + 23383*p^7
 + 54276*p^6 + 53436*p^5 - 8397*p^4 - 62973*p^3
 - 49606*p^2 - 14224*p - 1056)
 + 559872*J(0,5+p)*p*( - 7*p^10 - 151*p^9 - 1386*p^8
 - 7002*p^7 - 20859*p^6 - 35475*p^5 - 26276*p^4
 + 13252*p^3 + 41616*p^2 + 29376*p + 6912)
 + 23328*J(0,4+p)*p*( - 16*p^11 - 145*p^10 + 657*p^9
 + 15080*p^8 + 91206*p^7 + 283311*p^6 + 476125*p^5
 + 331666*p^4 - 199140*p^3 - 545816*p^2 - 368832*p - 84096)
 + 1944*J(0,3+p)*p*(200*p^11 + 2865*p^10 + 14981*p^9
 + 25008*p^8 - 73738*p^7 - 437223*p^6 - 855551*p^5
 - 635826*p^4 + 311548*p^3 + 911928*p^2 + 602560*p + 133248)
 + 1944*J(0,2+p)*p*( - 70*p^11 - 1015*p^10 - 5959*p^9
 - 17811*p^8 - 26156*p^7 - 6696*p^6 + 36010*p^5
 + 49732*p^4 + 13471*p^3 - 20274*p^2 - 17296*p - 3936)
 + 486*J(0,1+p)*p*(40*p^11 + 539*p^10 + 2915*p^9
 + 8077*p^8 + 11771*p^7 + 6819*p^6 - 4591*p^5
 - 11321*p^4 - 9191*p^3 - 4018*p^2 - 944*p - 96)
 )/972/p^5/(p^7 + 12*p^6 + 54*p^5 + 108*p^4
 + 69*p^3 - 72*p^2 - 124*p - 48)
 +(
 16*B(0,p+5)*(9569448*p^11 + 263172276*p^10
 + 2796333210*p^9 + 14979144042*p^8 + 45346896150*p^7
 + 81451195130*p^6 + 86698023173*p^5 + 50610254573*p^4
 + 11013851543*p^3 - 3191145997*p^2 - 1994622412*p - 245554896)
 + 4*B(0,p+4)*( - 61193772*p^11 - 1412170434*p^10
 - 12221193447*p^9 - 55077683067*p^8 - 145501643868*p^7
 - 235023215404*p^6 - 229902193451*p^5 - 124704441431*p^4
 - 24368641382*p^3 + 8307539200*p^2 + 4697326336*p + 557574080)
 + 3*B(0,p+3)*(2924316*p^11 + 54602601*p^10
 + 399544548*p^9 + 1560379173*p^8 + 3620624261*p^7
 + 5121679611*p^6 + 4177323684*p^5 + 1389082017*p^4
 - 624082689*p^3 - 825833082*p^2 - 322097816*p - 46392864)
 + 18*B(0,p+2)*( - 25866*p^11 - 271739*p^10
 - 1238954*p^9 - 3192049*p^8 - 4973923*p^7
 - 4312062*p^6 - 536457*p^5 + 3801002*p^4 + 5378048*p^3
 + 3758272*p^2 + 1396000*p + 217728)
)/486/(p+2)^2/(p+1)^4/p^5/(p-1);

J(3, - 6 + q) = (  8*B(0,q-1)*(2*q^3 - 7*q^2 - 3*q + 18)
 + 2*B(0,q-2)*( - 13*q^3 + 109*q^2 - 298*q + 264)
 + 1536*J(3,q-3)*( - q + 3)
 + 64*J(3,q-4)*(31*q - 87)
 + 8*J(3,q-5)*( - 6*q + 13)
 + 768*J(2,q-2)*q*(q - 3)
 + 32*J(2,q-3)*( - 25*q^2 + 61*q + 30)
 + 8*J(2,q-4)*(27*q^2 - 127*q + 148)
 + 4*J(2,q-5)*(12*q^2 - 65*q + 84)
 + 6*J(2,q-6)*(q^2 - 6*q + 8)
 + 48*J(1,q-2)*(2*q^2 - 3*q - 9)
 + 4*J(1,q-3)*( - 2*q^3 - 22*q^2 + 150*q - 201)
 + J(1,q-4)*(13*q^3 - 130*q^2 + 418*q - 420)
)/6/(q - 2);

J(2, - 6 + q) = ( 16*B(0,q-1)*( - 2034*q^4 + 13237*q^3
    - 18362*q^2 - 27483*q + 55062)
 + 4*B(0,q-2)*(13197*q^4 - 150272*q^3 + 634577*q^2 - 1175522*q + 803760)
 + 12*B(0,q-3)*(q^4 - 46*q^2 + 165*q - 162)
 + 9*B(0,q-4)*( - 7*q^4 + 95*q^3 - 470*q^2 + 1000*q - 768)
 + 3072*J(3,q-3)*(1017*q^2 - 6110*q + 9177)
 + 128*J(3,q-4)*( - 31698*q^2 + 184307*q - 267447)
 + 16*J(3,q-5)*(8463*q^2 - 44580*q + 56333)
 + 1536*J(2,q-2)*q*( - 1017*q^2 + 6110*q - 9177)
 + 64*J(2,q-3)*(25677*q^3 - 140330*q^2 + 160103*q + 88998)
 + 16*J(2,q-4)*( - 28941*q^3 + 222222*q^2 - 561953*q + 468596)
 + 16*J(2,q-5)*( - 5571*q^3 + 47436*q^2 - 132356*q + 120561)
 + 96*J(1,q-2)*( - 54*q^4 - 1494*q^3 + 7171*q^2 + 3216*q - 29475)
 + 8*J(1,q-3)*(2709*q^4 + 9326*q^3 - 193019*q^2 + 616413*q - 583827)
 + 2*J(1,q-4)*( - 14142*q^4 + 181553*q^3 - 859787*q^2 + 1768658*q - 1324320)
 + 24*J(1,q-5)*(11*q^4 - 141*q^3 + 664*q^2 - 1356*q+1008)
 + 18*J(1,q-6)*(q^4 - 14*q^3 + 71*q^2 - 154*q + 120)
)/12294/(q-2)/(q-3)/(q-4);

J(1, - 6 + q) = (  8392704*J(3,q-1)*(738162*q^7 - 13826958*q^6 + 107438526*q^5
  - 446532922*q^4 + 1064669687*q^3 - 1443484631*q^2
  + 1020689473*q - 289691337)
 + 349696*J(3,q-2)*( - 23305518*q^7 + 431941470*q^6
    - 3310148214*q^5 + 13505408608*q^4 - 31391328545*q^3
    + 41045263193*q^2 - 27524008819*q + 7242283425)
 + 64*J(3,q-3)*(5997560976*q^7 - 107967926286*q^6
    + 794611952772*q^5 - 3056894462088*q^4
    + 6480621271814*q^3 - 7204980963707*q^2
    + 3391870943690*q - 197859183171)
 + 288*J(3,q-4)*q*(71372070*q^6 - 1276658510*q^5
    + 9289274641*q^4 - 35007711040*q^3 + 71462127680*q^2
    - 73822280850*q + 29283876009)
 + 36*J(3,q-5)*q*( - 24179544*q^6 + 434455572*q^5
    - 3159380738*q^4 + 11827334899*q^3 - 23816366035*q^2
    + 24112882577*q - 9374746731)
 + 4196352*J(2,q)*(-738162*q^8 + 12350634*q^7 - 79784610*q^6 + 231655870*q^5
   - 171603843*q^4 - 685854743*q^3 + 1866279789*q^2
   -1751687609*q + 579382674)
 + 174848*J(2,q-1)*(18876546*q^8 - 305743830*q^7
    + 1860247722*q^6 - 4615222660*q^5 - 558487079*q^4
    + 27753054687*q^3 - 58619476249*q^2 + 49578489187*q - 15063949524)
 + 64*J(2,q-2)*( - 14810167224*q^8 + 270992616750*q^7 - 2041549790688*q^6
  + 8136869079315*q^5 - 18296821447304*q^4 + 22800932361744*q^3
  - 14321555406258*q^2 + 3748781983207*q - 395718366342)
 + 16*J(2,q-3)*( - 10824070743*q^8 + 207455511240*q^7 - 1660383188328*q^6
  + 7176976971729*q^5 - 18064650946958*q^4 + 26521521351577*q^3
  - 21365258254495*q^2 + 7986599358662*q - 791436732684)
 + 36*J(2,q-4)*q*( - 541141251*q^7 + 10676099500*q^6
    - 88342419106*q^5 + 396420066920*q^4 - 1038067530681*q^3
    + 1577944291122*q^2 - 1278897041426*q + 420807674922)
 + 36*J(2,q-5)*q*(15986898*q^7 - 324636501*q^6
    + 2767299563*q^5-12801558868*q^4 + 34573795993*q^3
    - 54197750750*q^2 + 45244769538*q - 15277905873)
 + 262272*J(1,q)*( - 1476324*q^8 + 22486782*q^7
    - 118088346*q^6+140996162*q^5 + 996391080*q^4
    - 4565718547*q^3 + 8063013471*q^2 - 6565443637*q + 2027839359)
 + 21856*J(1,q-1)*(1479240*q^9 - 718824*q^8 - 290505888*q^7
    + 3040697422*q^6 - 14276034470*q^5 + 36530607805*q^4
    - 52418121878*q^3 + 40019451544*q^2 - 13989602036*q + 1448456685)
 + 8*J(1,q-2)*q*( - 6672113604*q^8 + 143144560506*q^7 - 1317848051532*q^6
    + 6794557026194*q^5 - 21434683657627*q^4 + 42303073328518*q^3
    - 50886066609614*q^2 + 33974478049038*q - 9569982531879)
 + 6*J(1,q-3)*q*(221313528*q^8 - 5165620026*q^7  + 51969200474*q^6
    - 293881218901*q^5 + 1019347369493*q^4 - 2213379274564*q^3
    + 2923819741196*q^2 - 2132296331629*q + 649364820429)
 + 27*J(1,q-4)*q*( - 91046*q^8 + 2191855*q^7 - 22772339*q^6
    + 133314169*q^5 - 480172079*q^4 + 1085700520*q^3
    -1495620096*q^2 + 1136307096*q-358858080)
 + 324*J(1,q-5)*q*( - 3304*q^8 + 86533*q^7 - 972784*q^6
    + 6115624*q^5 - 23438146*q^4 + 55819747*q^3
    - 80159286*q^2 + 62865696*q - 20314080)
 + 63732096*J(0,q)*q*( - q^9 + 23*q^8 - 228*q^7
    + 1278*q^6 - 4461*q^5 + 10047*q^4 - 14582*q^3
    + 13132*q^2 - 6648*q + 1440)
 + 2655504*J(0,q-1)*q*(25*q^9 - 625*q^8 + 6812*q^7
    - 42490*q^6 + 167117*q^5 - 429505*q^4 + 720358*q^3
    - 758580*q^2 + 453528*q - 116640)
 + 663876*J(0,q-2)*q*( - 35*q^9 + 945*q^8 - 11182*q^7
    + 76062*q^6 - 327403*q^5  + 923145*q^4 - 1700348*q^3 + 1964808*q^2
    - 1284552*q + 358560)
 + 663876*J(0,q-3)*q*(5*q^9 - 145*q^8 + 1845*q^7 - 13500*q^6
    + 62476*q^5 - 189120*q^4 + 373040*q^3 - 459875*q^2 + 318994*q - 93720)
 + 165969*J(0,q-4)*q*( - q^9 + 31*q^8 - 421*q^7 + 3281*q^6 - 16130*q^5
    + 51704*q^4 - 107584*q^3 + 139264*q^2 - 100864*q + 30720)
)/25515/q/(q-1)/(q-2)/(q-3)/(q-4)/(q-4)/(q-5)/(q-5)/(7*q-24)
 +(87424*B(0,q+1)*( - 1476324*q^6 + 16581486*q^5 - 47333430*q^4
    - 98082016*q^3 + 746063306*q^2 - 1287219275*q + 675946453)
 + 43712*B(0,q)*(4862637*q^6 - 78129165*q^5 + 502175106*q^4
    - 1641458936*q^3 + 2853167747*q^2 - 2474965235*q + 830365446)
 + 24*B(0,q-1)*( - 187645068*q^6 + 3303988746*q^5 - 23822963042*q^4
    + 90075021407*q^3 - 188498675784*q^2 + 207202955323*q - 93562325614)
 + 108*B(0,q-2)*( - 2349349*q^6 + 47038978*q^5 - 386433111*q^4
    + 1668058574*q^3 - 3990763652*q^2 + 5015869656*q - 2585112480)
 + 324*B(0,q-3)*(125620*q^6 - 2697750*q^5 + 23873044*q^4
    - 111347697*q^3 + 288317245*q^2 - 392099154*q + 218030520)
 + 1701*B(0,q-4)*( - 631*q^6 + 14513*q^5 - 137798*q^4
    + 690128*q^3 - 1917728*q^2 + 2792192*q - 1653760)
)/51030/(q-2)/(q-4)/(q-4)/(q-5)/(q-5)/(7*q-24);
\end{verbatim}


{\Large \bf Appendix 6.}
\begin{verbatim}
J(p,1)= -( J(p+1,0)*(123*p^3 + 325*p^2 + 141*p + 8)
/(3*(8*p^3 + 12*p^2 + 6*p + 1))+
J(p+2,0)*( - 1610*p^4 - 7315*p^3 - 12784*p^2 - 10287*p - 3208)
/(3*(16*p^4 + 48*p^3 + 48*p^2 + 20*p + 3))+
J(p+3,0)*(2*( - 10100*p^5 - 96976*p^4 - 370679*p^3 - 704416*p^2 - 663875*p
 - 247102))/(3*(32*p^5 + 176*p^4 + 336*p^3 + 280*p^2 + 106*p + 15))+
J(p+4,0)*(64*(400*p^5 + 4432*p^4 + 19118*p^3 + 40113*p^2
 + 41003*p + 16386))/(32*p^5 + 176*p^4 + 336*p^3 + 280*p^2 + 106*p + 15)+
J(p+5,0)*(6144*(2*p^4 + 21*p^3 + 79*p^2 + 126*p + 72))
/(16*p^4 + 64*p^3 + 72*p^2 + 32*p + 5)+
J(p+1,-1)*(16*p^3 + 54*p^2 - 8*p - 65)/(4*(8*p^3 + 12*p^2 + 6*p + 1))+
J(p+2,-1)*( - 856*p^4 - 4474*p^3 - 7443*p^2 - 4839*p
 - 1014)/(6*(16*p^4 + 48*p^3 + 48*p^2 + 20*p + 3))+
J(p+3,-1)*(2*(1508*p^5 + 9252*p^4 + 19927*p^3 + 16451*
p^2 + 2040*p - 2084))/(3*(32*p^5 + 176*p^4 + 336*p^3
 + 280*p^2 + 106*p + 15))+
J(p+4,-1)*(8*(5908*p^5 + 62136*p^4 + 256375*p^3 + 
518814*p^2 + 515197*p + 201018))/(3*(32*p^5
 + 176*p^4 + 336*p^3 + 280*p^2 + 106*p + 15))+
J(p+5,-1)*(256*( - 2*p^4 - 21*p^3 - 79*p^2 - 126*p - 
72))/(16*p^4 + 64*p^3 + 72*p^2 + 32*p + 5)+
J(p+2,-2)*( - 61*p^3 - 156*p^2 - 56*p + 39)
/(3*(8*p^3 + 12*p^2 + 6*p + 1))+
J(p+3,-2)*(1794*p^4 + 8239*p^3 + 13400*p^2 + 9437*p + 
2482)/(6*(16*p^4 + 48*p^3 + 48*p^2 + 20*p + 3))+
J(p+4,-2)*(13756*p^5 + 133560*p^4 + 508645*p^3 +
953406*p^2 + 881659*p + 322086)/(3*(32*p^5 +
 176*p^4 + 336*p^3 + 280*p^2 + 106*p + 15))+
J(p+5,-2)*(608*( - 2*p^4 - 21*p^3 - 79*p^2 - 126*p -
72))/(16*p^4 + 64*p^3 + 72*p^2 + 32*p + 5)+
J(p+2,-3)*( - 8*p^3 - 13*p^2 + 8*p + 13)/(2*(8*p^3 + 12*p^2 + 6*p + 1))+
J(p+3,-3)*(3*(64*p^4 + 254*p^3 + 327*p^2 + 163*p + 26)
)/(4*(16*p^4 + 48*p^3 + 48*p^2 + 20*p + 3))+
J(p+4,-3)*(724*p^5 + 6504*p^4 + 22567*p^3 + 37962*p^2
 + 31033*p + 9858)/(32*p^5 + 176*p^4 + 336*p^3 + 280*p^2 + 106*p + 15)+
J(p+5,-3)*(96*( - 2*p^4 - 21*p^3 - 79*p^2 - 126*p - 72
))/(16*p^4 + 64*p^3 + 72*p^2 + 32*p + 5)+
J(p+1,1)*(4*(10*p^3 + 80*p^2 + 129*p + 61))/(8*p^3 + 12*p^2 + 6*p + 1)+
J(p+2,1)*(16*( - 540*p^4 - 2780*p^3 - 5285*p^2 - 
4411*p - 1384))/(3*(16*p^4 + 48*p^3 + 48*p^2 + 20*p + 3))+
J(p+3,1)*(128*( - 320*p^5 - 3220*p^4 - 12772*p^3 - 
24943*p^2 - 23993*p - 9110))/(32*p^5 +
 176*p^4 + 336*p^3 + 280*p^2 + 106*p + 15)+
J(p+4,1)*(6144*( - 4*p^4 - 40*p^3 - 145*p^2 - 225*p
- 126))/(16*p^4 + 64*p^3 + 72*p^2 + 32*p + 5)+
J(p,2)*(4*( - 10*p^2 - 14*p - 5))/(8*p^3 + 12*p^2 + 6*p + 1)+
J(p+1,2)*(16*( - 260*p^3 - 1052*p^2 - 1423*p - 649))/
(3*(16*p^4 + 48*p^3 + 48*p^2 + 20*p + 3))+
J(p+2,2)*(128*( - 40*p^3 - 238*p^2 - 457*p - 282))/(
16*p^4 + 64*p^3 + 72*p^2 + 32*p + 5)+
B(p+1,0)*( - 100*p^4 + 579*p^3 + 2723*p^2 + 2859*p + 
677)/(6*(8*p^5 + 36*p^4 + 58*p^3 + 43*p^2 + 15*p + 2))+
B(p+2,0)*(2920*p^5 + 14282*p^4 + 21517*p^3 + 7182*p^2
 - 8141*p - 5168)/(6*(16*p^6 + 128*p^5 + 384*
p^4 + 548*p^3 + 391*p^2 + 135*p + 18))+
B(p+3,0)*(4*(960*p^6 - 5788*p^5 - 128500*p^4 - 629377*
p^3 - 1383907*p^2 - 1447282*p - 585016))
/(3*(32*p^7 + 400*p^6 + 1952*p^5 + 4744*p^4
 + 6098*p^3 + 4117*p^2 + 1377*p + 180))+
B(p+4,0)*(3968*(2*p^3 + 13*p^2 + 27*p + 18))/(16*p^5 +
 128*p^4 + 328*p^3 + 320*p^2 + 133*p + 20)+
B(p+1,-1)*(44*p^3 + 244*p^2 + 440*p + 259)/(4*(8*p^5 +
 36*p^4 + 58*p^3 + 43*p^2 + 15*p + 2))+
B(p+2,-1)*(1600*p^5 + 8748*p^4 + 7904*p^3 - 25623*p^2
 - 49512*p - 23133)/(12*(16*p^6 + 128*p^5 + 
384*p^4 + 548*p^3 + 391*p^2 + 135*p + 18))+
B(p+3,-1)*( - 400*p^6 - 1728*p^5 - 6008*p^4 - 71868*p^3
 - 306657*p^2 - 490625*p - 262964)/(3*(32*p^7
 + 400*p^6 + 1952*p^5 + 4744*p^4 + 6098*p^3 + 4117*p^2 + 1377*p + 180))+
B(p+4,-1)*(5920*(2*p^3 + 13*p^2 + 27*p + 18))/(16*p^5
+ 128*p^4 + 328*p^3 + 320*p^2 + 133*p + 20)+
B(p+2,-2)*(60*p^3 - 163*p^2 - 777*p - 554)
/(6*(8*p^4 + 28*p^3 + 30*p^2 + 13*p + 2))+
B(p+3,-2)*( - 120*p^4 + 6946*p^3 + 27377*p^2 + 31658*p
 + 11347)/(12*(16*p^5 + 96*p^4 + 192*p^3 + 164*p^2 + 63*p + 9))+
B(p+4,-2)*(2816*(2*p^3 + 13*p^2 + 27*p + 18))/(16*p^5 
+ 128*p^4 + 328*p^3 + 320*p^2 + 133*p + 20)+
B(p+2,-3)*(3*( - 8*p^2 - 21*p - 13))
/(2*(8*p^4 + 28*p^3 + 30*p^2 + 13*p + 2))+
B(p+3,-3)*(6*(22*p^3 + 83*p^2 + 95*p + 34))
/(16*p^5 + 96*p^4 + 192*p^3 + 164*p^2 + 63*p + 9)+
B(p+4,-3)*(576*(2*p^3 + 13*p^2 + 27*p + 18))
/(16*p^5 + 128*p^4 + 328*p^3 + 320*p^2 + 133*p + 20)+
B(p,1)*(30*p^3 + 90*p^2 + 85*p + 29)
/(4*p*(4*p^4 + 16*p^3 + 21*p^2 + 11*p + 2))+
B(p+1,1)*( - 2400*p^6 - 17780*p^5 - 43716*p^4 -
32887*p^3 + 23271*p^2 + 43533*p + 15615)
/(6*(16*p^7 + 144*p^6 + 512*p^5 + 932*p^4 + 939*p^3
 + 526*p^2 + 153*p + 18))+
B(p+2,1)*(2*( - 3600*p^7 - 67792*p^6 - 501064*p^5 - 
1910808*p^4 - 4105397*p^3 - 5021161*p^2
 - 3276852*p - 894496))/(3*(32*p^8 + 464*p^7
 + 2752*p^6 + 8648*p^5 + 15586*p^4 + 16313*p^3
 + 9611*p^2 + 2934*p + 360))+
B(p+3,1)*(64*( - 1000*p^5 - 10760*p^4 - 45112*p^3 - 
92251*p^2 - 92195*p - 36122))/(32*p^7 + 400*p^6 + 1952*p^5 + 4744*p^4
 + 6098*p^3 + 4117*p^2 + 1377*p + 180)+
B(p+4,1)*(6144*( - 2*p^3 - 13*p^2 - 27*p - 18))
/(16*p^5 + 128*p^4 + 328*p^3 + 320*p^2 + 133*p + 20)+
B(p,2)*( - 440*p^4 - 1476*p^3 - 1390*p^2 - 255*p +
 87)/(3*p*(8*p^5 + 36*p^4 + 58*p^3 + 43*p^2+15*p+2))+
B(p+1,2)*(4*( - 160*p^5 - 2736*p^4 - 12072*p^3 - 
21842*p^2 - 17255*p - 4765))/(3*(16*p^7 
+ 144*p^6 + 512*p^5 + 932*p^4 + 939*p^3 + 526*p^2 + 153*p + 18))+
B(p+2,2)*(128*(40*p^5 + 404*p^4 + 1558*p^3
 + 2847*p^2 + 2426*p + 750))/(16*p^7 + 208*p^6
 + 1064*p^5 + 2728*p^4 + 3701*p^3 + 2605*p^2 + 898*p + 120)+
B(p-1,3)*( - 16*p - 29)/(p*(4*p^3 + 12*p^2 + 9*p + 2))+
B(p,3)*(4*(88*p^3 + 452*p^2 + 762*p + 447))
/(16*p^6 + 128*p^5 + 384*p^4 + 548*p^3 + 391*p^2 + 135*p + 18) +
B(p+1,3)*(768*(4*p^5 + 42*p^4 + 167*p^3 + 308*p^2 +
 251*p + 63))/(16*p^8 + 224*p^7 + 1272*p^6
 + 3792*p^5 + 6429*p^4 + 6306*p^3 + 3503*p^2 + 1018*p + 120)           );

J(p,2)= - (J(p+1,0)*(5*p - 1)/8 + 
J(p+2,0)*(302*p^3 + 591*p^2 - 69*p - 416)
 /(24*(2*p^2 + 5*p + 3))+ 
J(p+3,0)*( - 1178*p^3 - 6705*p^2 - 12285*p - 7174)
 /(12*(2*p^2 + 5*p + 3))+ 
J(p+4,0)*(8*( - 10*p^3 - 61*p^2 - 115*p - 66))/(2*p^2 + 5*p + 3)+ 
J(p+2,-1)*( - 20*p^3 - 50*p^2 - 84*p - 59)/(8*(2*p^2 + 5*p + 3))+ 
J(p+3,-1)*( - 590*p^3 - 2719*p^2 - 4057*p - 1958)
 /(12*(2*p^2 + 5*p + 3))+ 
J(p+4,-1)*(10*p^3 + 61*p^2 + 115*p + 66)/(3*(2*p^2 + 5*p + 3))+ 
J(p+2,-2)*( - 3*p)/8+ 
J(p+3,-2)*( - 742*p^3 - 2795*p^2 - 3167*p - 1090)
 /(48*(2*p^2 + 5*p + 3))+ 
J(p+4,-2)*(19*(10*p^3 + 61*p^2 + 115*p + 66))/(24*(2*p^2 + 5*p + 3))+ 
J(p+3,-3)*(p*( - 10*p^2 - 31*p - 22))/(4*(2*p^2 + 5*p + 3))+ 
J(p+4,-3)*(10*p^3 + 61*p^2 + 115*p + 66)/(8*(2*p^2 + 5*p + 3))+ 
J(p+1,1)*(20*p^2 + 42*p + 23)/(2*(p + 1))+ 
J(p+2,1)*(8*(60*p^3 + 299*p^2 + 495*p + 271))
 /(3*(2*p^2 + 5*p + 3))+ 
J(p+3,1)*(16*(10*p^3 + 59*p^2 + 110*p + 64))/(2*p^2 + 5*p + 3)+ 
J(p+1,2)*(2*( - 4*p^2 - 12*p - 11))/(3*(2*p^2 + 5*p + 3))+ 
J(p+2,2)*(48*( - p - 2))/(p + 1)+ 
B(p+1,0)*(5*p^2 + 20*p + 9)/(8*(p^2 + 3*p + 2))+ 
B(p+2,0)*( - 438*p^4 - 1663*p^3 - 1857*p^2 - 1002*p - 542)
 /(24*(2*p^4 + 15*p^3 + 40*p^2 + 45*p + 18))+ 
B(p+3,0)*( - 144*p^3 - 1246*p^2 - 2905*p - 1978)
 /(6*(2*p^3 + 11*p^2 + 18*p + 9))+ 
B(p+2,-1)*( - 40*p^4 - 300*p^3 - 708*p^2 - 528*p - 53)
 /(8*(2*p^4 + 15*p^3 + 40*p^2 + 45*p + 18))+ 
B(p+3,-1)*(5*(12*p^3 - 292*p^2 - 985*p - 706))
 /(24*(2*p^3 + 11*p^2 + 18*p + 9))+ 
B(p+2,-2)*(3*( - p^2 - 4*p - 2))/(8*(p^2 + 3*p + 2))+ 
B(p+3,-2)*(18*p^3 - 1643*p^2 - 5213*p - 3710)
 /(48*(2*p^3 + 11*p^2 + 18*p + 9))+ 
B(p+3,-3)*(3*( - 10*p^2 - 31*p - 22))
 /(4*(2*p^3 + 11*p^2 + 18*p + 9))+ 
B(p+1,1)*(120*p^5 + 1240*p^4 + 4674*p^3  + 8128*p^2 + 6639*p + 2082)
 /(8*(2*p^5 + 17*p^4 + 55*p^3 + 85*p^2 + 63*p + 18))+ 
B(p+2,1)*(540*p^4 + 6980*p^3 + 26255*p^2 + 38886*p + 20048)
 /(12*(2*p^4 + 15*p^3 + 40*p^2 + 45*p + 18))+ 
B(p+3,1)*(8*(10*p^2 + 31*p + 22))/(2*p^3 + 11*p^2 + 18*p + 9)+ 
B(p,2)*1/(p^2 + 3*p + 2)+ 
B(p+1,2)*( - 200*p^4 - 1168*p^3 - 2588*p^2 - 2525*p - 854)
 /(6*(2*p^5 + 17*p^4 + 55*p^3 + 85*p^2 + 63*p + 18))+ 
B(p,3)*( - 20*p^2 - 60*p - 57)
 /(2*(2*p^4 + 13*p^3 + 29*p^2 + 27*p + 9))        );

J(p,3)= (   12*B(p,3)*(4*p^3 + 21*p^2 + 36*p + 21)
 + 4*B(p+1, 2)*(20*p^4 + 92*p^3 + 125*p^2 + 36*p - 23)
 + 192*B(p+3,1)*( - p^4 - 6*p^3 - 13*p^2 - 12*p - 4)
 + 2*B(p+2,1)*( - 239*p^4 - 1382*p^3 - 2873*p^2 - 2556*p - 826)
 + 3*B(p+1,1)*( - 10*p^4 - 60*p^3 - 133*p^2 - 132*p - 51)
 + 18*B(p+3,-3)*(p^4 + 6*p^3 + 13*p^2 + 12*p + 4)
 + 88*B(p+3,-2)*(p^4 + 6*p^3 + 13*p^2 + 12*p + 4)
 + 185*B(p+3,-1)*(p^4 + 6*p^3 + 13*p^2 + 12*p + 4)
 + 3*B(p+2,-1)*( - 4*p^4 - 34*p^3 - 97*p^2 - 108*p - 41)
 + 124*B(p+3,0)*(p^4 + 6*p^3 + 13*p^2 + 12*p + 4)
 + 2*B(p+2,0)*( - 103*p^4 - 570*p^3 - 1123*p^2 -948*p -292)
 + 4*J(p+1,2)*( - 20*p^5 - 166*p^4 - 519*p^3 - 764*p^2 - 529*p - 138)
 + 192*J(p+3,1)*( - 2*p^6 - 23*p^5 - 107*p^4 - 257*p^3
	 - 335*p^2 - 224*p - 60)
 + 2*J(p+2,1)*( - 138*p^6 - 1417*p^5 - 5879*p^4 - 12645*p^3
	 - 14887*p^2 - 9098*p - 2256)
 + 3*J(p+1,1)*( - 4*p^6 - 38*p^5 - 142*p^4 - 265*p^3
	 - 256*p^2 - 117*p - 18)
 + 3*J(p+4,-3)*( - p^6 - 12*p^5 - 58*p^4 - 144*p^3 - 193*p^2 - 132*p - 36)
 + 6*J(p+3,-3)*p*(p^5 + 9*p^4 + 31*p^3	 + 51*p^2 + 40*p + 12)
 + 19*J(p+4,-2)*( - p^6 - 12*p^5 - 58*p^4 - 144*p^3
	 - 193*p^2 - 132*p - 36)
 + 2*J(p+3,-2)*(19*p^6 + 184*p^5 + 706*p^4 + 1372*p^3
	 + 1423*p^2 + 748*p + 156)
 + 8*J(p+4,-1)*( - p^6 - 12*p^5 - 58*p^4 - 144*p^3
	 - 193*p^2 - 132*p - 36)
 + J(p+3,-1)*(124*p^6 + 1303*p^5 + 5527*p^4 + 12121*p^3
	 + 14497*p^2 + 8968*p + 2244)
 + 3*J(p+2,-1)*( - 2*p^6 - 20*p^5 - 75*p^4 - 129*p^3 - 97*p^2 - 19*p + 6)
 + 192*J(p+4,0)*(p^6 + 12*p^5 + 58*p^4 + 144*p^3
	 + 193*p^2 + 132*p + 36)
 + 2*J(p+3,0)*(89*p^6 + 1006*p^5 + 4604*p^4 + 10894*p^3
	 + 14015*p^2 + 9268*p + 2460)
 + 2*J(p+2,0)*( - 37*p^6 - 341*p^5 - 1243*p^4 - 2303*p^3
 	- 2296*p^2 - 1172*p - 240)
)/12/(p+1)/(p+2)/(p+3)/(2*p+3);
\end{verbatim}

{\Large \bf Appendix 7.}

Here $B(0,q,n)$ is the coefficient of the expansion 
\[
\cB(0,q;\mu_B^2)=\sum_{n=0}^{q-2} {B(0,q,n)\over (\mu_B^2)^n } + D_0(0,q)(\ln \mu_B^2 + C),
\]
(see (\ref{DPcoeffOne}) and (\ref{BBandJJat_pleq0}))
and $D(p,q,r)$ is nothing but $D_{r}(p,q)$ defined in formula (\ref{DPcoeffOne}).

\begin{verbatim}
Z_0(q) = (( - 414720*q^2 + 5184000*q + 105460920/(q-1) + 22917120/(q-2)
   + 1462140/(q-3) - 337161300/(q+1) + 5257440)*B(0,q - 4,0)
 + (2903040*q^2 - 30274560*q - 350758440/(q-1) - 49733280/(q-2)
   - 1462140/(q-3) + 1236495780/(q+1) - 26287200)*B(0,q - 3,1)
 + (8294400*q^2 - 84602880*q + 426474720/(q-1) - 12792960/(q-2)
   - 4267440/(q-3) - 1713368880/(q+1) + 420923520)*B(0,q - 3,0)
 + ( - 8709120*q^2 + 72783360*q + 388058400/(q-1) + 26816160/(q-2)
   - 1679831040/(q+1) + 52574400)*B(0,q - 2,2)
 + ( - 50595840*q^2 + 413061120*q - 681805440/(q-1) + 32651520/(q-2)
   + 4409120640/(q+1) - 1598987520)*B(0,q - 2,1)
 + ( - 58060800*q^2 + 443750400*q - 1814227200/(q-1) - 29468160/(q-2)
   + 8956742400/(q+1) - 2136844800)*B(0,q - 2,0)
 + (14515200*q^2 - 91238400*q - 142760880/(q-1) + 1003590000/(q+1)
   - 52574400)*B(0,q - 1,3)
 + (128563200*q^2 - 787968000*q + 234077760/(q-1) - 4178364480/(q+1)
   + 2271421440)*B(0,q - 1,2)
 + (306892800*q^2 - 1758412800*q + 1767352320/(q-1) - 17987028480/(q+1)
   + 6567989760)*B(0,q - 1,1)
 + (165888000*q^2 - 729907200*q + 388177920/(q-1) - 2986813440/(q+1)
   + 1305262080)*B(0,q - 1,0)
 + (414720*q^2 + 829440*q)*B(0,q + 3,7)
 + (9123840*q^2 + 18247680*q)*B(0,q + 3,6)
 + (74649600*q^2 + 149299200*q)*B(0,q + 3,5)
 + (282009600*q^2 + 564019200*q)*B(0,q + 3,4)
 + (491028480*q^2 + 982056960*q)*B(0,q + 3,3)
 + (318504960*q^2 + 637009920*q)*B(0,q + 3,2)
 + ( - 2903040*q^2 + 207360*q)*B(0,q + 2,6)
 + ( - 53913600*q^2 + 1658880*q)*B(0,q + 2,5)
 + ( - 356659200*q^2 - 20736000*q)*B(0,q + 2,4)
 + ( - 1011916800*q^2 - 265420800*q)*B(0,q + 2,3)
 + ( - 1141309440*q^2 - 915701760*q)*B(0,q + 2,2)
 + ( - 318504960*q^2 - 955514880*q)*B(0,q + 2,1)
 + (8709120*q^2 - 18662400*q + 5257440/(q+1) - 5257440)*B(0,q + 1,5)
 + (132710400*q^2 - 273715200*q - 336216960/(q+1) + 336216960)*B(0,q + 1,4)
 + (680140800*q^2 - 1277337600*q - 2232368640/(q+1) + 2232368640)*B(0,q + 1,3)
 + (1343692800*q^2 - 1891123200*q - 2815395840/(q+1)
   + 2815395840)*B(0,q + 1,2)
 + (809533440*q^2 + 212336640*q + 3530096640/(q+1) - 3530096640)*B(0,q + 1,1)
 + (955514880*q + 4140564480/(q+1) - 4140564480)*B(0,q + 1,0)
 + ( - 14515200*q^2 + 61171200*q - 228350880/(q+1) + 26287200)*B(0,q,4)
 + ( - 174182400*q^2 + 713318400*q + 1824491520/(q+1) - 1429574400)*B(0,q,3)
 + ( - 646963200*q^2 + 2463436800*q + 11243888640/(q+1) - 6663513600)*B(0,q,2)
 + ( - 779673600*q^2 + 2322432000*q + 6705192960/(q+1) - 4209131520)*B(0,q,1)
 + ( - 159252480*q^2 - 278691840*q - 11515944960/(q+1) + 5381406720)*B(0,q,0)
 + (3034260/(q-1) + 706560/(q-2) + 32940/(q-3) + 604860/(q-5)
   - 3963900/(q+1))*D(3,q - 8,1)
 + ( - 22299660/(q-1) - 4327200/(q-2) - 164700/(q-3) - 2903328/(q-4)
   - 1814580/(q-5) + 28191708/(q+1))*D(3,q - 7,2)
 + (2175210/(q-1) + 974320/(q-2) + 1169730/(q-3) - 1752336/(q-4)
   + 3125110/(q-5) - 12880514/(q+1))*D(3,q - 7,1)
 + (70144380/(q-1) + 11037600/(q-2) + 5773140/(q-3) + 8709984/(q-4)
   + 1814580/(q-5) - 85867524/(q+1))*D(3,q - 6,3)
 + ( - 36941910/(q-1) - 11226160/(q-2) + 1505790/(q-3) - 14098512/(q-4)
   - 6149410/(q-5) + 112114682/(q+1))*D(3,q - 6,2)
 + ( - 59868660/(q-1) + 3139840/(q-2) + 3407220/(q-3) - 11231232/(q-4)
   + 10685860/(q-5) + 86076892/(q+1))*D(3,q - 6,1)
 + ( - 122426100/(q-1) - 19848480/(q-2) - 16660620/(q-3) - 8709984/(q-4)
   - 604860/(q-5) + 145025724/(q+1))*D(3,q - 5,4)
 + (98060730/(q-1) + 17506320/(q-2) + 24591150/(q-3) + 28615152/(q-4)
   + 2721870/(q-5) - 303376182/(q+1))*D(3,q - 5,3)
 + (252260220/(q-1) - 25022880/(q-2) + 30365100/(q-3) - 48235296/(q-4)
   - 604860/(q-5) - 349490604/(q+1))*D(3,q - 5,2)
 + (239320080/(q-1) + 52675200/(q-2) + 9501840/(q-3) + 87594624/(q-4)
   - 16936080/(q-5) - 306353424/(q+1))*D(3,q - 5,1)
 + (129869280/(q-1) + 25993440/(q-2) + 16495920/(q-3) + 2903328/(q-4)
   - 146231568/(q+1))*D(3,q - 4,5)
 + ( - 112768560/(q-1) - 32463600/(q-2) - 49388040/(q-3) - 12764304/(q-4)
   + 425803704/(q+1))*D(3,q - 4,4)
 + ( - 543929760/(q-1) - 33583200/(q-2) + 64962720/(q-3) + 1550304/(q-4)
   + 828260736/(q+1))*D(3,q - 4,3)
 + (350288640/(q-1) - 32133120/(q-2) - 93375360/(q-3) + 81593856/(q-4)
   - 526175616/(q+1))*D(3,q - 4,2)
 + ( - 841824000/(q-1) + 77160960/(q-2) - 173082240/(q-3) + 8418816/(q-4)
   + 13071744/(q+1))*D(3,q - 4,1)
 + ( - 85719600/(q-1) - 19195200/(q-2) - 5476680/(q-3)
   + 87167160/(q+1))*D(3,q - 3,6)
 + (66969720/(q-1) + 42256800/(q-2) + 22104900/(q-3)
   - 350303580/(q+1))*D(3,q - 3,5)
 + (662632560/(q-1) - 9288000/(q-2) + 9513720/(q-3)
   - 1072048680/(q+1))*D(3,q - 3,4)
 + ( - 897347520/(q-1) - 51736320/(q-2) - 171905760/(q-3)
   + 1579946400/(q+1))*D(3,q - 3,3)
 + ( - 347765760/(q-1) + 318689280/(q-2) - 67564800/(q-3)
   + 886268160/(q+1))*D(3,q - 3,2)
 + (1177804800/(q-1) - 349470720/(q-2) + 99671040/(q-3)
   + 1938539520/(q+1))*D(3,q - 3,1)
 + (33372000/(q-1) + 5633280/(q-2) - 27393120/(q+1))*D(3,q - 2,7)
 + ( - 23814000/(q-1) - 17400960/(q-2) + 173510640/(q+1))*D(3,q - 2,6)
 + ( - 389590560/(q-1) - 33592320/(q-2) + 729384480/(q+1))*D(3,q - 2,5)
 + (592565760/(q-1) + 186209280/(q-2) - 1316528640/(q+1))*D(3,q - 2,4)
 + (1726202880/(q-1) + 196715520/(q-2) - 2554398720/(q+1))*D(3,q - 2,3)
 + ( - 2683514880/(q-1) - 91791360/(q-2) + 5111009280/(q+1))*D(3,q - 2,2)
 + ( - 1648926720/(q-1) + 139345920/(q-2) - 2179768320/(q+1))*D(3,q - 2,1)
 + ( - 5974560/(q-1) + 2656800/(q+1))*D(3,q - 1,8)
 + (4801680/(q-1) - 49314960/(q+1))*D(3,q - 1,7)
 + (108125280/(q-1) - 232264800/(q+1))*D(3,q - 1,6)
 + ( - 136183680/(q-1) + 456347520/(q+1))*D(3,q - 1,5)
 + ( - 1061475840/(q-1) + 1740510720/(q+1))*D(3,q - 1,4)
 + (173352960/(q-1) - 4119275520/(q+1))*D(3,q - 1,3)
 + (3423928320/(q-1) - 10377953280/(q+1))*D(3,q - 1,2)
 + (2229534720/(q-1) - 1592524800/(q+1))*D(3,q - 1,1)
 + 414720/(q+1)*D(3,q,9) + 6428160/(q+1)*D(3,q,8)
 + 21150720/(q+1)*D(3,q,7) - 75479040/(q+1)*D(3,q,6)
 - 296939520/(q+1)*D(3,q,5) + 1074954240/(q+1)*D(3,q,4)
 + 4804116480/(q+1)*D(3,q,3) + 4459069440/(q+1)*D(3,q,2)
 + ( - 55219230/(q-1) - 8610480/(q-2) - 247050/(q-3) - 1451664/(q-4)
   - 907290/(q-5) + 102412674/(q+1) + 1866240)*D(2,q - 7,1)
 + (172115550/(q-1) + 21855600/(q-2) + 3215970/(q-3) + 4354992/(q-4)
   + 907290/(q-5) - 327435642/(q+1) - 7464960)*D(2,q - 6,2)
 + ( - 183874230/(q-1) - 31308000/(q-2) - 3285090/(q-3) - 2452320/(q-4)
   - 2117010/(q-5) + 549853290/(q+1) - 21841920)*D(2,q - 6,1)
 + ( - 297787050/(q-1) - 31999440/(q-2) - 8659710/(q-3) - 4354992/(q-4)
   - 302430/(q-5) + 590587782/(q+1) + 17418240)*D(2,q - 5,3)
 + (477897210/(q-1) + 66902640/(q-2) + 7588890/(q-3) + 9710640/(q-4)
   + 1512150/(q-5) - 1517830410/(q+1) + 75479040)*D(2,q - 5,2)
 + (246864120/(q-1) - 42446400/(q-2) - 7535160/(q-3) + 20407872/(q-4)
   - 6048600/(q-5) - 286075752/(q+1) + 33315840)*D(2,q - 5,1)
 + (309786840/(q-1) + 29772720/(q-2) + 8412660/(q-3) + 1451664/(q-4)
   - 654424524/(q+1) - 26127360)*D(2,q - 4,4)
 + ( - 663575040/(q-1) - 71833200/(q-2) - 18628200/(q-3) - 7107984/(q-4)
   + 2314426344/(q+1) - 149022720)*D(2,q - 4,3)
 + ( - 710805600/(q-1) + 40844160/(q-2) - 36370800/(q-3) + 28281600/(q-4)
   + 1218275280/(q+1) - 73681920)*D(2,q - 4,2)
 + (1596136320/(q-1) - 11470080/(q-2) - 28428480/(q-3) - 89899776/(q-4)
   - 4522732224/(q+1) + 201000960)*D(2,q - 4,1)
 + ( - 194803920/(q-1) - 16395840/(q-2) - 2771280/(q-3)
   + 452746080/(q+1) + 26127360)*D(2,q - 3,5)
 + (518633640/(q-1) + 48381600/(q-2) + 13039380/(q-3)
   - 2107865820/(q+1) + 183859200)*D(2,q - 3,4)
 + (1012815360/(q-1) + 40938240/(q-2) - 50103360/(q-3)
   - 2232649920/(q+1) + 76723200)*D(2,q - 3,3)
 + ( - 2834017920/(q-1) + 23185920/(q-2) + 140374080/(q-3)
   + 10110626880/(q+1) - 742348800)*D(2,q - 3,2)
 + (731934720/(q-1) - 143769600/(q-2) + 135060480/(q-3)
   - 1296829440/(q+1) + 511488000)*D(2,q - 3,1)
 + (69033600/(q-1) + 3964320/(q-2) - 188030880/(q+1) - 17418240)*D(2,q - 2,6)
 + ( - 219587760/(q-1) - 17556480/(q-2) + 1152483120/(q+1)
   - 145152000)*D(2,q - 2,5)
 + ( - 728297280/(q-1) + 29093760/(q-2) + 2024331840/(q+1)
   - 28339200)*D(2,q - 2,4)
 + (2019525120/(q-1) - 57300480/(q-2) - 10609781760/(q+1)
   + 1346457600)*D(2,q - 2,3)
 + (1086750720/(q-1) - 235653120/(q-2) - 4727715840/(q+1)
   - 199065600)*D(2,q - 2,2)
 + ( - 2711162880/(q-1) + 273162240/(q-2) + 19434885120/(q+1)
   - 5240954880)*D(2,q - 2,1)
 + ( - 10711440/(q-1) + 41089680/(q+1) + 7464960)*D(2,q - 1,7)
 + (40821840/(q-1) - 357097680/(q+1) + 71608320)*D(2,q - 1,6)
 + (183746880/(q-1) - 876052800/(q+1) - 13409280)*D(2,q - 1,5)
 + ( - 500601600/(q-1) + 5522446080/(q+1) - 1304985600)*D(2,q - 1,4)
 + ( - 908236800/(q-1) + 7611770880/(q+1) - 818380800)*D(2,q - 1,3)
 + (1542205440/(q-1) - 20560711680/(q+1) + 7406346240)*D(2,q - 1,2)
 + ( - 145981440/(q-1) - 16137584640/(q+1) + 7843184640)*D(2,q - 1,1)
 + ( - 207360/(q+1) + 207360)*D(2,q + 1,9)
 + ( - 2488320/(q+1) + 2488320)*D(2,q + 1,8)
 + (3732480/(q+1) - 3732480)*D(2,q + 1,7)
 + (131880960/(q+1) - 131880960)*D(2,q + 1,6)
 + (343388160/(q+1) - 343388160)*D(2,q + 1,5)
 + ( - 806215680/(q+1) + 806215680)*D(2,q + 1,4)
 + ( - 3808788480/(q+1) + 3808788480)*D(2,q + 1,3)
 + ( - 3503554560/(q+1) + 3503554560)*D(2,q + 1,2)
 + ( - 2864160/(q+1) - 1866240)*D(2,q,8)
 + (53412480/(q+1) - 20183040)*D(2,q,7)
 + (136045440/(q+1) + 15068160)*D(2,q,6)
 + ( - 1351987200/(q+1) + 651110400)*D(2,q,5)
 + ( - 3259975680/(q+1) + 1014681600)*D(2,q,4)
 + (7019274240/(q+1) - 4265533440)*D(2,q,3)
 + (18685624320/(q+1) - 9900195840)*D(2,q,2)
 + (5414584320/(q+1) - 3503554560)*D(2,q,1)
 + (5806080*q + 525346245/(q-1) + 60022440/(q-2) + 271485/(q-3)
   + 3629160/(q-4) + 756075/(q-5) - 1441292205/(q+1) + 56987280)*D(1,q -5,1)
 + ( - 20321280*q - 1095709320/(q-1) - 90395640/(q-2) - 7076160/(q-3)
   - 3629160/(q-4) + 3353470920/(q+1) - 170961840)*D(1,q - 4,2)
 + ( - 88542720*q + 128255400/(q-1) + 20111040/(q-2) + 4887540/(q-3)
   - 3279168/(q-4) - 2442027132/(q+1) + 755926560)*D(1,q - 4,1)
 + (40642560*q + 1140907140/(q-1) + 66553680/(q-2) + 6895170/(q-3)
   - 4146890790/(q+1) + 284936400)*D(1,q - 3,3)
 + (269982720*q + 169401240/(q-1) + 995040/(q-2) + 5354100/(q-3)
   + 3574083060/(q+1) - 1710309600)*D(1,q - 3,2)
 + (385689600*q - 2864138400/(q-1) - 61000320/(q-2) + 5040720/(q-3)
   + 16590738000/(q+1) - 3805142400)*D(1,q - 3,1)
 + ( - 50803200*q - 595411200/(q-1) - 21235200/(q-2) + 2887281600/(q+1)
   - 284936400)*D(1,q - 2,4)
 + ( - 457228800*q - 378488160/(q-1) - 15764640/(q-2) - 2653613760/(q+1)
   + 2041070400)*D(1,q - 2,3)
 + ( - 1026432000*q + 2902803840/(q-1) + 28934400/(q-2) - 25745139840/(q+1)
   + 7837274880)*D(1,q - 2,2)
 + ( - 120268800*q + 479473920/(q-1) + 80524800/(q-2) - 2808011520/(q+1)
   + 539619840)*D(1,q - 2,1)
 + (40642560*q + 125452800/(q-1) - 1090694160/(q+1) + 170961840)*D(1,q - 1,5)
 + (464486400*q + 170800560/(q-1) + 1167529680/(q+1)
   - 1351296000)*D(1,q - 1,4)
 + (1451520000*q - 940357440/(q-1) + 18635400000/(q+1)
   - 7927856640)*D(1,q - 1,3)
 + (481075200*q - 833817600/(q-1) + 9558397440/(q+1)
   - 3023516160)*D(1,q - 1,2)
 + ( - 2202992640*q + 830822400/(q-1) - 27323412480/(q+1)
   + 13092710400)*D(1,q - 1,1)
 - 725760*q*D(1,q + 2,8)
 - 13893120*q*D(1,q + 2,7)
 - 85017600*q*D(1,q + 2,6)
 - 120268800*q*D(1,q + 2,5)
 + 550748160*q*D(1,q + 2,4)
 + 1897758720*q*D(1,q + 2,3)
 + 1592524800*q*D(1,q + 2,2)
 + (5806080*q - 8141040/(q+1) + 8141040)*D(1,q + 1,7)
 + (95800320*q + 66152160/(q+1) - 66152160)*D(1,q + 1,6)
 + (485222400*q + 769979520/(q+1) - 769979520)*D(1,q + 1,5)
 + (481075200*q + 1112209920/(q+1) - 1112209920)*D(1,q + 1,4)
 + ( - 2202992640*q - 3838924800/(q+1) + 3838924800)*D(1,q + 1,3)
 + ( - 4591779840*q - 9196830720/(q+1) + 9196830720)*D(1,q + 1,2)
 + ( - 1592524800*q - 3822059520/(q+1) + 3822059520)*D(1,q + 1,1)
 + ( - 20321280*q + 189367200/(q+1) - 56987280)*D(1,q,6)
 + ( - 283046400*q - 362849760/(q+1) + 468715680)*D(1,q,5)
 + ( - 1150848000*q - 6244715520/(q+1) + 3941015040)*D(1,q,4)
 + ( - 721612800*q - 6432307200/(q+1) + 3339394560)*D(1,q,3)
 + (3304488960*q + 18636963840/(q+1) - 12536709120)*D(1,q,2)
 + (3490283520*q + 19292774400/(q+1) - 13052067840)*D(1,q,1)
 + (2903040*q^2 - 27786240*q - 41109120/(q-1) - 3663360/(q-2) + 285120/(q-3)
   - 8389440/(q+1) + 81907200)*D(0,q - 3,1)
 + ( - 8709120*q^2 + 66562560*q + 43856640/(q-1) + 1451520/(q-2)
   + 74960640/(q+1) - 163814400)*D(0,q - 2,2)
 + ( - 50595840*q^2 + 396472320*q + 249246720/(q-1) + 14929920/(q-2)
   + 498908160/(q+1) - 999475200)*D(0,q - 2,1)
 + (14515200*q^2 - 82944000*q - 15344640/(q-1) - 119439360/(q+1)
   + 163814400)*D(0,q - 1,3)
 + (128563200*q^2 - 754790400*q - 135613440/(q-1) - 1170754560/(q+1)
   + 1530316800)*D(0,q - 1,2)
 + (306892800*q^2 - 1924300800*q - 282839040/(q-1) - 3881779200/(q+1)
   + 4227655680)*D(0,q - 1,1)
 + (414720*q^2 + 829440*q)*D(0,q + 3,7)
 + (9123840*q^2 + 18247680*q)*D(0,q + 3,6)
 + (74649600*q^2 + 149299200*q)*D(0,q + 3,5)
 + (282009600*q^2 + 564019200*q)*D(0,q + 3,4)
 + (491028480*q^2 + 982056960*q)*D(0,q + 3,3)
 + (318504960*q^2 + 637009920*q)*D(0,q + 3,2)
 + ( - 2903040*q^2 - 207360*q)*D(0,q + 2,6)
 + ( - 53913600*q^2 - 1658880*q)*D(0,q + 2,5)
 + ( - 356659200*q^2 + 20736000*q)*D(0,q + 2,4)
 + ( - 1011916800*q^2 + 265420800*q)*D(0,q + 2,3)
 + ( - 1141309440*q^2 + 915701760*q)*D(0,q + 2,2)
 + ( - 318504960*q^2 + 955514880*q)*D(0,q + 2,1)
 + (8709120*q^2 - 16174080*q - 16381440/(q+1) + 16381440)*D(0,q + 1,5)
 + (132710400*q^2 - 257126400*q - 265420800/(q+1) + 265420800)*D(0,q + 1,4)
 + (680140800*q^2 - 1443225600*q - 1575106560/(q+1) + 1575106560)*D(0,q + 1,3)
 + (1343692800*q^2 - 3483648000*q - 4283228160/(q+1)
   + 4283228160)*D(0,q + 1,2)
 + (809533440*q^2 - 3450470400*q - 5202247680/(q+1) + 5202247680)*D(0,q + 1,1)
 + ( - 14515200*q^2 + 54950400*q + 74649600/(q+1) - 81907200)*D(0,q,4)
 + ( - 174182400*q^2 + 680140800*q + 970444800/(q+1) - 1040947200)*D(0,q,3)
 + ( - 646963200*q^2 + 2712268800*q + 4494735360/(q+1) - 4476487680)*D(0,q,2)
 + ( - 779673600*q^2 + 3914956800*q + 8944680960/(q+1) - 7697203200)*D(0,q,1)
 + (1517130/(q-1) + 353280/(q-2) + 16470/(q-3) + 302430/(q-5)
   - 1981950/(q+1))*J(3,q - 8)
 + (6662520/(q-1) + 1568960/(q-2) + 626040/(q-3) - 150336/(q-4)
   + 2016200/(q-5) - 13488184/(q+1))*J(3,q - 7)
 + ( - 144679680/(q-1) - 25155840/(q-2) + 607680/(q-3) - 1002240/(q-4)
   - 19355520/(q-5) + 177696960/(q+1))*J(3,q - 6)
 + (129669120/(q-1) + 65802240/(q-2) - 44904960/(q-3)
   + 9621504/(q-4) + 215824896/(q+1))*J(3,q - 5)
 + (2016645120/(q-1) - 189112320/(q-2) + 113909760/(q-3)
   - 4272721920/(q+1))*J(3,q - 4)
 + ( - 4910284800/(q-1) + 159252480/(q-2) + 10590289920/(q+1))*J(3,q - 3)
 + (7585650/(q-1) + 1413120/(q-2) + 49410/(q-3) + 302430/(q-5)
   - 13873650/(q+1) - 207360)*J(2,q - 8)
 + (29684340/(q-1) + 5413440/(q-2) + 1285020/(q-3) - 150336/(q-4)
   + 604860/(q-5) - 84893004/(q+1) + 2764800)*J(2,q - 7)
 + ( - 34566480/(q-1) + 12218240/(q-2) + 3280320/(q-3) - 300672/(q-4)
   - 4032400/(q-5) + 14409632/(q+1) - 5944320)*J(2,q - 6)
 + ( - 311880960/(q-1) + 21319680/(q-2) + 2508480/(q-3) + 2004480/(q-4)
   + 19355520/(q-5) + 728736960/(q+1) - 19353600)*J(2,q - 5)
 + ( - 546186240/(q-1) - 48936960/(q-2) + 51586560/(q-3) - 9621504/(q-4)
   + 1238552064/(q+1) - 165335040)*J(2,q - 4)
 + ( - 498216960/(q-1) + 199065600/(q-2) - 113909760/(q-3)
   - 3636817920/(q+1) + 1293926400)*J(2,q - 3)
 + (4817387520/(q-1) - 159252480/(q-2) - 5613649920/(q+1)
   - 1751777280)*J(2,q - 2)
 + ( - 725760*q - 100585665/(q-1) - 14945280/(q-2) - 90495/(q-3)
   - 756075/(q-5) + 256898475/(q+1) - 8141040)*J(1,q - 6)
 + (12441600*q - 88154460/(q-1) - 10180320/(q-2) - 4797900/(q-3)
   + 375840/(q-4) + 604860/(q-5) + 650604780/(q+1) - 137954880)*J(1,q - 5)
 + ( - 60134400*q + 913299840/(q-1) + 20880000/(q-2) - 2079360/(q-3)
   - 300672/(q-4) - 4009344768/(q+1) + 724688640)*J(1,q - 4)
 + (280108800/(q-1) - 11750400/(q-2) - 9106560/(q-3) - 1342730880/(q+1)
   + 256711680)*J(1,q - 3)
 + (550748160*q - 391219200/(q-1) - 36495360/(q-2) + 11039016960/(q+1)
   - 4306452480)*J(1,q - 2)
 + ( - 796262400*q - 905748480/(q-1) - 1403412480/(q+1)
   + 2826731520)*J(1,q - 1)
 + ( - 414720*q^2 + 4769280*q + 12597120/(q-1) + 2211840/(q-2) - 285120/(q-3)
   - 5400000/(q+1) - 16381440)*J(0,q - 4)
 + (8294400*q^2 - 81285120*q - 111352320/(q-1) - 17971200/(q-2)
   + 1140480/(q-3) - 33557760/(q+1) + 244684800)*J(0,q - 3)
 + ( - 58060800*q^2 + 485222400*q + 236390400/(q-1) + 32071680/(q-2)
   + 976527360/(q+1) - 1326274560)*J(0,q - 2)
 + (165888000*q^2 - 1260748800*q - 49766400/(q-1) - 4439162880/(q+1)
   + 3413975040)*J(0,q - 1)
 + ( - 159252480*q^2 + 1552711680*q + 6210846720/(q+1) - 4379443200)*J(0,q)
  ) / 19906560;

Z_1(q) = (( - 41148*q - 194304/(q-1) - 26292/(q-2) + 246888)*B(0,q - 3,0)
 + (164592*q + 414900/(q-1) + 26292/(q-2) - 740664)*B(0,q - 2,1)
 + ( - 72000*q - 318960/(q-1) + 34320/(q-2) + 324000)*B(0,q - 2,0)
 + ( - 246888*q - 220596/(q-1) + 740664)*B(0,q - 1,2)
 + (98208*q + 178656/(q-1) - 294624)*B(0,q - 1,1)
 + (1127232*q + 1319616/(q-1) - 3381696)*B(0,q - 1,0)
 - 41148*q*B(0,q + 1,4)
 - 45792*q*B(0,q + 1,3)
 + 848448*q*B(0,q + 1,2)
 + 2377728*q*B(0,q + 1,1)
 + 1327104*q*B(0,q + 1,0)
 + (164592*q - 246888)*B(0,q,3)
 + (19584*q - 29376)*B(0,q,2)
 + ( - 2003328*q + 3004992)*B(0,q,1)
 + ( - 2529792*q + 3794688)*B(0,q,0)
 + ( - 8832/(q-1) - 1476/(q-2) - 5892/(q-4))*D(3,q - 7,1)
 + (54468/(q-1) + 7380/(q-2) + 17676/(q-3) + 17676/(q-4))*D(3,q - 6,2)
 + (2482/(q-1) - 11622/(q-2) + 12150/(q-3) - 30442/(q-4))*D(3,q - 6,1)
 + ( - 139860/(q-1) - 32436/(q-2) - 53028/(q-3) - 17676/(q-4))*D(3,q - 5,3)
 + (40598/(q-1) + 20394/(q-2) + 81390/(q-3) + 59902/(q-4))*D(3,q - 5,2)
 + (37216/(q-1) - 35580/(q-2) + 76032/(q-3) - 104092/(q-4))*D(3,q - 5,1)
 + (197292/(q-1) + 67788/(q-2) + 53028/(q-3) + 5892/(q-4))*D(3,q - 4,4)
 + ( - 74622/(q-1) - 101910/(q-2) - 169770/(q-3) - 26514/(q-4))*D(3,q - 4,3)
 + ( - 137916/(q-1) - 92724/(q-2) + 278604/(q-3) + 5892/(q-4))*D(3,q - 4,2)
 + ( - 469008/(q-1) - 116208/(q-2) - 507120/(q-3) + 164976/(q-4))*D(3,q - 4,1)
 + ( - 164916/(q-1) - 60408/(q-2) - 17676/(q-3))*D(3,q - 3,5)
 + (60978/(q-1) + 163356/(q-2) + 76230/(q-3))*D(3,q - 3,4)
 + (475212/(q-1) - 144264/(q-2) - 2772/(q-3))*D(3,q - 3,3)
 + ( - 351744/(q-1) + 64896/(q-2) - 498240/(q-3))*D(3,q - 3,2)
 + (1008576/(q-1) + 993024/(q-2) - 92736/(q-3))*D(3,q - 3,1)
 + (78048/(q-1) + 19152/(q-2))*D(3,q - 2,6)
 + ( - 36576/(q-1) - 69480/(q-2))*D(3,q - 2,5)
 + ( - 448848/(q-1) - 73296/(q-2))*D(3,q - 2,4)
 + (733248/(q-1) + 630720/(q-2))*D(3,q - 2,3)
 + (1196928/(q-1) + 450432/(q-2))*D(3,q - 2,2)
 + ( - 705024/(q-1) - 483840/(q-2))*D(3,q - 2,1)
 - 16200/(q-1)*D(3,q - 1,7)
 + 11556/(q-1)*D(3,q - 1,6)
 + 204552/(q-1)*D(3,q - 1,5)
 - 266112/(q-1)*D(3,q - 1,4)
 - 1772928/(q-1)*D(3,q - 1,3)
 - 1548288/(q-1)*D(3,q - 1,2)
 + (108198/(q-1) + 11070/(q-2) + 8838/(q-3) + 8838/(q-4) - 56700)*D(2,q - 6,1)
 + ( - 276030/(q-1) - 30978/(q-2) - 26514/(q-3) - 8838/(q-4)
   + 170100)*D(2,q - 5,2)
 + (256764/(q-1) + 58086/(q-2) + 12708/(q-3) + 20622/(q-4)
   - 218592)*D(2,q - 5,1)
 + (378366/(q-1) + 48654/(q-2) + 26514/(q-3) + 2946/(q-4)
   - 283500)*D(2,q - 4,3)
 + ( - 560046/(q-1) - 91206/(q-2) - 56898/(q-3) - 14730/(q-4)
   + 522720)*D(2,q - 4,2)
 + ( - 110904/(q-1) + 96552/(q-2) - 129432/(q-3) + 58920/(q-4)
   - 180144)*D(2,q - 4,1)
 + ( - 295938/(q-1) - 37584/(q-2) - 8838/(q-3) + 283500)*D(2,q - 3,4)
 + (605046/(q-1) + 101820/(q-2) + 42534/(q-3) - 665280)*D(2,q - 3,3)
 + (530352/(q-1) + 80592/(q-2) - 168480/(q-3) - 181440)*D(2,q - 3,2)
 + ( - 825312/(q-1) + 180864/(q-2) + 532512/(q-3) + 1518336)*D(2,q - 3,1)
 + (125892/(q-1) + 11052/(q-2) - 170100)*D(2,q - 2,5)
 + ( - 336048/(q-1) - 51912/(q-2) + 475200)*D(2,q - 2,4)
 + ( - 752208/(q-1) + 156240/(q-2) + 777024)*D(2,q - 2,3)
 + (1178112/(q-1) - 340992/(q-2) - 2460672)*D(2,q - 2,2)
 + ( - 177408/(q-1) - 767232/(q-2) - 1442304)*D(2,q - 2,1)
 + ( - 22824/(q-1) + 56700)*D(2,q - 1,6)
 + (80244/(q-1) - 180576)*D(2,q - 1,5)
 + (246960/(q-1) - 713232)*D(2,q - 1,4)
 + ( - 468288/(q-1) + 1347840)*D(2,q - 1,3)
 + ( - 582912/(q-1) + 3303936)*D(2,q - 1,2)
 + (497664/(q-1) + 110592)*D(2,q - 1,1)
 - 8100*D(2,q,7) + 28512*D(2,q,6) + 212976*D(2,q,5) - 139968*D(2,q,4)
   - 1575936*D(2,q,3) - 1603584*D(2,q,2)
 + ( - 136404*q - 761517/(q-1) - 42759/(q-2) - 22095/(q-3) - 7365/(q-4)
   + 842490)*D(1,q - 4,1)
 + (341010*q + 1163655/(q-1) + 64854/(q-2) + 22095/(q-3)
   - 1684980)*D(1,q - 3,2)
 + (144684*q + 297576/(q-1) - 56556/(q-2) + 21816/(q-3) - 366696)*D(1,q - 3,1)
 + ( - 454680*q - 797388/(q-1) - 36348/(q-2) + 1684980)*D(1,q - 2,3)
 + ( - 489456*q - 575604/(q-1) - 9540/(q-2) + 1253448)*D(1,q - 2,2)
 + (853056*q + 1411440/(q-1) - 15888/(q-2) - 3081312)*D(1,q - 2,1)
 + (341010*q + 208434/(q-1) - 842490)*D(1,q - 1,4)
 + (689544*q + 302484/(q-1) - 1304568)*D(1,q - 1,3)
 + ( - 1067616*q - 716640/(q-1) + 2909088)*D(1,q - 1,2)
 + ( - 1528128*q - 794304/(q-1) + 3600576)*D(1,q - 1,1)
 + 22734*q*D(1,q + 1,6) + 108972*q*D(1,q + 1,5)
 - 41472*q*D(1,q + 1,4) - 931392*q*D(1,q + 1,3)
 - 1520640*q*D(1,q + 1,2) - 663552*q*D(1,q + 1,1)
 + ( - 136404*q + 168498)*D(1,q,5) + ( - 444816*q + 443376)*D(1,q,4)
   + (482688*q - 816768)*D(1,q,3)
 + (2328192*q - 2827584)*D(1,q,2) + (815616*q - 1472256)*D(1,q,1)
 + ( - 8640*q^2 + 60480*q + 53568/(q-1) - 1728/(q-2) - 129600)*D(0,q - 2,1)
 + (17280*q^2 - 86400*q - 25920/(q-1) + 129600)*D(0,q - 1,2)
 + (138240*q^2 - 691200*q - 207360/(q-1) + 1036800)*D(0,q - 1,1)
 + ( - 1728*q^2 - 1728*q)*D(0,q + 2,5)
 + ( - 34560*q^2 - 34560*q)*D(0,q + 2,4)
 + ( - 241920*q^2 - 241920*q)*D(0,q + 2,3)
 + ( - 691200*q^2 - 691200*q)*D(0,q + 2,2)
 + ( - 663552*q^2 - 663552*q)*D(0,q + 2,1)
 + (8640*q^2 - 8640*q)*D(0,q + 1,4)
 + (138240*q^2 - 138240*q)*D(0,q + 1,3)
 + (725760*q^2 - 725760*q)*D(0,q + 1,2)
 + (1382400*q^2 - 1382400*q)*D(0,q + 1,1)
 + ( - 17280*q^2 + 51840*q - 43200)*D(0,q,3)
 + ( - 207360*q^2 + 622080*q - 518400)*D(0,q,2)
 + ( - 725760*q^2 + 2177280*q - 1810944)*D(0,q,1)
 + ( - 4416/(q-1) - 738/(q-2) - 2946/(q-4))*J(3,q - 7)
 + ( - 12376/(q-1) - 7656/(q-2) + 1656/(q-3) - 19640/(q-4))*J(3,q - 6)
 + (377376/(q-1) + 37632/(q-2) + 11040/(q-3) + 188544/(q-4))*J(3,q - 5)
 + ( - 1218816/(q-1) + 232704/(q-2) - 105984/(q-3))*J(3,q - 4)
 + (1216512/(q-1) - 552960/(q-2))*J(3,q - 3)
 + ( - 17664/(q-1) - 2214/(q-2) - 2946/(q-4) + 8100)*J(2,q - 7)
 + ( - 45960/(q-1) - 16788/(q-2) + 1656/(q-3) - 5892/(q-4)
   + 38016)*J(2,q - 6)
 + ( - 12400/(q-1) - 38784/(q-2) + 3312/(q-3) + 39280/(q-4)
  + 84816)*J(2,q - 5)
 + (60672/(q-1) - 60864/(q-2) - 22080/(q-3) - 188544/(q-4)
   - 265536)*J(2,q - 4)
 + (1200384/(q-1) - 221184/(q-2) + 105984/(q-3) + 36864)*J(2,q - 3)
 + ( - 1714176/(q-1) + 552960/(q-2) + 110592)*J(2,q - 2)
 + (22734*q + 186816/(q-1) + 14253/(q-2) + 7365/(q-4) - 168498)*J(1,q - 5)
 + ( - 8928*q - 18564/(q-1) + 48420/(q-2) - 4140/(q-3) - 5892/(q-4)
   - 25560)*J(1,q - 4)
 + ( - 226656*q - 636624/(q-1) + 14304/(q-2) + 3312/(q-3) + 988992)*J(1,q - 3)
 + (158976*q + 25344/(q-1) + 41472/(q-2) - 663552)*J(1,q - 2)
 + (857088*q + 290304/(q-1) - 635904)*J(1,q - 1)
 + (1728*q^2 - 15552*q - 27648/(q-1) + 1728/(q-2) + 43200)*J(0,q - 3)
 + ( - 34560*q^2 + 241920*q + 214272/(q-1) - 6912/(q-2) - 518400)*J(0,q - 2)
 + (241920*q^2 - 1209600*q - 359424/(q-1) + 1810944)*J(0,q - 1)
 + ( - 691200*q^2 + 2073600*q - 1714176)*J(0,q)
) / 41472;

Z_2(q) = ((900*q^2 - 3764*q + 1964/(q-1) + 1964)*B(0,q - 2,0)
 + ( - 2700*q^2 + 7528*q - 1964/(q-1) - 1964)*B(0,q - 1,1)
 + ( - 1776*q^2 + 3616*q - 368/(q-1) - 368)*B(0,q - 1,0)
 - 900*q^2*B(0,q + 1,3) - 3408*q^2*B(0,q + 1,2)
 - 3072*q^2*B(0,q + 1,1) + (2700*q^2 - 3764*q)*B(0,q,2)
 + (5184*q^2 - 6112*q)*B(0,q,1) + (1152*q^2 - 192*q)*B(0,q,0)
 + (156/(q-1) + 468/(q-3) + 312)*D(3,q - 6,1)
 + ( - 780/(q-1) - 624/(q-2) - 1404/(q-3) - 1560)*D(3,q - 5,2)
 + (554/(q-1) - 504/(q-2) + 2418/(q-3) + 1108)*D(3,q - 5,1)
 + (1716/(q-1) + 1872/(q-2) + 1404/(q-3) + 3120)*D(3,q - 4,3)
 + ( - 1986/(q-1) - 2648/(q-2) - 4758/(q-3) - 4896)*D(3,q - 4,2)
 + (1604/(q-1) - 3072/(q-2) + 8268/(q-3) + 2824)*D(3,q - 4,1)
 + ( - 2028/(q-1) - 1872/(q-2) - 468/(q-3) - 3120)*D(3,q - 3,4)
 + (2894/(q-1) + 5768/(q-2) + 2106/(q-3) + 6480)*D(3,q - 3,3)
 + (316/(q-1) - 9072/(q-2) - 468/(q-3) - 4376)*D(3,q - 3,2)
 + (6480/(q-1) + 16576/(q-2) - 13104/(q-3) + 10400)*D(3,q - 3,1)
 + (1248/(q-1) + 624/(q-2) + 1560)*D(3,q - 2,5)
 + ( - 2176/(q-1) - 2616/(q-2) - 3484)*D(3,q - 2,4)
 + ( - 2416/(q-1) - 240/(q-2) - 2536)*D(3,q - 2,3)
 + (12416/(q-1) + 17664/(q-2) + 21248)*D(3,q - 2,2)
 + ( - 21888/(q-1) + 5376/(q-2) - 19200)*D(3,q - 2,1)
 + ( - 312/(q-1) - 312)*D(3,q - 1,6) + (636/(q-1) + 636)*D(3,q - 1,5)
   + (3144/(q-1) + 3144)*D(3,q - 1,4)
 + ( - 9504/(q-1) - 9504)*D(3,q - 1,3) + ( - 21504/(q-1) - 21504)*D(3,q - 1,2)
 + (936*q - 1170/(q-1) - 312/(q-2) - 702/(q-3) - 1560)*D(2,q - 5,1)
 + ( - 2340*q + 2418/(q-1) + 936/(q-2) + 702/(q-3) + 3120)*D(2,q - 4,2)
 + (4496*q - 4362/(q-1) - 336/(q-2) - 1638/(q-3) - 5076)*D(2,q - 4,1)
 + (3120*q - 2574/(q-1) - 936/(q-2) - 234/(q-3) - 3120)*D(2,q - 3,3)
 + ( - 8544*q + 6642/(q-1) + 1896/(q-2) + 1170/(q-3) + 7980)*D(2,q - 3,2)
 + (6176*q - 4824/(q-1) + 4832/(q-2) - 4680/(q-3) - 3968)*D(2,q - 3,1)
 + ( - 2340*q + 1404/(q-1) + 312/(q-2) + 1560)*D(2,q - 2,4)
 + (8096*q - 4832/(q-1) - 1464/(q-2) - 5564)*D(2,q - 2,3)
 + ( - 544*q + 656/(q-1) + 5760/(q-2) + 3536)*D(2,q - 2,2)
 + ( - 22144*q - 3264/(q-1) - 18048/(q-2) - 12288)*D(2,q - 2,1)
 + (936*q - 312/(q-1) - 312)*D(2,q - 1,5) + ( - 3824*q + 1452/(q-1)
   + 1452)*D(2,q - 1,4)
 + ( - 5984*q - 480/(q-1) - 480)*D(2,q - 1,3) + (27904*q - 4800/(q-1)
   - 4800)*D(2,q - 1,2)
 + (16896*q + 18432/(q-1) + 18432)*D(2,q - 1,1)
   - 156*q*D(2,q,6) + 720*q*D(2,q,5)
 + 3216*q*D(2,q,4) - 6528*q*D(2,q,3) - 26880*q*D(2,q,2) - 18432*q*D(2,q,1)
 + (2010*q^2 - 9136*q + 5025/(q-1) + 780/(q-2) + 585/(q-3)
   + 5610)*D(1,q - 3,1)
 + ( - 4020*q^2 + 13704*q - 5220/(q-1) - 780/(q-2) - 5610)*D(1,q - 2,2)
 + ( - 1428*q^2 - 2316*q + 2532/(q-1) - 864/(q-2) + 2100)*D(1,q - 2,1)
 + (4020*q^2 - 9136*q + 1870/(q-1) + 1870)*D(1,q - 1,3)
 + (4716*q^2 - 3096*q - 12/(q-1) - 12)*D(1,q - 1,2)
 + ( - 15696*q^2 + 27232*q - 1840/(q-1) - 1840)*D(1,q - 1,1)
 + 402*q^2*D(1,q + 1,5) + 1644*q^2*D(1,q + 1,4)
 - 1776*q^2*D(1,q + 1,3) - 16128*q^2*D(1,q + 1,2) - 18432*q^2*D(1,q + 1,1)
 + ( - 2010*q^2 + 2284*q)*D(1,q,4)
 + ( - 4860*q^2 + 2836*q)*D(1,q,3)
 + (11328*q^2 - 13856*q)*D(1,q,2)
 + (37248*q^2 - 19776*q)*D(1,q,1)
 + (78/(q-1) + 234/(q-3) + 156)*J(3,q - 6)
 + (472/(q-1) - 96/(q-2) + 1560/(q-3) + 944)*J(3,q - 5)
 + ( - 5312/(q-1) - 640/(q-2) - 14976/(q-3) - 10624)*J(3,q - 4)
 + (3072/(q-1) + 6144/(q-2) + 6144)*J(3,q - 3)
 + ( - 156*q + 234/(q-1) + 234/(q-3) + 312)*J(2,q - 6)
 + ( - 944*q + 1100/(q-1) - 96/(q-2) + 468/(q-3) + 1208)*J(2,q - 5)
 + ( - 2864*q + 2048/(q-1) - 192/(q-2) - 3120/(q-3) + 912)*J(2,q - 4)
 + (1536*q + 5056/(q-1) + 1280/(q-2) + 14976/(q-3) + 10688)*J(2,q - 3)
 + (23808*q - 7680/(q-1) - 6144/(q-2) - 10752)*J(2,q - 2)
 + ( - 402*q^2 + 2284*q - 1675/(q-1) - 585/(q-3) - 1870)*J(1,q - 4)
 + ( - 72*q^2 + 2576*q - 2364/(q-1) + 240/(q-2) + 468/(q-3) - 2088)*J(1,q - 3)
 + (6144*q^2 - 11648*q + 448/(q-1) - 192/(q-2) + 352)*J(1,q - 2)
 + ( - 19200*q^2 + 9984*q + 1152/(q-1) + 1152)*J(1,q - 1)
) / 1152;

Z_3(q) = (( - 26*q^3 + 88*q^2 - 62*q)*B(0,q - 1,0)
 + ( - 26*q^3 + 26*q^2)*B(0,q + 1,2)
 + ( - 64*q^3 + 64*q^2)*B(0,q + 1,1)
 + (52*q^3 - 114*q^2 + 62*q)*B(0,q,1)
 + (16*q^3 + 24*q^2 - 40*q)*B(0,q,0)
 + ( - 12*q - 24/(q-2) - 12)*D(3,q - 5,1)
 + (48*q + 72/(q-2) + 36)*D(3,q - 4,2)
 + ( - 72*q - 124/(q-2) - 62)*D(3,q - 4,1)
 + ( - 72*q - 72/(q-2) - 36)*D(3,q - 3,3)
 + (200*q + 244/(q-2) + 122)*D(3,q - 3,2)
 + ( - 324*q - 424/(q-2) - 212)*D(3,q - 3,1)
 + (48*q + 24/(q-2) + 12)*D(3,q - 2,4)
 + ( - 160*q - 108/(q-2) - 54)*D(3,q - 2,3)
 + (184*q + 24/(q-2) + 12)*D(3,q - 2,2)
 + (1056*q + 672/(q-2) + 336)*D(3,q - 2,1)
 - 12*q*D(3,q - 1,5) + 38*q*D(3,q - 1,4) + 92*q*D(3,q - 1,3)
 - 832*q*D(3,q - 1,2) - 1536*q*D(3,q - 1,1)
 + ( - 30*q^2 + 54*q + 36/(q-2) + 18)*D(2,q - 4,1)
 + (60*q^2 - 96*q - 36/(q-2) - 18)*D(2,q - 3,2)
 + ( - 176*q^2 + 254*q + 84/(q-2) + 42)*D(2,q - 3,1)
 + ( - 60*q^2 + 84*q + 12/(q-2) + 6)*D(2,q - 2,3)
 + (240*q^2 - 336*q - 60/(q-2) - 30)*D(2,q - 2,2)
 + ( - 312*q^2 + 568*q + 240/(q-2) + 120)*D(2,q - 2,1)
 + (30*q^2 - 36*q)*D(2,q - 1,4) + ( - 144*q^2 + 182*q)*D(2,q - 1,3)
 + ( - 8*q^2 - 200*q)*D(2,q - 1,2)
 + (1472*q^2 - 1536*q)*D(2,q - 1,1) + ( - 6*q^2 + 6*q)*D(2,q,5)
 + (32*q^2 - 32*q)*D(2,q,4)
 + (104*q^2 - 104*q)*D(2,q,3) + ( - 352*q^2 + 352*q)*D(2,q,2)
 + ( - 768*q^2 + 768*q)*D(2,q,1)
 + ( - 52*q^3 + 208*q^2 - 186*q - 30/(q-2) - 15)*D(1,q - 2,1)
 + (78*q^3 - 234*q^2 + 171*q)*D(1,q - 1,2)
 + (74*q^3 + 96*q^2 - 194*q)*D(1,q - 1,1)
 + (13*q^3 - 13*q^2)*D(1,q + 1,4)
 + (58*q^3 - 58*q^2)*D(1,q + 1,3)
 + (64*q^3 - 64*q^2)*D(1,q + 1,2)
 + ( - 52*q^3 + 104*q^2 - 52*q)*D(1,q,3)
 + ( - 124*q^3 + 90*q^2 + 34*q)*D(1,q,2)
 + ( - 16*q^3 - 344*q^2 + 360*q)*D(1,q,1)
 + ( - 6*q - 12/(q-2) - 6)*J(3,q - 5)
 + ( - 48*q - 80/(q-2) - 40)*J(3,q - 4)
 + (1984*q + 768/(q-2) + 384)*J(3,q - 3)
 + (6*q^2 - 12*q - 12/(q-2) - 6)*J(2,q - 5)
 + (48*q^2 - 68*q - 24/(q-2) - 12)*J(2,q - 4)
 + (216*q^2 - 152*q + 160/(q-2) + 80)*J(2,q - 3)
 + ( - 800*q^2 - 1248*q - 768/(q-2) - 384)*J(2,q - 2)
 + (13*q^3 - 65*q^2 + 67*q + 30/(q-2) + 15)*J(1,q - 3)
 + ( - 8*q^3 - 128*q^2 + 144*q - 24/(q-2) - 12)*J(1,q - 2)
 + (96*q^2 + 240*q)*J(1,q - 1)
) / 96;
\end{verbatim}


{\Large \bf Appendix 8.}

Here and below, the symbol {\tt JJ(p,q)} in any Appendix
designates $\cJ(p,q)$, whereas {\tt J(p,q)} designates $J(p,q)$.
In computations at $q\leq 0$, $\mu_B=${\tt muB}$=0$ and
{\tt JJ(p,q)} should be replaced with {\tt J(p,q)}.

\begin{verbatim}
JJ(p,q)=(( 24*BB(p,q)*(1/(p+3)+1/(p+2)+1/(p+1))*( - q^3 + 6*q^2 - 11*q + 6) 
 + 24*BB(p+1,q-1)*(1/(p+3)+1/(p+2))*( - 5*(p+1)*q^2
    + 25*(p+1)*q - 30*(p+1) + 2*q^2 - 10*q + 12)
 + 48*BB(p+1,q-1)/(p+1)*(q^2 - 5*q + 6) 
 + 2*BB(p+1,q-2)/(p+3)*( - 8*(p+2)*(p+1)*q + 24*(p+2)*(p+1)
    - 21*(p+1)*q^2 + 163*(p+1)*q - 300*(p+1) + 6*q^3 - 42*q^2 + 96*q - 72)
 + 2*BB(p+1,q-2)/(p+2)*( - 12*(p+3)*(p+1)*q + 36*(p+3)*(p+1)
    - 21*(p+1)*q^2 + 163*(p+1)*q - 300*(p+1) + 6*q^3 - 42*q^2 + 96*q - 72) 
 + 12*BB(p+1,q-2)/(p+1)*(q^3 - 7*q^2 + 16*q - 12) 
 + 24*BB(p+2,q-2)/(p+3)*(p+1)*( - 11*(p+2)*q + 33*(p+2) + 8*q - 24) 
 + 192*BB(p+2,q-2)/(p+2)*(p+1)*(q - 3) 
 + 4*BB(p+2,q-3)/(p+3)*(p+1)*( - 20*(p+2)*q + 118*(p+2) + 15*q^2 - 81*q + 83) 
 + 4*BB(p+2,q-3)/(p+2)*(p+1)*(12*(p+3) + 15*q^2 - 81*q + 83)
 + BB(p+2,q-4)/(p+3)*(p+1)*( - 4*(p+2)*q + 46*(p+2) + 21*q^2 - 172*q + 352) 
 + BB(p+2,q-4)/(p+2)*(p+1)*(12*(p+3)*q-48*(p+3)+21*q^2 - 172*q + 352) 
 + 336*BB(p+3,q-3)/(p+3)*(p+2)*(p+1) 
 + 4*BB(p+3,q-4)/(p+3)*(p+2)*(p+1)*(33*q - 152) 
 + 4*BB(p+3,q-5)/(p+3)*(p+2)*(p+1)*(10*q - 53) 
 + 6*BB(p+3,q-6)/(p+3)*(p+2)*(p+1)*(q - 6) 
 + 24*JJ(p+1,q-1)*(5*(p+1)*q^2 - 25*(p+1)*q + 30*(p+1)
    - 2*q^2 + 10*q - 12) 
 + 2*JJ(p+1,q-2)*(12*(p+3)*(p+1)*q - 36*(p+3)*(p+1)
     + 8*(p+2)*(p+1)*q - 24*(p+2)*(p+1) + 21*(p+1)*q^2
    - 163*(p+1)*q + 300*(p+1) - 6*q^3 + 42*q^2-96*q+72)
 + 24*JJ(p+2,q-2)*(p+1)*(11*(p+2)*q - 33*(p+2) - 8*q + 24) 
 + 4*JJ(p+2,q-3)*(p+1)*(30*(p+3)*(p+2)-12*(p+3)+20*(p+2)*q
    - 118*(p+2) - 15*q^2 + 81*q - 83) 
 + JJ(p+2,q-4)*(p+1)*(10*(p+3)*(p+2) - 12*(p+3)*q+48*(p+3)
    + 4*(p+2)*q - 46*(p+2) - 21*q^2 + 172*q - 352) 
 + 24*JJ(p+3,q-3)*(p+2)*(p+1)*(15*(p+3) - 14) 
 + 2*JJ(p+3,q-4)*(p+2)*(p+1)*( - 73*(p+3) - 66*q + 304) 
 + 4*JJ(p+3,q-5)*(p+2)*(p+1)*( - 10*(p+3) - 10*q + 53)
 + 6*JJ(p+3,q-6)*(p+2)*(p+1)*( - (p+3) - q + 6) 
 - 192*JJ(p+4,q-4)*(p+3)*(p+2)*(p+1) 
 + 20*JJ(p+4,q-5)*(p+3)*(p+2)*(p+1) 
 + 3*JJ(p+4,q-6)*(p+3)*(p+2)*(p+1)
 )/12+muB^2*(
  24*BB(p+1,q)*(1/(p+3)+1/(p+2)+1/(p+1))*(q^3 - 6*q^2 + 11*q - 6) 
 + 6*BB(p+1,q-1)*(1/(p+3)+1/(p+2))*(7*(p+1)*q^2-35*(p+1)*q
     + 42*(p+1) - 4*q^3 + 26*q^2 - 54*q + 36) 
 + 12*BB(p+1,q-1)/(p+1)*( - 2*q^3 + 13*q^2 - 27*q + 18)
 + 120*BB(p+2,q-1)*(1/(p+3)+1/(p+2))*(p+1)*(q^2 - 5*q + 6) 
 + 2*BB(p+2,q-2)/(p+3)*(p+1)*(68*(p+2)*q - 204*(p+2)
    - 39*q^2 + 149*q - 96) 
 + 2*BB(p+2,q-2)/(p+2)*(p+1)*(12*(p+3)*q - 36*(p+3) - 39*q^2 + 149*q - 96) 
 + BB(p+2,q-3)/(p+3)*(p+1)*(8*(p+2)*q - 58*(p+2) - 63*q^2 + 449*q - 805) 
 + BB(p+2,q-3)/(p+2)*(p+1)*( - 24*(p+3)*q + 84*(p+3)
    - 63*q^2 + 449*q - 805) 
 + 264*BB(p+3,q-2)/(p+3)*(p+2)*(p+1)*(q - 3) 
 + 4*BB(p+3,q-3)/(p+3)*(p+2)*(p+1)*( - 46*q + 121) 
 + 2*BB(p+3,q-4)/(p+3)*(p+2)*(p+1)*( - 64*q + 273) 
 + 6*BB(p+3,q-5)/(p+3)*(p+2)*(p+1)*( - 4*q + 21) 
 + 24*JJ(p+1,q)*( - q^3 + 6*q^2 - 11*q + 6) 
 + 6*JJ(p+1,q-1)*( - 7*(p+1)*q^2 + 35*(p+1)*q
    - 42*(p+1) + 4*q^3 - 26*q^2 + 54*q - 36) 
 + 120*JJ(p+2,q-1)*(p+1)*( - q^2 + 5*q - 6) 
 + 2*JJ(p+2,q-2)*(p+1)*(-12*(p+3)*q+36*(p+3) - 68*(p+2)*q
 + 204*(p+2) + 39*q^2 - 149*q + 96)
 + JJ(p+2,q-3)*(p+1)*( - 10*(p+3)*(p+2) + 24*(p+3)*q
    -84*(p+3)-8*(p+2)*q+58*(p+2) + 63*q^2 - 449*q + 805)
 + 264*JJ(p+3,q-2)*(p+2)*(p+1)*( - q + 3)
 + 2*JJ(p+3,q-3)*(p+2)*(p+1)*( - 47*(p+3) + 92*q - 242) 
 + 2*JJ(p+3,q-4)*(p+2)*(p+1)*(30*(p+3) + 64*q - 273) 
 + 6*JJ(p+3,q-5)*(p+2)*(p+1)*(3*(p+3) + 4*q - 21) 
 - 168*JJ(p+4,q-3)*(p+3)*(p+2)*(p+1) 
 + 106*JJ(p+4,q-4)*(p+3)*(p+2)*(p+1) 
 + 31*JJ(p+4,q-5)*(p+3)*(p+2)*(p+1) 
 + 6*JJ(p+4,q-6)*(p+3)*(p+2)*(p+1)
 )/12+muB^4*(
  12*BB(p+1,q)*(1/(p+3)+1/(p+2)+1/(p+1))*(q^3 - 6*q^2 + 11*q - 6) 
 + 18*BB(p+2,q-1)*(1/(p+3)+1/(p+2))*(p+1)*(q^2 - 5*q + 6) 
 + BB(p+2,q-2)/(p+3)*(p+1)*( - 4*(p+2)*q + 12*(p+2) + 63*q^2 - 382*q + 579) 
 + BB(p+2,q-2)/(p+2)*(p+1)*(12*(p+3)*q - 36*(p+3) + 63*q^2 - 382*q + 579) 
 + 12*BB(p+3,q-2)/(p+3)*(p+2)*(p+1)*(q - 3) 
 + 2*BB(p+3,q-3)/(p+3)*(p+2)*(p+1)*(68*q - 239) 
 + 18*BB(p+3,q-4)/(p+3)*(p+2)*(p+1)*(2*q - 9) 
 + 12*JJ(p+1,q)*( - q^3 + 6*q^2 - 11*q + 6) 
 + 18*JJ(p+2,q-1)*(p+1)*( - q^2 + 5*q - 6) 
 + JJ(p+2,q-2)*(p+1)*(-12*(p+3)*q + 36*(p+3) + 4*(p+2)*q
    - 12*(p+2) - 63*q^2 + 382*q - 579) 
 + 12*JJ(p+3,q-2)*(p+2)*(p+1)*( - q + 3) 
 + 2*JJ(p+3,q-3)*(p+2)*(p+1)*( - 10*(p+3) - 68*q + 239) 
 + 18*JJ(p+3,q-4)*(p+2)*(p+1)*( - (p+3) - 2*q + 9) 
 - 6*JJ(p+4,q-3)*(p+3)*(p+2)*(p+1) 
 - 61*JJ(p+4,q-4)*(p+3)*(p+2)*(p+1) 
 - 18*JJ(p+4,q-5)*(p+3)*(p+2)*(p+1)
 )/12+muB^6*(
  7*BB(p+2,q-1)*(1/(p+3)+1/(p+2))*(p+1)*( - q^2 + 5*q - 6) 
 + 16*BB(p+3,q-2)/(p+3)*(p+2)*(p+1)*( - q + 3) 
 + 2*BB(p+3,q-3)/(p+3)*(p+2)*(p+1)*( - 4*q + 15) 
 + 7*JJ(p+2,q-1)*(p+1)*(q^2 - 5*q + 6) 
 + 16*JJ(p+3,q-2)*(p+2)*(p+1)*(q - 3) 
 + 2*JJ(p+3,q-3)*(p+2)*(p+1)*((p+3) + 4*q - 15) 
 + 9*JJ(p+4,q-3)*(p+3)*(p+2)*(p+1) 
 + 6*JJ(p+4,q-4)*(p+3)*(p+2)*(p+1)
 )/4+muB^8*(
  BB(p+3,q-2)/(p+3)*(p+2)*(p+1)*(q - 3) 
 + JJ(p+3,q-2)*(p+2)*(p+1)*( - q + 3) 
 - JJ(p+4,q-3)*(p+3)*(p+2)*(p+1)
 )/2)/(-2)/(q-1)/(q-2)/(q-3);

JJ(-3,q) = (   6*BB(0,q - 6)
 + ( - 18*muB^2 + 40)*BB(0,q - 5)
 + (18*muB^4 - 60*muB^2 + 146)*BB(0,q - 4)
 + ( - 6*muB^6 + 20*muB^4 + 94*muB^2 - 360)*BB(0,q - 3)
 + ( - 21*q^2 + 160*q - 314)*BB(-1,q - 4)
 + ((63*q^2 - 425*q + 731)*muB^2 - 60*q^2 + 324*q - 500)*BB(-1,q - 3)
 + (( - 63*q^2 + 370*q - 543)*muB^4 + (78*q^2 - 322*q + 264)*muB^2
   - 192*q + 576)*BB(-1,q - 2)
 + ((21*q^2 - 105*q + 126)*muB^6 + ( - 18*q^2 + 90*q - 108)*muB^4
   + ( - 120*q^2 + 600*q - 720)*muB^2)*BB(-1,q - 1)
 + (9*q^3 - 21*q^2 - 158*q + 420)*BB(-2,q - 2)
 + (( - 18*q^3 + 75*q^2 - 33*q - 90)*muB^2
   + 156*q^2 - 780*q + 936)*BB(-2,q - 1)
 + ((9*q^3 - 54*q^2 + 99*q - 54)*muB^4
   + (18*q^3 - 108*q^2 + 198*q - 108)*muB^2)*BB(-2,q)
 + ( - 18*q^3 + 108*q^2 - 198*q + 108)*BB(-3,q)
 + ( - 6*muB^2 - 3)*JJ(1,q - 6)
 + (18*muB^4 - 31*muB^2 - 20)*JJ(1,q - 5)
 + ( - 18*muB^6 + 61*muB^4 - 106*muB^2 + 192)*JJ(1,q - 4)
 + (6*muB^8 - 27*muB^6 + 6*muB^4 + 168*muB^2)*JJ(1,q - 3)
 + (6*q - 36)*JJ(0,q - 6)
 + (( - 24*q + 126)*muB^2 + 40*q - 212)*JJ(0,q - 5)
 + ((36*q - 162)*muB^4 + ( - 128*q + 546)*muB^2
   + 132*q - 608)*JJ(0,q - 4)
 + (( - 24*q + 90)*muB^6 + (136*q - 478)*muB^4
   + ( - 184*q + 484)*muB^2 + 336)*JJ(0,q - 3)
 + ((6*q - 18)*muB^8 + ( - 48*q + 144)*muB^6 + (12*q - 36)*muB^4
   + (264*q - 792)*muB^2)*JJ(0,q - 2)
 + ( - 21*q^2 + 168*q - 306)*JJ(-1,q - 4)
 + ((63*q^2 - 441*q + 747)*muB^2  - 60*q^2 + 244*q + 140)*JJ(-1,q - 3)
 + (( - 63*q^2 + 378*q - 567)*muB^4
   + (78*q^2 - 162*q - 216)*muB^2 - 456*q + 1368)*JJ(-1,q - 2)
 + ((21*q^2 - 105*q + 126)*muB^6 + ( - 18*q^2 + 90*q - 108)*muB^4
   + ( - 120*q^2 + 600*q - 720)*muB^2)*JJ(-1,q - 1)
 + (6*q^3 - 246*q + 576)*JJ(-2,q - 2)
 + (( - 12*q^3 + 36*q^2 + 48*q - 144)*muB^2
   + 144*q^2 - 720*q + 864)*JJ(-2,q - 1)
 + ((6*q^3 - 36*q^2 + 66*q - 36)*muB^4
   + (12*q^3 - 72*q^2 + 132*q - 72)*muB^2)*JJ(-2,q))
/ (12*q^3 - 72*q^2 + 132*q - 72);

JJ(-2,q) = ((21*q^2 - 168*q + 316)*BB(0,q - 4)
 + (( - 63*q^2 + 441*q - 757)*muB^2 + 60*q^2 - 244*q - 20)*BB(0,q - 3)
 + ((63*q^2 - 378*q + 567)*muB^4
   + ( - 78*q^2 + 162*q + 216)*muB^2 + 456*q - 1368)*BB(0,q - 2)
 + (( - 21*q^2 + 105*q - 126)*muB^6 + (18*q^2 - 90*q + 108)*muB^4
   + (120*q^2 - 600*q + 720)*muB^2)*BB(0,q - 1)
 + ( - 42*q^2 + 342*q - 648)*BB(-1,q - 2)
 + ((42*q^2 - 210*q + 252)*muB^2 - 120*q^2 + 600*q - 720)*BB(-1,q - 1)
 + (6*muB^2 + 3)*JJ(2,q - 6)
 + ( - 18*muB^4 + 31*muB^2 + 20)*JJ(2,q - 5)
 + (18*muB^6 - 61*muB^4 + 106*muB^2 - 192)*JJ(2,q - 4)
 + ( - 6*muB^8 + 27*muB^6 - 6*muB^4 - 168*muB^2)*JJ(2,q - 3)
 + ( - 6*q + 30)*JJ(1,q - 6)
 + ((24*q - 108)*muB^2 - 40*q + 172)*JJ(1,q - 5)
 + (( - 36*q + 144)*muB^4 + (128*q - 486)*muB^2 - 132*q + 462)*JJ(1,q - 4)
 + ((24*q - 84)*muB^6 + ( - 136*q + 458)*muB^4
   + (184*q - 578)*muB^2 + 24)*JJ(1,q - 3)
 + (( - 6*q + 18)*muB^8 + (48*q - 144)*muB^6
   + ( - 12*q + 36)*muB^4 + ( - 264*q + 792)*muB^2)*JJ(1,q - 2)
 + ( - 21*q^2 + 160*q - 304)*JJ(0,q - 4)
 + ((63*q^2 - 425*q + 721)*muB^2 - 60*q^2 + 324*q - 380)*JJ(0,q - 3)
 + (( - 63*q^2 + 370*q - 543)*muB^4 + (78*q^2 - 322*q + 264)*muB^2
   - 192*q + 576)*JJ(0,q - 2)
 + ((21*q^2 - 105*q + 126)*muB^6 + ( - 18*q^2 + 90*q - 108)*muB^4
   + ( - 120*q^2 + 600*q - 720)*muB^2)*JJ(0,q - 1)
 + (12*q^3 - 42*q^2 - 110*q + 384)*JJ(-1,q - 2)
 + (( - 24*q^3 + 114*q^2 - 114*q - 36)*muB^2
   + 168*q^2 - 840*q + 1008)*JJ(-1,q - 1)
 + ((12*q^3 - 72*q^2 + 132*q - 72)*muB^4
   + (24*q^3 - 144*q^2 + 264*q - 144)*muB^2)*JJ(-1,q))
/ (24*q^3 - 144*q^2 + 264*q - 144);

JJ(-1,q) = (  ( - 42*q^2 + 262*q - 408)*BB(0,q - 2)
 + ((42*q^2 - 210*q + 252)*muB^2 - 120*q^2 + 600*q - 720)*BB(0,q - 1)
 + ( - 12*muB^2 - 6)*JJ(3,q - 6)
 + (36*muB^4 - 62*muB^2 - 40)*JJ(3,q - 5) 
 + ( - 36*muB^6 + 122*muB^4 - 212*muB^2 + 384)*JJ(3,q - 4)
 + (12*muB^8 - 54*muB^6 + 12*muB^4 + 336*muB^2)*JJ(3,q - 3)
 + (6*q - 24)*JJ(2,q - 6)
 + (( - 24*q + 90)*muB^2 + 40*q - 132)*JJ(2,q - 5)
 + ((36*q - 126)*muB^4 + ( - 128*q + 426)*muB^2 + 132*q - 316)*JJ(2,q - 4)
 + (( - 24*q + 78)*muB^6 + (136*q - 438)*muB^4
   + ( - 184*q + 672)*muB^2 - 384)*JJ(2,q - 3)
 + ((6*q - 18)*muB^8 + ( - 48*q + 144)*muB^6
   + (12*q - 36)*muB^4 + (264*q - 792)*muB^2)*JJ(2,q - 2)
 + (21*q^2 - 152*q + 282)*JJ(1,q - 4)
 + (( - 63*q^2 + 409*q - 675)*muB^2 + 60*q^2 - 404*q + 660)*JJ(1,q - 3)
 + ((63*q^2 - 362*q + 519)*muB^4
   + ( - 78*q^2 + 482*q - 744)*muB^2 - 72*q + 216)*JJ(1,q - 2)
 + (( - 21*q^2 + 105*q - 126)*muB^6 + (18*q^2 - 90*q + 108)*muB^4
   + (120*q^2 - 600*q + 720)*muB^2)*JJ(1,q - 1)
 + (12*q^3 - 84*q^2 + 192*q - 144)*JJ(0,q - 2)
 + (( - 24*q^3 + 156*q^2 - 324*q + 216)*muB^2
   + 48*q^2 - 240*q + 288)*JJ(0,q - 1)
 + ((12*q^3 - 72*q^2 + 132*q - 72)*muB^4
   + (24*q^3 - 144*q^2 + 264*q - 144)*muB^2)*JJ(0,q))
 / (24*q^3 - 144*q^2 + 264*q - 144);
\end{verbatim}

{\Large \bf Appendix 9.}

\begin{verbatim}
J(p,q) = ( 24*J(p-4,q+4)*(q^3 + 6*q^2 + 11*q + 6) 
+ 2*J(p-3,q+2)*( - 6*q^3 + 21*q^2*(p-3) - 30*q^2
   + 12*q*(p-1)*(p-3) + 8*q*(p-2)*(p-3) + 5*q*(p-3)
   - 48*q + 12*(p-1)*(p-3) + 8*(p-2)*(p-3) - 16*(p-3) - 24) 
+ 24*J(p-3,q+3)*(5*q^2*(p-3) - 2*q^2 + 15*q*(p-3) - 6*q + 10*(p-3) - 4) 
+ J(p-2,q)*(p-3)*( - 21*q^2 - 12*q*(p-1) + 4*q*(p-2)
    + 4*q + 10*(p-1)*(p-2) - 30*(p-2)) 
+ 4*J(p-2,q+1)*(p-3)*( - 15*q^2 + 20*q*(p-2) - 39*q
    + 30*(p-1)*(p-2) - 12*(p-1) - 38*(p-2) + 1) 
+ 24*J(p-2,q+2)*(p-3)*(11*q*(p-2) - 8*q + 11*(p-2) - 8) 
+ 6*J(p-1,q-2)*(p-2)*(p-3)*( - q - (p-1) + 2) 
+ 4*J(p-1,q-1)*(p-2)*(p-3)*( - 10*q - 10*(p-1) + 13) 
+ 2*J(p-1,q)*(p-2)*(p-3)*( - 66*q - 73*(p-1) + 40) 
+ 24*J(p-1,q+1)*(p-2)*(p-3)*(15*(p-1) - 14) 
+ (3*J(p,q-2) + 20*J(p,q-1))*(p-1)*(p-2)*(p-3) 
+ 6*D(p-3,q+3,1)*(4*q^3 - 7*q^2*(p-3) + 22*q^2
    - 21*q*(p-3) + 38*q - 14*(p-3) + 20) 
+ 24*D(p-3,q+4,1)*( - q^3 - 6*q^2 - 11*q - 6) 
+ D(p-2,q+1,1)*(p-3)*(63*q^2 + 24*q*(p-1) - 8*q*(p-2)
    + 55*q - 10*(p-1)*(p-2) + 12*(p-1) + 26*(p-2) + 17) 
+ 2*D(p-2,q+2,1)*(p-3)*(39*q^2-12*q*(p-1) - 68*q*(p-2)
    + 163*q - 12*(p-1) - 68*(p-2) + 124) 
+ 120*D(p-2,q+3,1)*(p-3)*( - q^2 - 3*q - 2) 
+ 6*D(p-1,q-1,1)*(p-2)*(p-3)*(4*q + 3*(p-1) - 5) 
+ 2*D(p-1,q,1)*(p-2)*(p-3)*(64*q + 30*(p-1) - 17) 
+ 2*D(p-1,q+1,1)*(p-2)*(p-3)*(92*q - 47*(p-1) + 126) 
- 264*D(p-1,q+2,1)*(p-2)*(p-3)*(q + 1) 
+ (6*D(p,q-2,1) + 31*D(p,q-1,1) + 106*D(p,q,1)
    - 168*D(p,q+1,1))*(p-1)*(p-2)*(p-3) 
- 12*D(p-3,q+4,2)*(q + 1)*(q + 2)*(q + 3)
+ D(p-2,q+2,2)*(p-3)*(-63*q^2 - 12*q*(p-1) + 4*q*(p-2)
    - 122*q - 12*(p-1) + 4*(p-2) - 59) 
+ 18*D(p-2,q+3,2)*(p-3)*( - q^2 - 3*q - 2) 
+ 18*D(p-1,q,2)*(p-2)*(p-3)*( - 2*q - (p-1) + 1) 
+ 2*D(p-1,q+1,2)*(p-2)*(p-3)*( - 68*q - 10*(p-1) - 33) 
- 12*D(p-1,q+2,2)*(p-2)*(p-3)*(q + 1) 
- (18*D(p,q-1,2) + 61*D(p,q,2) + 6*D(p,q+1,2))*(p-1)*(p-2)*(p-3) 
+ 3*(p-3)*(  7*D(p - 2,q + 3,3)*(q + 1)*(q + 2)
- 2*D(p - 1,q + 2,4)*(q+1)*(p-2) 
+ 16*D(p - 1,q + 2,3)*(q+1)*(p-2) 
+ 2*D(p - 1,q + 1,3)*(4*q+p)*(p-2) 
+(- 2*D(p,q + 1,4) + 9*D(p,q + 1,3)
    + 6*D(p,q,3))*(p-1)*(p-2)
))/192/(p-1)/(p-2)/(p-3);
\end{verbatim}

{\Large \bf Appendix 10.}
\bea\hspace*{-12 mm}
&&  J(4,-1) = - 523385/2976768 \; Y_{11} - 104677/91392 \; Y_{10} - 17130665/26790912 \; Y_9 \\ \nonumber && 
 + 523385/459648 \; Y_8 + 11014687/5515776 \; Y_7 - 104677/31256064 \; Y_5 \\ \nonumber && 
 - 53544991/62512128 \; Y_4-3 \; Y_0 + 11316787/7354368 \;  (2\pi)^{-2}-3/4 \; F_0 \;  (2\pi)^{-2}; \\ \nonumber && 
  J(5,-1) =    881852823875/3876752130048 \; Y_{11} + 176370564775/119023091712 \; Y_{10} \\ \nonumber && 
 + 28894631548079/34890769170432 \; Y_9 - 881852823875/598616137728 \; Y_8 \\ \nonumber && 
 - 17115981580297/7183393652736 \; Y_7 + 176370564775/40705897365504 \; Y_5 \\ \nonumber && 
 + 86132421399433/81411794731008 \; Y_4 + 25/8 \; Y_0  \\ \nonumber && 
 + (25/32\; F_0 - 20532421134805/9577858203648 ) (2\pi)^{-2};\\ \nonumber && 
  J(6,-1) =  - 28542210469686553/93496923371077632 \; Y_{11} \\ \nonumber && 
 - 28542210469686553/14352597885911040 \; Y_{10} - 2087037111021330529/4207361551698493440 \; Y_9 \\ \nonumber && 
 + 28542210469686553/14437024932298752 \; Y_8 + 2217244724616422759/866221495937925120 \; Y_7 \\ \nonumber && 
 - 28542210469686553/4908588476981575680 \; Y_5 - 12694496481519981287/9817176953963151360 \; Y_4  \\ \nonumber && 
 - 3163/960 \; Y_0 + (3108094517023801819/1154961994583900160 - 3163/3840 \; F_0 ) (2\pi)^{-2}; \\ \nonumber && 
  J(7,-1) = 318815426788929230845117/821909762068921372901376 \; Y_{11} \\ \nonumber && 
 + 318815426788929230845117/126170358212334421278720 \; Y_{10} \\ \nonumber &&
 + 11542347686773160604861797/36985939293101461780561920 \; Y_9 \\ \nonumber && 
 - 318815426788929230845117/126912536790054035521536 \; Y_8 \\ \nonumber && 
 - 22134243295873177135745923/7614752207403242131292160 \; Y_7 \\ \nonumber && 
 + 318815426788929230845117/43150262508618372077322240 \; Y_5 \\ \nonumber && 
 + 135786390066815073864737539/86300525017236744154644480 \; Y_4 + 14099/3840 \; Y_0 \\ \nonumber && 
 - 34028993605943331376331719/10153002943204322841722880 \;  (2\pi)^{-2} + 14099/15360 \; F_0 \;  (2\pi)^{-2}; \\ \nonumber && 
  J(8,-1) = - 9235026172226013567109728738337/18732046758141338739230512250880 \; Y_{11} \\ \nonumber && 
 - 9235026172226013567109728738337/2875533493574328315232754073600 \; Y_{10} \\ \nonumber && 
 - 55876713828271512392254716439529/842942104116360243265373051289600 \; Y_9 \\ \nonumber && 
 + 9235026172226013567109728738337/2892448396477706717087064391680 \; Y_8 \\ \nonumber && 
 + 582625738707581471281711807374367/173546903788662403025223863500800 \; Y_7 \\ \nonumber && 
 - 9235026172226013567109728738337/983432454802420283809601893171200 \; Y_5 \\ \nonumber && 
 - 3802005063193739791001963690889631/1966864909604840567619203786342400 \; Y_4 \\ \nonumber && 
 - 150949/35840 \; Y_0 + 965476530505339393064578836261427/ \\ \nonumber && 
   231395871718216537366965151334400 \;  (2\pi)^{-2} - 150949/143360 \; F_0 \;  (2\pi)^{-2}; \\ \nonumber && 
  J(9,-1) =  \\ \nonumber && 
   10189310117043029375946591507272718551/16263612564540505685529672268508037120 \; Y_{11} \\ \nonumber && 
 + 10189310117043029375946591507272718551/2496607191925077627164642672797286400 \; Y_{10} \\ \nonumber && 
 - 150087551903588632763072238483785668913/731862565404322755848835252082861670400 \; Y_9 \\ \nonumber && 
 - 10189310117043029375946591507272718551/2511293116583460436736199394401976320 \; Y_8 \\ \nonumber && 
 - 598626882055818268392045217522764699881/150677586995007626204171963664118579200 \; Y_7 \\ \nonumber && 
 + 10189310117043029375946591507272718551/853839659638376548490307794096671948800 \; Y_5 \\ \nonumber && 
 + 4095937462034066852819132682453485174633/1707679319276753096980615588193343897600 \; Y_4 \\ \nonumber && 
 + 8535263/1720320 \; Y_0 + 8535263/6881280 \; F_0 \;  (2\pi)^{-2} \\ \nonumber && 
 - 1051450108870580974979233285323814292021/200903449326676834938895951552158105600 \; (2\pi)^{-2} ; \\ \nonumber 
&& J(4,-2) =  775/124032 \; Y_{11} + 155/3808 \; Y_{10} + 2315881/1116288 \; Y_9 - 775/19152 \; Y_8 \\ \nonumber && 
 - 535775/229824 \; Y_7 + 155/1302336 \; Y_5 + 1268159/2604672 \; Y_4 + 4 \; Y_0 \\ \nonumber && 
 - 256019/306432 \;  (2\pi)^{-2} + \; F_0 \;  (2\pi)^{-2}; \\ \nonumber && 
  J(5,-2) = - 35198207035/161531338752 \; Y_{11} - 7039641407/4959295488 \; Y_{10} \\ \nonumber && 
 - 1590044736319/1453782048768 \; Y_9 + 35198207035/24942339072 \; Y_8 \\ \nonumber && 
 + 824164813817/299308068864 \; Y_7 - 7039641407/1696079056896 \; Y_5 \\ \nonumber && 
 - 3786323755001/3392158113792 \; Y_4 - 4 \; Y_0 + (886011822725/399077425152 - \; F_0 )\; (2\pi)^{-2}; \\ \nonumber && 
  J(6,-2) = 11638328158211917/35061346264154112 \; Y_{11} \\ \nonumber && 
 + 11638328158211917/5382224207216640 \; Y_{10} + 1835049614849615189/1577760581886935040 \; Y_9 \\ \nonumber && 
 - 11638328158211917/5413884349612032 \; Y_8 - 1118098215139974931/324833060976721920 \; Y_7 \\ \nonumber && 
 + 11638328158211917/1840720678868090880 \; Y_5 + 5661105033256044691/3681441357736181760 \; Y_4 \\ \nonumber && 
 + 181/40 \; Y_0 + (181/160 \; F_0 - 1412418713096881687/433110747968962560 )\; (2\pi)^{-2} ; \\ \nonumber && 
  J(7,-2) = - 16095494538985786829003/34246240086205057204224 \; Y_{11} \\ \nonumber && 
 - 16095494538985786829003/5257098258847267553280 \; Y_{10} \\ \nonumber && 
 - 1220878670422477200084451/1541080803879227574190080 \; Y_9 \\ \nonumber && 
 + 16095494538985786829003/5288022366252251480064 \; Y_8 \\ \nonumber && 
 + 1258484886934090324074869/317281341975135088803840 \; Y_7 \\ \nonumber && 
 - 16095494538985786829003/1797927604525765503221760 \; Y_5 \\ \nonumber && 
 - 1435246543741057501042993/719171041810306201288704 \; Y_4 - 2447/480 \; Y_0  \\ \nonumber &&  
 + (1817967522495901075073489/423041789300180118405120 - 2447/1920 \; F_0 )\;(2\pi)^{-2}; \\ \nonumber && 
  J(8,-2) = 97496623093573646692443907729/156100389651177822826920935424 \; Y_{11} \\ \nonumber && 
 + 97496623093573646692443907729/23962779113119402626939617280 \; Y_{10} \\ \nonumber && 
 + 703823889446906512753183459525/1404903506860600405442288418816 \; Y_9 \\ \nonumber && 
 - 97496623093573646692443907729/24103736637314222642392203264 \; Y_8 \\ \nonumber && 
 - 6768519700158561511510642407343/1446224198238853358543532195840 \; Y_7 \\ \nonumber && 
 + 97496623093573646692443907729/8195270456686835698413349109760 \; Y_5 \\ \nonumber && 
 + 41525287801127597648514834574639/16390540913373671396826698219520 \; Y_4 \\ \nonumber && 
 + 79489/13440 \; Y_0 + 79489/53760 \; F_0 \;  (2\pi)^{-2} \\ \nonumber && 
 - 2132153450344177531160317231847/385659786197027562278275252224 \;  (2\pi)^{-2} ; \\ \nonumber && 
  J(9,-2) = - 184720162027043790012174961950856949/225883507840840356743467670395944960 \; Y_{11} \\ \nonumber && 
 - 184720162027043790012174961950856949/34675099887848300377286703788851200 \; Y_{10} \\ \nonumber && 
 - 1182390536568106894749732117915566173/10164757852837816053456045167817523200 \; Y_9 \\ \nonumber && 
 + 184720162027043790012174961950856949/34879071063659172732447213811138560 \; Y_8 \\ \nonumber && 
 + 11666341875097592955381754264025444939/2092744263819550363946832828668313600 \; Y_7 \\ \nonumber && 
 - 184720162027043790012174961950856949/11858884161644118729032052695787110400 \; Y_5 \\ \nonumber && 
 - 76075734513749258577040135726310510027/23717768323288237458064105391574220800 \; Y_4 \\ \nonumber && 
 - 501267/71680 \; Y_0 - 501267/286720 \; F_0 \;  (2\pi)^{-2} \\ \nonumber && 
 + 19693107617724772864308858550443517679/2790325685092733818595777104891084800 \; (2\pi)^{-2}; \nonumber 
\eea
\newpage

{\Large \bf Appendix 11.}
\begin{verbatim}
J(p,q) = -(  6*D(p - 1,q + 4,4)*( - p^2*q - 3*p^2 + 5*p*q + 15*p - 6*q - 18) 
+ 6*D(p,q + 3,4)*( - p^3 + 6*p^2 - 11*p + 6)
 + 21*D(p-2,q+5,3)*(p*q^2 + 7*p*q + 12*p - 3*q^2 - 21*q - 36) 
 + 48*D(p - 1,q + 4,3)*(p^2*q + 3*p^2 - 5*p*q - 15*p + 6*q + 18) 
 + 6*D(p-1,q+3,3)*(p-2)*(p-3)*(p+4*q+8) 
 + 27*D(p,q + 3,3)*(p^3 - 6*p^2 + 11*p - 6)
 + 18*D(p,q+ 2,3)*(p^3 - 6*p^2 + 11*p - 6) 
 + 12*D(p-3,q+6,2)*( - q^3 - 12*q^2 - 47*q - 60) 
 + 18*D(p-2,q+5,2)*( - p*q^2 - 7*p*q - 12*p + 3*q^2 + 21*q + 36) 
 + D(p - 2,q + 4,2)*( - 8*p^2*q - 24*p^2
    - 63*p*q^2 - 346*p*q - 471*p + 189*q^2 + 1110*q + 1629) 
 + 12*D(p-1,q+4,2)*( - p^2*q - 3*p^2 + 5*p*q + 15*p - 6*q - 18) 
 + 2*D(p-1,q+3,2)*( - 10*p^3 - 68*p^2*q - 109*p^2
    + 340*p*q + 735*p - 408*q - 954) 
 + 6*D(p,q + 3,2)*( - p^3 + 6*p^2 - 11*p + 6) 
 + 18*D(p - 1,q + 2,2)*( - p^3 - 2*p^2*q + 3*p^2 + 10*p*q + 4*p - 12*q - 12) 
 + 61*D(p,q + 2,2)*( - p^3 + 6*p^2 - 11*p + 6) 
 + 18*D(p,q + 1,2)*( - p^3 + 6*p^2 - 11*p + 6) 
 + 24*D(p-3,q+6,1)*( - q^3 - 12*q^2 - 47*q - 60) 
 + 6*D(p-3,q+5,1)*( - 7*p*q^2 - 49*p*q - 84*p
 + 4*q^3 + 67*q^2 + 321*q + 468) 
 + 120*D(p - 2,q + 5,1)*( - p*q^2 - 7*p*q - 12*p + 3*q^2 + 21*q + 36) 
 + 2*D(p - 2,q + 4,1)*( - 80*p^2*q - 240*p^2
    + 39*p*q^2 + 707*p*q + 1770*p - 117*q^2 - 1401*q - 3150) 
 + 264*D(p - 1,q + 4,1)*( - p^2*q - 3*p^2 + 5*p*q + 15*p - 6*q - 18) 
 + D(p - 2,q + 3,1)*( - 10*p^3 + 16*p^2*q + 130*p^2
    + 63*p*q^2 + 251*p*q - 21*p - 189*q^2 - 897*q - 837) 
 + 2*D(p - 1,q + 3,1)*( - 47*p^3 + 92*p^2*q
    + 592*p^2 - 460*p*q - 2067*p + 552*q + 2142) 
 + 168*D(p,q + 3,1)*( - p^3 + 6*p^2 - 11*p + 6) 
 + 2*D(p - 1,q + 2,1)*(30*p^3 + 64*p^2*q - 69*p^2
    - 320*p*q - 225*p + 384*q + 486) 
 + 106*D(p,q + 2,1)*(p^3 - 6*p^2 + 11*p - 6) 
 + 6*D(p - 1,q + 1,1)*(3*p^3 + 4*p^2*q - 15*p^2 - 20*p*q + 18*p + 24*q) 
 + 31*D(p,q + 1,1)*(p^3 - 6*p^2 + 11*p - 6) 
 + 6*D(p,q,1)*(p^3 - 6*p^2 + 11*p - 6) 
 + 24*J(p - 4,q + 6)*(q^3 + 12*q^2 + 47*q + 60) 
 + 24*J(p - 3,q + 5)*(5*p*q^2 + 35*p*q + 60*p - 17*q^2 - 119*q - 204) 
 + 2*J(p - 3,q + 4)*(20*p^2*q + 60*p^2 + 21*p*q^2
    + p*q - 186*p - 6*q^3 - 129*q^2 - 423*q - 270) 
 + 24*J(p - 2,q + 4)*(11*p^2*q + 33*p^2 - 63*p*q - 189*p + 90*q + 270) 
 + 4*J(p - 2,q + 3)*(30*p^3 + 20*p^2*q - 190*p^2
    - 15*p*q^2 - 199*p*q + 231*p + 45*q^2 + 417*q + 207) 
 + 24*J(p - 1,q + 3)*(15*p^3 - 104*p^2 + 235*p - 174) 
 + J(p - 2,q + 2)*(10*p^3 - 8*p^2*q - 106*p^2
    - 21*p*q^2 - 52*p*q + 240*p+63*q^2+228*q-36)
 + 2*J(p - 1,q + 2)*( - 73*p^3 - 66*p^2*q + 346*p^2
    + 330*p*q - 343*p - 396*q - 114) 
 + 192*J(p,q + 2)*( - p^3 + 6*p^2 - 11*p + 6) 
 + 4*J(p-1,q+1)*( - 10*p^3 - 10*p^2*q + 53*p^2 + 50*p*q - 75*p - 60*q + 18) 
 + 20*J(p,q+1)*(p^3 - 6*p^2 + 11*p - 6) 
 + 6*J(p-1,q)*( - p^3 - p^2*q + 6*p^2 + 5*p*q - 11*p - 6*q + 6) 
 )/3/(p-1)/(p-2)/(p-3);
\end{verbatim}

\end{document}